\documentclass[prb,twocolumn,floatfix]{revtex4}
\usepackage{amsmath}
\usepackage{graphicx}
\usepackage{latexsym}
\usepackage{amsfonts}
\usepackage{amssymb}
\usepackage{bm}
\usepackage{longtable}

\begin{document}
\title{Self and Tracer Diffusion of Polymers in Solution }
\author{George D. J. Phillies}
\email[To whom inquiries should be sent ] {phillies@wpi.edu}
\affiliation{Department of Physics, Worcester Polytechnic Institute,Worcester, MA 01609}

\begin{abstract}

The literature on self-diffusion of polymers in solution, and on tracer 
diffusion of probe polymers through solutions of matrix polymers, is 
systematically reviewed.  Virtually the entirety of the published 
experimental data on the concentration dependence of polymer self--- and probe--diffusion is represented well by the same functional form. This form is the stretched exponential $\exp(- \alpha c^{\nu})$, where  $c$ is 
polymer concentration, $\alpha$ is a scaling prefactor, and $\nu$ is a scaling 
exponent.  

Correlations of the scaling parameters with polymer molecular weight, 
concentration, and size are examined, and compared with predictions based on the form's hydrodynamic and renormalization-group derivations.  $\alpha$ increases markedly with polymer 
molecular weight, namely $\alpha \sim M^{x}$ for $x \approx 1$.  $\nu$ is 
$\approx 0.5$ for large polymers ($M$ larger than 400 kDa or so), but increases 
toward 1.0 or so at smaller $M$.  Scaling parameters for the diffusion of star 
polymers do not differ significantly from scaling parameters for the diffusion 
of linear chains of equal size. 

\end{abstract} 

\maketitle

\section{Introduction}

The objective of this review is to present the experimental phenomenology for
polymer self-diffusion and for diffusion of tracer polymers through polymer 
solutions.  We proceed by examining how the diffusion 
coefficient depends on  
polymer concentration $c$, polymer molecular weight $M$, solvent quality, and other variables.  
We examine how rapidly a probe polymer of molecular 
weight $P$ diffuses through a matrix polymer of molecular weight $M$, perhaps 
as the concentrations of the matrix and probe polymers are varied. 
After establishing the phenomenological behavior, we compare the phenomenology with various classes of theoretical  model.  
 
This is not a review of the extremely extensive theoretical literature on 
polymer diffusion in solution.  For such reviews, note works of Graessley\cite{graessley1974a}, 
Tirrell\cite{tirrell1984a}, Pearson\cite{pearson1987a}, 
Skolnick\cite{skolnick1989a}, Lodge\cite{lodge1990a}, and (more recently but less directly) McLeish\cite{mcleish2002}.  Recent 
papers by Schweizer and collaborators\cite{fuchs1996a,schweizer1989a,schweizer1995a} include 
extensive references to the more recent literature. We will note classes of models, but not their underlying derivations.  We also do not consider melt systems, or polymer mutual diffusion.

In comparing phenomenology with model predictions, it will remain critical to  
distinguish between properties that are consistent with a particular model, but do not actually 
prove it, and properties that require or refute the correctness of a particular model.  
For example, Skolnick\cite{skolnick1987a} has shown that the nominal signature of reptation
$D_{s} \sim M^{-2}$ is found in computer models of polymers, in a model system in which the chains are very certainly not reptating.  A finding $D_{s} \sim M^{-2}$ is thus consistent with many tube-type solution models, but does not prove their correctness.  On the other hand, most tube-type models require that polymer chains move via constrained diffusion for shorter times.  In these models, the shorter-time mean-square displacement of a single chain must scale more slowly than linearly with time.  If the shorter-time mean-square displacement were instead linear in elapsed time, many constrained-diffusion models would be rejected by experiment.

Section 
II presents a theoretical background for the experimental papers reviewed here.  
Sections III and IV treat, respectively,  data on (i) self-diffusion 
coefficients of polymers of solution and (ii) data on probe diffusion coefficients in
polymer matrix solutions.
Section V briefly remarks upon other papers on polymer self- and 
probe-diffusion that do not lend themselves to the analytic approach applied 
here. A systematic study of the phenomenological parameters obtained 
by our analysis appears in Section VI.  Section VII summarizes 
conclusions.  Tables of fitting parameters appear as Appendices.  

\section{Theoretical Background}

This Section presents a consistent nomenclature for polymer diffusion coefficients and 
describes methods whereby which such diffusion coefficients are measured.  A 
short description is given of the literature on diffusion in multimacrocomponent solutions.  
The literature 
in question provides the fundamental basis for interpreting the 
experimental measurements.  Finally, phenomenological classes of models for polymer 
dynamics are identified.  

\subsection{Nomenclature for Diffusion Coefficients}

This section sets out a consistent nomenclature
for the diffusion coefficients that we are reviewing.  In a solution containing a solvent and one 
macromolecular species, two physically distinct diffusion coefficients 
usefully characterize macromolecule diffusion.  One of these, the 
two-particle or {\em mutual} 
diffusion coefficient $D_{m}$, describes via Fick's law
\begin{equation}
   \vec{J} = D_{m} \vec{\nabla} c
   \label{eq:fickslaw}
\end{equation}
the relaxation of a concentration gradient.  Here $\vec{J}$ is the diffusion 
current and $c$ is the local instantaneous macromolecule concentration.  The 
other diffusion coefficient, the single-particle or {\em self} diffusion 
coefficient $D_{s}$, describes the diffusion of a single macromolecule through 
a uniform solution of elsewise identical macromolecules.  In a simple macromolecule 
solution, quasi-elastic light scattering spectroscopy (QELSS) measures the 
mutual diffusion coefficient of the macromolecules\cite{phillies1974a,phillies1974b}.  Pulsed-field-gradient 
nuclear magnetic resonance (PFGNMR) measures the self-diffusion coefficient.  

Solutions containing a solvent and two macromolecule species show a more 
complicated diffusive behavior.  Theoretical
treatments of this issue are described in the next section.
In general, in two-macrocomponent solutions the temporal evolution of transient concentration 
fluctuations is described by two relaxation times, each describing the 
relaxation of a coupled mode involving the concentrations of both macromolecule 
species.  An interesting special case arises if one macrocomponent, the {\em 
probe}, is adequately dilute, while the other macrocomponent, the {\em matrix}, 
may be either dilute or concentrated.  In this special case, the diffusion of 
the probe species 
is governed by a single-particle diffusion coefficient, namely the 
{\em probe} diffusion coefficient $D_{p}$.  In the special case that the matrix 
species and solvent are isorefractive, so that the index of refraction is 
independent of the relative amounts of matrix polymer and solvent,  the matrix 
species scatters next to no light.  If in this special case light scattering 
by a dilute probe species dominates scattering by the solution (which 
further requires 
that solvent scattering be sufficiently weak), the diffusion 
coefficient measured by QELSS is\cite{phillies1974b,phillies1975a} the probe diffusion 
coefficient $D_{p}$.  

A variety of physical techniques have been used to measure probe 
diffusion through a polymer solution.  In the fluorescence recovery after 
photobleaching (FRAP) technique, a small fluorescent label is attached 
to the probes.  An intense pulse of light (the `pump' laser beam) is then used 
to destroy (``bleach") the fluorescent labels in some regions of the solution, the 
regions sometimes being defined via a holographic grating.  Much weaker laser 
illumination (the `probe' beam) is then used to monitor the recovery of the 
fluorescence intensity as unbleached fluorescent molecules diffuse back into 
the regions(s) in which the bleaching occurred.   

An alternative to FRAP is Forced Rayleigh Scattering (FRS). In an FRS 
experiment, an intense laser pulse is used to generate a holographic brightness 
grating in solution.  The grating selectively alters a photosensitive part of 
the probe molecule, thereby creating an index of refraction grating in the 
solution.  A much weaker probe beam is then used to monitor the diffusive 
relaxation of the induced grating.  Significant complications may 
arise if the 
photomodified and non-photomodified forms of the probe species differ 
significantly in their diffusive properties.  If the probe species is dilute, 
with either FRAP or FRS the time dependence of the recovery profile is 
determined by the single-particle (probe-) diffusion coefficient of the probe 
molecules\cite{phillies1975a}.  

A physically-distinct alternative to FRAP and FRS is fluorescence correlation 
spectroscopy (FCS), in which optical methods are used to measure 
fluctuations in the number of fluorescent molecules in a small volume of space.  
 In nondilute solution the 
diffusion coefficient measured by fluorescence correlation spectroscopy (FCS) 
varies from $D_{p}$ to $D_{m}$ as the fraction of macromolecules that bear  
fluorescent tags is varied from small to large\cite{phillies1975a}.  

When is a probe species dilute?  In general, the careful experimenter makes an 
adequate control study of the effect of varying the probe concentration 
$c_{p}$.  $D_{p}$ is the extrapolation of the measured probe diffusion 
coefficient to zero probe concentration.  In some systems $D_{p}$ is 
substantially independent of $c_{p}$, and no extrapolation is needed.  In other 
systems (see below) $D_{p}$ depends significantly on $c_{p}$, and 
extrapolation to $c_{p} \rightarrow 0$ 
must be performed.   Several authors have 
measured the initial slope 
$k_{d}$ of the dependence of $D_{p}$ on probe concentration, 
and the effect of matrix concentration on $k_{d}$.  
It is also possible to study--most published work has used QELSS--the 
relaxation of concentration fluctuations in a ternary 
solvent:macromolecule:macromolecule system in which neither macrocomponent is 
dilute.  As reviewed below, such studies give information on the diagonal--- and 
cross-diffusion coefficients of the two species and on their thermodynamic 
interactions.  

Strictly speaking, if FRAP or FRS is used to study the diffusion of a dilute 
labelled polymer through a 
matrix polymer solution, the labelled polymer 
differing from the 
matrix in that a fluorescent small-molecule label has been attached, 
the measured diffusion coefficient should be termed 
a probe diffusion coefficient, because the molecules being tracked are chemically distinguishable from the molecules not being tracked.  However, for the purposes of this review, we 
distinguish between (i) the case in which the probe and matrix polymers are 
substantially distinct in molecular weight or chemical nature, and (ii) 
the case in which the probe and matrix polymers are very nearly the same except 
for the presence of the label.  The phrase 'probe diffusion coefficient' is 
reserved for case (i), while measurements under case (ii) are here treated as 
determinations of the self diffusion coefficient.

The term {\em tracer} diffusion coefficient is synonymous with 
{\em single-particle} diffusion coefficient, and includes both the self and 
probe diffusion coefficients as special cases.  The {\em interdiffusion} and 
{\em cooperative diffusion} coefficients characterize  the relaxation times in 
a ternary system in which neither macrocomponent is dilute.  In the following 
the latter two terms are almost never used.  

\subsection{General Theory of Diffusion in Multimacrocomponent Solutions}

What is the basis for using spectroscopic methods such as QELSS, FRAP, FRS, or FCS to measure $D_{s}$ or $D_{p}$?  In each case, there is a physical theory that links the correlation function determined spectroscopically to fluctuations in microscopic variables that describe the liquid.  For example, for FCS the fluctuations in the fluorescent light intensity $I_{f}(t)$ arise from fluctuations in the number of fluorescent molecules in the scattering volume.  The time correlation function $C_{f}(\tau) = \langle I_{f}(t) I_{f}(t + \tau) \rangle$ of the fluorescent intensity follows faithfully the time correlation function $C_{N}(\tau) = \langle N(t) N(t+\tau) \rangle$ of the (fluctuating) number $N(t)$ of labelled molecules in the scattering volume, so to determine the time dependence of $C_{f}(\tau)$ one only needs to determine $C_{N}(t)$ and vice versa.

For QELSS, the spectrum is determined\cite{berne} by the field correlation function $g^{(1)}(q,t)$ of the scattered light, which is in turn determined by the motion of the scattering molecules via
\begin{equation}
     g^{(1)}(q,\tau) = \left\langle \sum_{i,j=1}^{N} \alpha_{i} \alpha_{j} \exp[i \mathbf{q} \cdot (\mathbf{r}_{i}(t+\tau) - \mathbf{r}_{j}(t))] \right\rangle,
     \label{eq:g1def}
\end{equation}
where $i$ and $j$ label two of the $N$ scatterers, $\mathbf{r}_{i}(t)$ is the position of scatterer $i$ at time $t$, $\alpha_{j}$ is the scattering cross-section of particle $j$, and $\mathbf{q}$ is the scattering vector.  For the special case in which the scattering particles are dilute (non-scattering particles do not need to be dilute), correlations between the positions of two distinct scatterers at two times vanish.  The correlation function for probe scattering then reduces to  
\begin{equation}
     g^{(1)}_{P}(q,\tau) = \left\langle \sum_{i=1}^{N} \alpha_{i}^{2} \exp[i \mathbf{q} \cdot \Delta \mathbf{r}_{i}(\tau)] \right\rangle,
     \label{eq:g1defP}
\end{equation}
in which $\Delta \mathbf{r}_{i}(\tau) = \mathbf{r}_{i}(t+\tau) - \mathbf{r}_{i}(t))$.

There are three general approaches to calculating $g^{(1)}_{P}(q,\tau)$.  First, scatterers can be treated as objects having hydrodynamic and direct, e.g., excluded volume, interactions, and the effect of the intermacromolecular interactions on $d  g^{(1)}_{P}(q,\tau)/d \tau$ can be calculated.  Second, it can be recognized that $\sum_{i}^{N} \exp[i \mathbf{q} \cdot (\mathbf{r}_{i}(t)]$ is the $q^{\rm th}$ spatial Fourier component of the scatterer concentration at time $t$, and semicontinuum hydrodynamic methods and the Onsager regression hypothesis can be used calculate the average temporal evolution of a fluctuation.  Third, it can be proposed that the scatterers perform simple Brownian motion, with the probability distribution function for $\Delta \mathbf{r}_{i}(\tau)$ being applied to calculate $g^{(1)}_{P}(q,\tau)$.

Light scattering spectra of non-ideal single-component macromolecule solutions, including direct interactions and hydrodynamic interactions at the Oseen level, were initially calculated 
by Altenberger and Deutsch\cite{altenberger1973a}. Calculations of light scattering spectra 
of non-dilute many-component macromolecule solutions soon 
followed\cite{phillies1973a,phillies1974a,phillies1974b}.  In the latter
papers, diffusion coefficients were modeled on the lines of Kirkwood, et 
al.\cite{kirkwood1960a}, the diffusion coefficients being written as products 
of thermodynamic derivatives $(\partial \mu_{i}/\partial c_{j})$ obtained from the intermacromolecular forces, and 
phenomenological transport coefficients $\Omega_{ij}$ (here $i$ and $j$ label 
chemical species).  If neither macromolecular species was dilute, the light 
scattering spectrum obtained by QELSS is
predicted\cite{phillies1973a,phillies1974a} to contain two relaxation modes.  
Even if only one of the two macromolecule species scattered light, both modes 
are predicted to be visible, in general, in the light scattering 
spectrum\cite{phillies1974b}.  

A subsequent paper\cite{phillies1975a} extended these findings 
to the special case explored e.g., by fluorescence correlation spectroscopy, in 
which some macromolecules are tagged so that their movements can be tracked.  
For the case that the tagged macromolecules are dilute, it was 
shown\cite{phillies1975a} that fluorescence correlation spectroscopy and other 
equivalent techniques measure the self-diffusion coefficient of the tagged probe macromolecules diffusing through the unlabelled matrix macromolecules. 

A Smoluchowski approach was used by Jones\cite{jones1979a} to examine interacting spherical polymers, including the special case that one of the polymer species is dilute and tagged so that its motions could be observed.
In the special case, the measured diffusion coefficient was predicted by Jones 
to be determined only by the hydrodynamic interactions of the tracer polymers 
with their (predominantly matrix-polymer) neighbors.  While Jones does not use 
the same nomenclature, the diffusion coefficient of the tracers calculated in ref.\ \onlinecite{jones1979a} 
may be recognized as 
the single-particle diffusion coefficient.  The hydrodynamic approach culminates in analyses of Carter, et al.\cite{carter1985a} and Phillies\cite{phillies1995c} of mutual and tracer diffusion coefficients, including full hydrodynamic and direct interactions between interacting diffusing particles and reference frame 
issues.  

Extensive studies of diffusion of interacting polymers using semicontinuum and related means were made by Akcasu, Benmouna, Cohen and others. In 1987, Benmouna, et 
al.\cite{benmouna1987a} calculated dynamic scattering from a solution 
containing two polymer species, including excluded volume terms with a Flory 
interaction parameter but ignoring hydrodynamic interactions between polymer 
chains.  Hydrodynamic interactions were neglected in the sense that 
cross-species transport coefficients were assumed to vanish.  The assumption 
that the hydrodynamic parts $\Omega_{ij}$ of the cross-diffusion tensor vanish  
cannot be true simultaneously in the solvent- and volume-fixed reference 
frames\cite{kirkwood1960a}.  Polymer solutions can be very concentrated, so 
neglect of reference frame issues in entirely correct calculations may lead to complications in the physical interpretation of 
an assumption that hydrodynamics has been neglected.  

Benmouna, et al.\cite{benmouna1987a} examined the special case of two species identical except for 
their optical scattering cross-sections, showing results consistent with those 
of Phillies\cite{phillies1974b,phillies1975a}, notably:
If 
the matrix species scatters weakly, an interdiffusion 
mode describing single-chain motion is dominant at low probe concentrations.
There is a thermodynamic regime 
relative to the spinodal where the diffusion equations describe phase 
separation.  The Benmouna model was then extended to treat solutions of 
copolymers\cite{benmouna1987b}, still with neglect of hydrodynamic 
interactions, and to treat homopolymer: copolymer mixtures\cite{benmouna1987c}.  

In the same period, Foley and Cohen\cite{foley1987a} analyzed concentration 
fluctuations in polymer: polymer: solvent mixtures, using an ornate 
Flory-Huggins form for the thermodynamic free energy of the mixture, again
neglecting interchain hydrodynamic interactions.   The case of a solvent 
isorefractive with the matrix polymer, in the presence of a dilute scattering 
species, was treated.  Foley and Cohen also examined systems in which both 
polymer species were nondilute, predicting that in this case the relaxation 
spectrum is characterized by three distinct relaxation times. In contrast, other calculations on similar models predict only two relaxation times.

Roby and Joanny\cite{roby1992a} improved the model of Benmouna, et 
al.\cite{benmouna1987a} by incorporating interchain hydrodynamic interactions 
and by improving the model for direct chain-chain interactions.  At elevated 
concentrations, the correctness of the reptation model was assumed.  The effect 
of reptation dynamics on concentration fluctuations was estimated with the
approximation that a solution is effectively a polymer melt in which 
mesoscopic polymer-solvent blobs play the role taken in polymer melts by
monomer units. For simplicity, the model 
was restricted to systems containing equal amounts of two species having the 
same molecular weight.  These restrictions are nontrivial to remove but exclude 
analysis of tagged-tracer experiments.  A model calculation incorporating a
similar picture, but without the restrictions, was reported by 
Hammouda\cite{hammouda1993a}.  

An alternative analysis of a ternary solution of two polymers and a solvent was
presented by Wang\cite{wang1997a}, who followed much 
of the earlier work on polymer 
solutions in assuming that the cross terms in the mobility matrix vanish.  Wang 
systematically analysed a variety of general and special cases, showing a range 
of interesting parameters that can be determined from light scattering spectra 
if the accuracy of his model is assumed.  In addition to the tracer case (one 
polymer:solvent pair isorefractive, visible polymer dilute), Wang analysed 
the special case ``zero average contrast", in which the polymer refractive 
increments are of opposite sign, so that concentration fluctuations, that change 
the total polymer concentration without changing the local polymer composition, 
scatter no light.  Wang showed that the QELSS spectrum of a 
zero-average-contrast system is almost always bimodal, though it may happen 
that one of the modes is much weaker than the other. 

Treatments of the field correlation function based on a Brownian motion description can be traced back to Berne and Pecora\cite{berne}, who treat the light scattering spectrum of a solution of dilute, \emph{noninteracting} Brownian particles.  The motion of such particles is described as a series of random, uncorrelated steps, in which case from the Central Limit Theorem the probability distribution for particle displacements is
\begin{displaymath}
    G_{s}(\Delta \mathbf{R}, t) = 
\end{displaymath}
\begin{equation}
    \left[ \frac{2 \pi}{3} \langle (\Delta \mathbf{R})^{2} \rangle\right]^{-3/2} \exp\left[ - 3 (\Delta R)^2/ 2 \langle (\Delta \mathbf{R})^{2} \rangle\right],
     \label{eq:deltaR}
\end{equation}
where the mean-square particle displacement is related to the diffusion coefficient by 
\begin{equation}
   \langle (\Delta \mathbf{R})^{2} \rangle = 6 D t.
   \label{eq:deltaRval}
\end{equation}

Combining eqs \ref{eq:g1def} and \ref{eq:deltaR},
\begin{displaymath}
     g^{(1)}_{P}(q,\tau) \sim \int d \Delta \mathbf{r} \exp[i \mathbf{q} \cdot \Delta \mathbf{r}]  
\end{displaymath}     
\begin{equation}     
     \times \left[ \frac{2 \pi}{3} \langle (\Delta \mathbf{R})^{2} \rangle\right]^{-3/2} \exp[ - 3 (\Delta R)^2/ 2 \langle (\Delta \mathbf{R})^{2} \rangle ],
     \label{eq:g1calc}
\end{equation}
which using eq \ref{eq:deltaRval} leads to
\begin{equation}
     g^{(1)}_{P}(q,\tau) \sim \exp(- q^{2} D \tau).
     \label{eq:g1calc2}
\end{equation}

For Brownian particles in low-viscosity small-molecule solvents, the Stokes-Einstein equation 
\begin{equation}
     D = \frac{k_{B}T}{6 \pi \eta R},
     \label{eq:SEeq}
\end{equation}
in which $k_{B}$ is Boltzmann's constant, $T$ is the absolute temperature, $\eta$ is the solution viscosity, and $R$ is the particle radius, generally predicts accurately the particle diffusion coefficient.  For diffusion in polymer solutions, two \emph{ad hoc} extensions of this form are encountered:  

First, it might be the case that $D$ is not predicted accurately by eq \ref{eq:SEeq}.  In this case, one could formally define a microviscosity $\eta_{\mu}$ as
\begin{equation}
     \eta_{\mu} = \frac{k_{B}T}{6 \pi D R},
     \label{eq:etamudef}
\end{equation}
and compares $\eta_{\mu}$ with the macroscopically-measured viscosities $\eta$ of the solution and $\eta_{o}$ of the solvent.  The microviscosity is more commonly applied to describe diffusion of mesoscopic globular probe particles\cite{kiril1}, rather than to treat the diffusion of random-coil polymers in solution.

Second, light scattering and other relaxation spectra are not always single exponentials.  No matter what functional form $g^{(1)}_{P}(q,\tau)$ has, one may formally write
\begin{equation}
   g^{(1)}_{P}(q,\tau) \sim \exp(- q^{2} D(\tau) \tau)
     \label{eq:g1calcDt}
\end{equation} 
as the \emph{definition} of $D(\tau)$.
In this equation, $D(\tau)$ formally appears to be a time-dependent diffusion coefficient, which equally formally defines a frequency-dependent diffusion coefficient such as
\begin{equation}
     D(\omega) = \int_{0}^{\infty}d\tau \ \exp(- i \omega \tau) D(\tau),
     \label{eq:Domega}
\end{equation}
and via several slightly different paths a frequency-dependent microviscosity
\begin{equation}
     \eta_{\mu}({\omega}) =\frac{k_{B}T}{6 \pi D(\omega) R}.
     \label{eq:etaomega}
\end{equation}
The real and imaginary parts of $\eta_{\mu}(\omega)$ can be brought into correspondence with the storage and loss moduli $G'(\omega)$ and $G''(\omega)$.

The second extension has serious physical difficulties.  In particular, comparing eqs \ref{eq:deltaRval}, \ref{eq:g1calc2}, and \ref{eq:g1calcDt}, and noting from reflection symmetry that terms odd in $q$ vanish, one finds that the extension assumes that
\begin{equation}
   \left \langle \exp[- q^{2} (\Delta \mathbf{r})^{2} ] \right\rangle = \exp[- q^{2} \left\langle  (\Delta \mathbf{r})^{2} \right\rangle ].
     \label{eq:rsquaregaussian}
\end{equation} 

Equation \ref{eq:rsquaregaussian} would be correct if sequential random changes in $\sum_{j=1}^{N} \exp[i \mathbf{q} \cdot \mathbf{r}_{j}]$ were described by a Gaussian random process, because in that case (and only in that case) the average on the lhs of eq \ref{eq:rsquaregaussian} would be entirely described by the mean-square average displacement via the expression $\left\langle  (\Delta \mathbf{r})^{2} \right\rangle$ seen on the rhs of this equation.   Brownian motion, in which particle displacements are described by the Langevin equation, is an example of a dynamics that generates a Gaussian random process for which eq \ref{eq:rsquaregaussian} is correct.  For extensive details, see ref \onlinecite{berne}.

Omitted from, but critical to, the above discussion are the consequences of Doob's First Theorem\cite{doob}.  Doob treated the joint expectation value--what we would now call the correlation function--of random variables including those following Langevin dynamics.  If eq \ref{eq:rsquaregaussian} is correct, it is an \emph{inescapable} consequence of Doob's Theorem that the relaxation spectrum is a \emph{single} exponential.  Conversely, if the spectrum is not a single exponential, then from Doob's theorem 
\begin{equation}
     g^{(1)}(\mathbf{q}, \tau) \neq \exp[- q^{2}  \langle  (\Delta \mathbf{r})^{2} \rangle ].
     \label{eq:doob}
\end{equation}  

Berne and Pecora's text\cite{berne} is sometimes incorrectly cited as asserting that eq \ref{eq:rsquaregaussian} is uniformly correct for light-scattering spectra.
The analysis in Berne and Pecora\cite{berne}, which correctly obtains eq \ref{eq:rsquaregaussian}, refers only to a special-case system.  In the system that these authors correctly analyzed, particle displacements are governed by the simple Langevin equation, and particle displacements in successive moments are uncorrelated.  That is, this analysis refers to systems in which particle motion is observed only over times \emph{much longer} than any viscoelastic relaxation times. 

If particle motion were observed over times shorter than the viscoelastic relaxation times, which in a concentrated polymer solution might be 1 second or more: Particle displacements in successive moments would be correlated.  Equation \ref{eq:rsquaregaussian} would not be correct. $\log[g^{(1)}(\mathbf{q},\tau)]/q^{2}$ would not be proportional to the mean-square particle displacement during $\tau$.  The path from eq \ref{eq:g1def} to eq \ref{eq:etaomega} might be an interesting heuristic, but would not be consistent with the properties of particles executing Brownian motion.  Of course, with respect to polymer dynamics the interest in eqs \ref{eq:g1calcDt}--\ref{eq:rsquaregaussian} has been exactly the study of particle motions at short times, during which viscoelastic effects are apparent in $D(\tau)$, but these are precisely the circumstances under which eq \ref{eq:rsquaregaussian} is incorrect.

In addition to the above, there are analyses of melt systems such as the very 
interesting work of Akcasu and collaborators\cite{akcasu1993a,akcasu1995a}.  
Melt systems are not considered here.  

\subsection{Phenomenological Forms for \\ Comparison with Experiment}

The approach here is to compare experimental measurements of 
$D_{s}$ and $D_{p}$ with the functional forms and $c$, $P$, and $M$ 
dependences predicted by various treatments of polymer dynamics.  There are a very large number of proposed models.  Most 
models fall into two major phenomenological classes, 
distinguished by the functional forms taken to describe $D_{s}(c)$.  This section sketches predictions of classes of model, not including their underlying physical rationales, in preparation for the comparison.

(1) In \emph{scaling-law} models\cite{degennes1979a}, the relationship between
$D_{s}$, $D_{p}$, and polymer properties is described by scaling laws such
as 
\begin{equation}
    D_{s} = D_{1} M^{\gamma} c^{-x},
     \label{eq:Dsscaling}
\end{equation}
where here $\gamma$ and $x$ are scaling exponents, and $D_{1}$ is a scaling 
prefactor, namely the nominal diffusion 
coefficient at unit molecular weight and concentration.  In some cases, scaling 
laws are proposed to be true only over some range of their variables, or only 
to be true asymptotically in some limit.  On moving away from the limit, corrections 
to scaling then arise.   For melts, some models derive a scaling 
law for $D_{s}(M)$ from model dynamics, and then predict numerical values for 
$\gamma$.  (In melts, $D_{s}$ has no concentration 
dependence because $c$ is constant.) For polymer solutions, more typically a scaling-law form is 
postulated.  The theoretical objective is then to calculate the exponents 
$\gamma$ and $x$.  

Many scaling-type models propose a transition in solution behavior between a 
lower-concentration dilute regime and a higher-concentration semidilute regime.
Scaling arguments do not usually supply numerical coefficients, so there is 
no guarantee that an interesting transition 
actually occurs at unit value of a predicted
transition concentration $c_{t}$
rather than at, say, $2 c_{t}$.  Correspondingly, the observation that a 
transition is found at $2 c_{t}$ rather than $c_{t}$ is generally in no sense a 
disproof of a scaling model, because in most cases
scaling models do not supply numerical 
prefactors adequate to make a disproof.  (Some level of rationality must be preserved.  If a physical 
model leads to $c_{t}$ as the transition concentration, and the 
nominally corresponding transition is only observed for 30--150 $c_{t}$, and 
then only in some systems, one is entitled to question if the observed 
transition corresponds to the transition implied by the model.)

Two transition concentrations are often identified in the literature. 
The first transition concentration is the overlap concentration $c^{*}$, 
formally defined as the concentration $c^{*} = N/V$ at which $4 \pi R_{g}^3 N/ 
(3 V) = 1$.  Here $N$ is the number of macromolecules in a solution having 
volume $V$ and $R_{g}$ is the macromolecule radius of gyration.  In many cases, 
$c^{*}$ is obtained from the intrinsic viscosity via  $c^{*} = n/[\eta]$ for 
some $n$ in the range 1--4.  The second transition concentration is the entanglement 
concentration $c_{e}$.  
In some papers, the entanglement concentration is obtained from 
a log-log plot of viscosity against concentration by extrapolating an assumed 
low-concentration linear behavior and an assumed higher-concentration
power-law behavior ($c^{x}$ for, e.g., $x = 4$) to an intermediate 
concentration at which the two forms predict the same viscosity; this 
intermediate concentration is taken to be $c_{e}$.  In other papers, the 
entanglement concentration is inferred from the behavior of the viscoelastic 
moduli; in particular, an onset of viscous recovery in the melt or solution is 
taken to mark $c_{e}$.

(2) In {\em exponential} models\cite{adler1980a,phillies1987a}, the 
concentration dependence of $D_{s}$ becomes an exponential or stretched 
exponential in concentration 
\begin{equation} 
    D_{s} = D_{o} \exp(-\alpha c ^{\nu}).
    \label{eq:dsseeq}
\end{equation}
Here $D_{o}$ is the diffusion coefficient in the limit of infinite dilution of 
the polymer, $\alpha$ is a scaling prefactor, and $\nu$ is a scaling exponent; 
$\nu =1$ for simple exponentials.  Under the circumstance that the 
probe and matrix molecular weights $P$ and $M$ differ, an elaborated form of 
the stretched exponential is    
\begin{equation}
   D_{p} = D_{o} P^{-a} \exp( - \alpha c^{\nu} P^{\gamma} M^{\delta}),
   \label{eq:Dsseeq2}
\end{equation}
where $\gamma$, $a$, and $\delta$ are additional scaling exponents, $D_{o}$ now 
represents the diffusion coefficient in the limit of zero matrix concentration 
of a hypothetical probe polymer having unit molecular weight, and $P^{-a}$ describes the 
dependence, on probe molecular weight, of the diffusion coefficient of a dilute probe 
molecule.  On setting $a = \gamma = 0$ and freeing $D_{o}$, eq.\ 
\ref{eq:Dsseeq2} becomes a parameterization of the matrix molecular weight 
dependence of $D_{p}$ for a fixed probe molecular weight.

In the derivations\cite{adler1980a,phillies1987a,phillies1988a,phillies1998a} of the 
stretched-exponential models, functional forms and numerical values for exponents and 
pre-factors were both obtained, subject to various approximations.   
The latter two derivations assume that chain motion is adequately approximated 
by whole-body translation and rotation, which may be appropriate if $P 
\approx 
M$, but which is not obviously appropriate if $P$ and $M$ are substantially 
unequal, since in this case whole-body motion of one chain and local modes of 
the other chain occur over the same distance scale.

Some exponential models\cite{phillies1992a,phillies1995a} also include a 
transition concentration, namely a transition between a lower-concentration 
regime in which some transport coefficients show stretched-exponential 
concentration dependences and a higher-concentration regime in which the same 
transport coefficients show power-law concentration dependences.  The 
transition concentration, which experimentally is sharply 
defined\cite{phillies1992a,phillies1995a}, 
is here denoted $c^{+}$.  The lower-concentration regime is the {\em 
solutionlike} regime; the higher-concentration regime is the {\em meltlike} 
regime. Such transitions are seen in some but not all viscosity 
data\cite{phillies1992a}, generally but not always\cite{phillies1995b} at very 
high concentrations $c[\eta] > 35$.  A solutionlike-meltlike transition is 
very rarely apparent in measurements of $D_{s}$ or $D_{p}$.  

There has been interest in derivations of concentration and other dependences 
of transport coefficients
from renormalization group approaches.  Power-law and exponential forms 
can\cite{altenberger1996a} both follow from a renormalization-group approach, 
depending on the location of the supporting fixed point.  The 
stretched-exponential form is\cite{phillies1998a} an invariant of the 
Altenberger-Dahler\cite{altenberger1996a} Positive-Function Renormalization 
Group.  

Our analysis will examine whether either of these functions describe experiment.  While a power law and a stretched exponential both can represent a range of measurements, on a log-log plot a power law is always seen as a straight line, while a stretched exponential is always seen as smooth curve of nonzero curvature.  Neither form can fit well data that is described well by the other form, except in the sense that in real measurements with experimental scatter a data set that is described well by either function is tangentially approximated over a narrow region by the other function.  

It would also be possible to divide theories of polymer dynamics into classes 
based on assumptions as to the nature of the dominant forces in solution. Assertions
as to the dominant forces are independent of assertions as to the 
functional form used to describe $D_{s}(c)$.  The major forces common to all polymer solutions are the 
excluded-volume force that prevents polymer chains from interpenetrating each 
other, and the hydrodynamic force that creates correlations in the 
displacements of nearby chains.  In some models, excluded-volume forces 
(topological constraints) are assumed to dominate, hydrodynamic interactions 
serving primarily to dress the monomer diffusion coefficient.  In other models, 
hydrodynamic interactions between nearly chains are assumed to dominate, while 
chain-crossing constraints provide at most secondary corrections. 

In addition to the generic forces, chains have chemically-specific interactions 
including in various cases van der Waals, hydrogen bonding, and 
electrostatic forces.  These interactions substantially modulate the 
properties of particular polymers.  However, diffusion coefficients and 
viscoelastic parameters of most neutral polymers show highly-characteristic ``polymeric" 
behaviors, almost independent of the chemical identity of the polymer, implying 
that general polymer properties do not arise from chemically-specific 
interactions.  

In the following, eqs.\ \ref{eq:Dsscaling}--\ref{eq:Dsseeq2}
will be systematically compared with the literature on $D_{s}$ and $D_{p}$.
The following largely treats diffusion by neutral polymers in good and theta solvents. There is rather little data on self-diffusion of random coil polyelectrolytes.  The measurements reviewed here determine diffusion coefficients, not the physical nature of intermolecular forces, so our emphasis is on identifying the class of model, not the type of force, that is significant for solutions.

Comparisons were made via non-linear least-squares fits.  The quantity minimized by the fitting 
algorithm was the mean-square difference between the data and the 
fitting function, expressed as a fraction of the value of the fitting function.  
This quantity is the appropriate choice for minimization if the error in the measurement is some 
constant fraction of the value of the quantity being measured, e.g., if 
regardless of the value of $D_{s}$, $D_{s}$ was measured to 
within 1\%. In some cases, one or more potentially free parameters were held 
constant (``frozen'') during the fitting process. For each fit, the corresponding Table reports the
final fitting error.

\section{Self-Diffusion}

This Section presents measurements of the true self-diffusion coefficient, 
which describes the motion of a labelled chain through a solution of elsewise 
identical chains.  Measurements were primarily made with pulsed-field gradient 
nuclear magnetic resonance (PFGNMR) and forced Rayleigh scattering (FRS).  
Results are presented alphabetically by first author, showing for each paper 
the data and fits to stretched-exponential forms.  
Fitting parameters appear in Table
\ref{table1}.  A more detailed analysis of the fitting coefficients appears in 
Section VI.  Whenever possible, polymer concentrations have been converted to 
grams/liter.  

\begin{figure}

\includegraphics{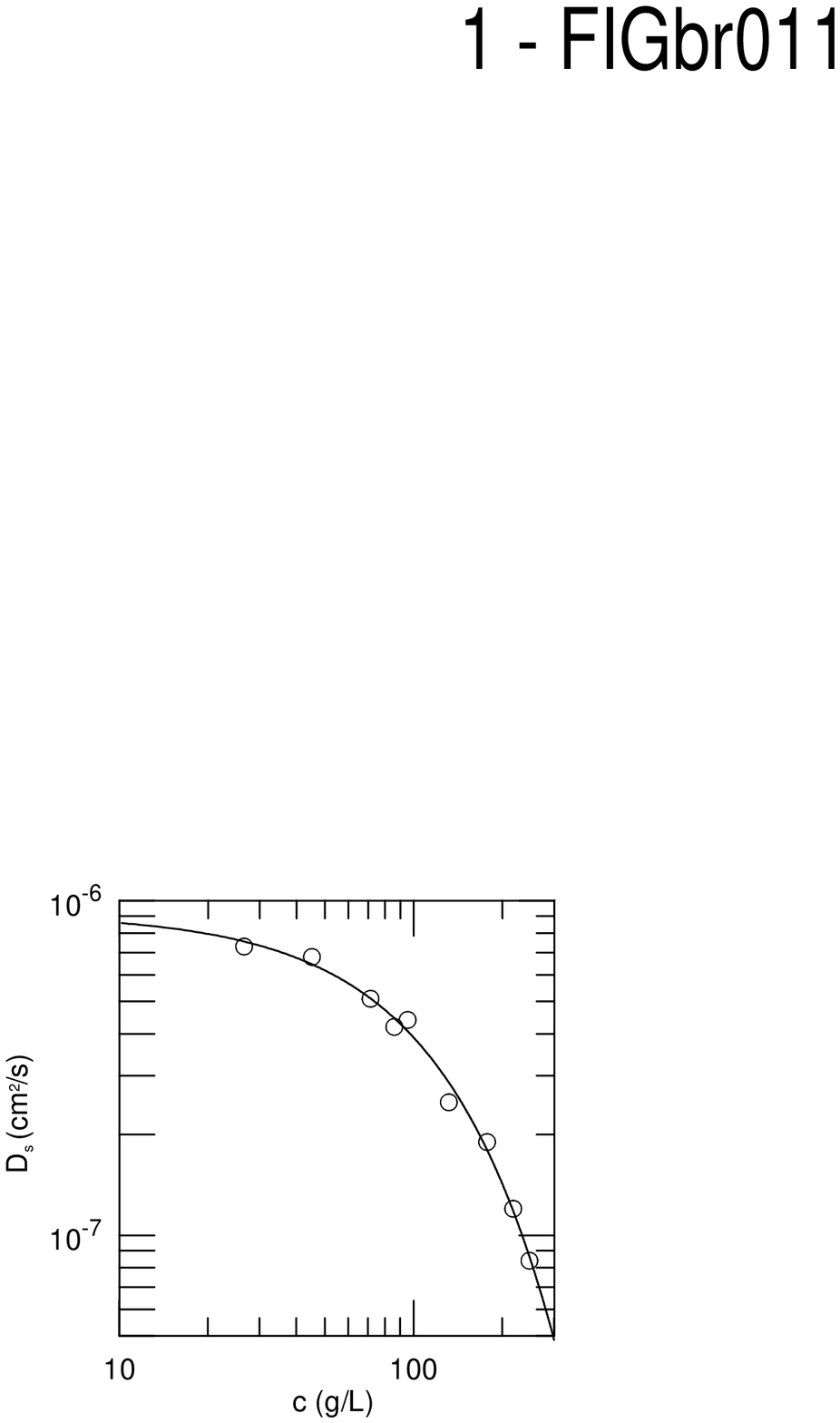} %br011

\caption{\label{figure1} 
$D_{s}$ of 64.2 kDa dextran in water, 
as obtained 
with PFGNMR by Brown, et al.\cite{brown1982a},
and a stretched-exponential fit.}
\end{figure}

\begin{figure}

\includegraphics{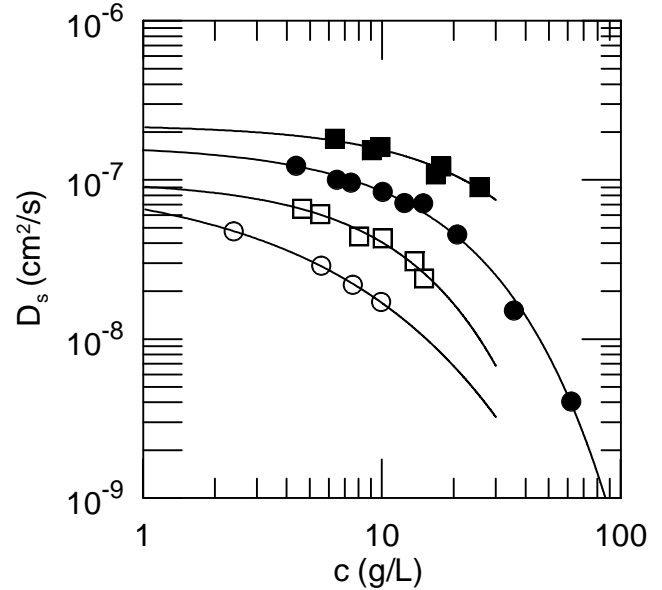} %BR021

\caption{\label{figure3} 
Self-diffusion coefficients of (top to bottom) 
73, 148, 278, and 661 kDa polyethylene oxides
in water, 
using data of Brown, et al.\cite{brown1983a},
and exponential fits.}
\end{figure}

\begin{figure} 

\includegraphics{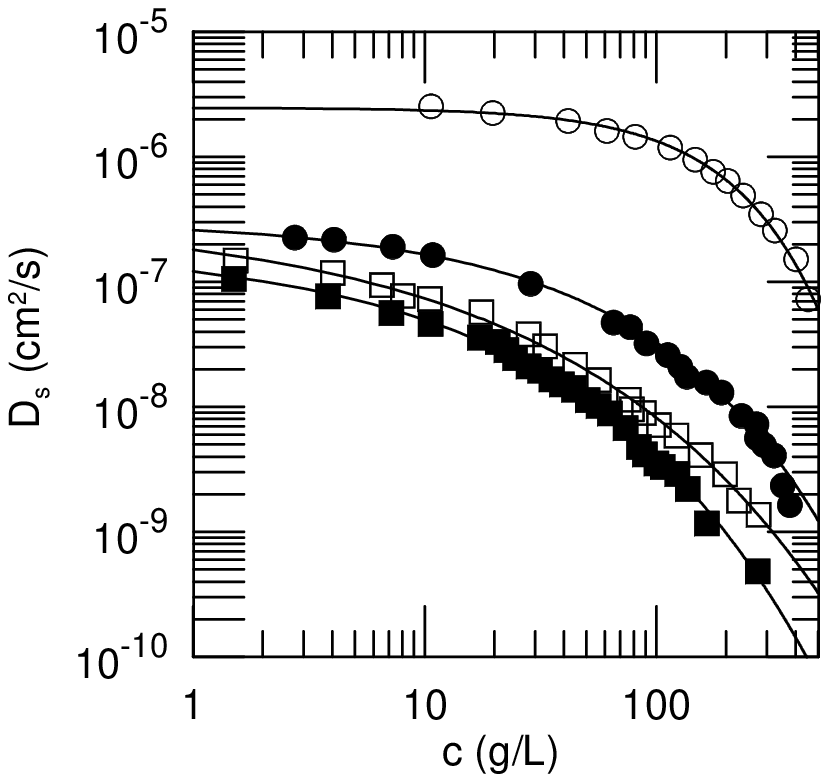} %CA03

\caption{\label{figure4} $D_{s}$ of [top to bottom]  
2, 110, 233, and 350 kDa polystyrenes in CCl$_{4}$ as obtained 
with PFGNMR by Callaghan and 
Pinder\cite{callaghan1980a,callaghan1981a,callaghan1984a}, 
and fits to stretched exponentials.}
\end{figure}

\begin{figure} 

\includegraphics{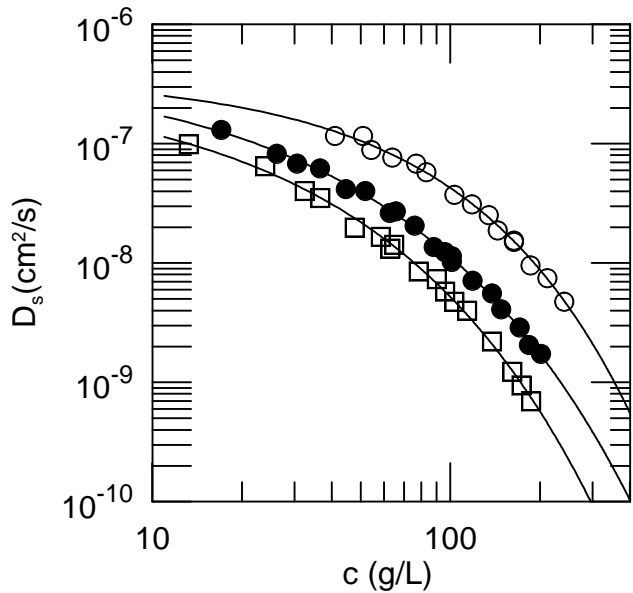} %CA022

\caption{\label{figure5}  
$D_{s}$ of [top to bottom]
110, 233, and 350 kDa polystyrenes in C$_{6}$D$_{6}$ as obtained 
with PFGNMR by Callaghan and 
Pinder\cite{callaghan1980a,callaghan1981a,callaghan1984a}, 
and fits to stretched exponentials.}
\end{figure}

Brown, et al.\cite{brown1982a} report self-diffusion and sedimentation 
coefficients for dextran ($M_{n} = 44$ kDa, $M_{w} = 64.2$ kDa) in water, using 
PFGNMR to determine $D_{s}$.  They report $D_{s}$ and $s$ as functions of $c$ 
for concentrations as large as 250 g/L.  Figures \ref{figure1} shows 
their data and the corresponding stretched-exponential fit.  The 
self-diffusion data are described accurately by the stretched-exponential form.

Brown and the same collaborators\cite{brown1983a} 
used PFGNMR to measure $D_{s}$ of of narrow 
($M_{w}/M_{n}$ of 1.02-1.20) polyethylene oxides in water.  Polymer molecular 
weights were in the range 73-661 kDa; polymer concentrations ranged up to 70 
g/l.  The same paper reports measurements on these systems of the mutual diffusion 
coefficient (from QELSS) and the sedimentation coefficient.   Figure \ref{figure3}
shows the measurements of $D_{s}$.
Because measurements were only reported over 
a limited concentration range, the data were fit both to a pure ($\nu = 1$)
and to a stretched 
exponential in $c$.
There is
excellent agreement between $D_{s}$ and a pure exponential in $c$, and almost 
no improvement in the quality of the fit attendant to allowing $\nu$ to be a 
floating parameter in the fit.

\begin{figure} 

\includegraphics{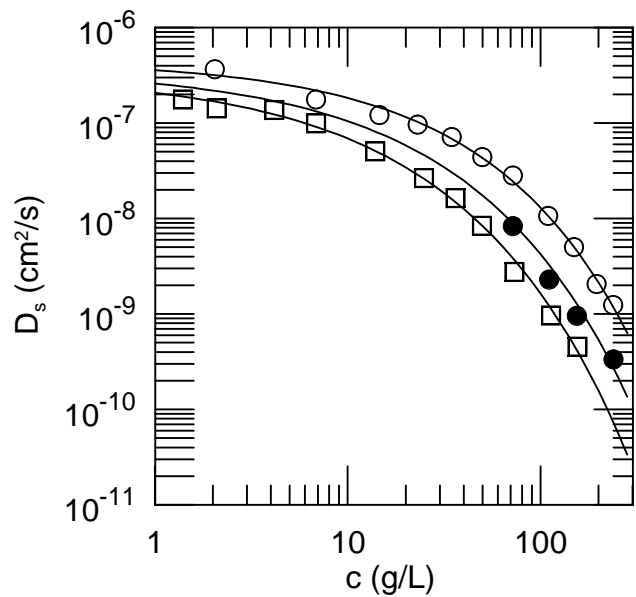} %DE021

\caption{\label{figure6}
$D_{s}$ of [top to bottom]
262, 657, and 861 kDa polystyrenes in cyclopentane near the theta temperature,
as obtained with FRS by Deschamps and Leger\cite{deschamps1986a}, 
and fits to stretched exponentials.}
\end{figure}

\begin{figure} 

\includegraphics{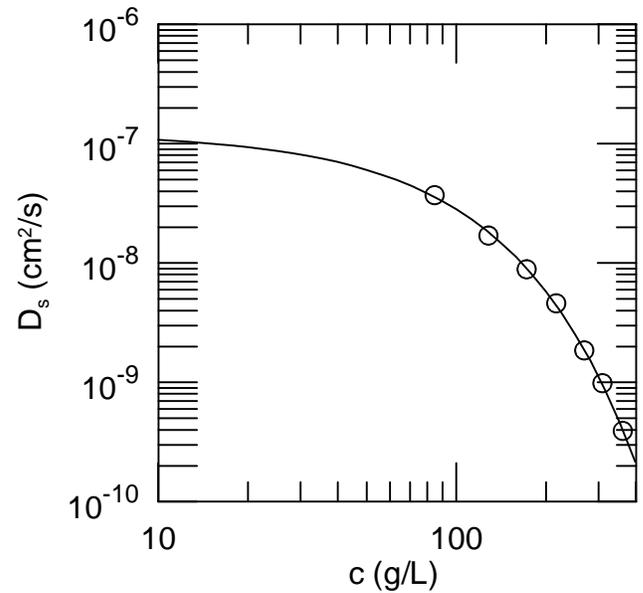} %FL01

\caption{\label{figure7}  
$D_{s}$ of 125 kDa polystyrene in toluene
as obtained with PFGNMR by Fleischer\cite{fleischer1999a}, 
and a fit to a stretched exponential.}
\end{figure}

Callaghan and Pinder\cite{callaghan1980a,callaghan1981a,callaghan1984a} used 
PFGNMR to study $D_{s}$ of 2, 110, 233, and 350 kDa polystyrenes with 
$M_{w}/M_{n}$ in the range 1.06-1.10. Polymer concentrations were
as large as several hundred g/L.  The solvents were CCl$_{4}$ 
and hexadeuterobenzene.  
In CCl$_{4}$ concentrations well into the dilute regime were 
observed; in C$_{6}$D$_{6}$ the measurements of $D_{s}$ were not extended to 
low concentration.  
The data and corresponding stretched-exponential fits appear in figs.\ \ref{figure4}
and \ref{figure5}.  For each polymer:solvent pair, a stretched exponential with 
constant parameters describes $D_{s}(c)$ well all the way from the lowest to 
the largest concentrations studied.    The prefactor $\alpha$ tends to increase 
with increasing polymer $M$, while $D_{o}$ and the exponent $\nu$ generally 
decrease with increasing polymer $M$. 

Deschamps, et al.\cite{deschamps1986a} used FRS to study self-diffusion of 
polystyrenes in cyclopentane in the vicinity of the theta point.  The polymer 
molecular weights were 262, 657, and 861 kDa; polymer concentrations ranged 
from 1 to 240 g/L.  The polymer polydispersity was in the range $M_{w}/M_{n} 
\approx 1.1 - 1.3$.  Figure \ref{figure6} shows their data and fits to stretched 
exponentials.  For the 657 kDa polymer, $D_{s}$ was only reported over a 
narrow, elevated concentration range.  For the 657 kDa polymer, the
stretched-exponential fit was therefore made by interpolating $D_{o}$ and $\nu$ 
from their values for the 262 and 861 kDa polymers, leaving $\alpha$ as the 
only free parameter.  For all three polymer molecular weights, the 
stretched-exponential forms are in good agreement with experiment. The stretched-exponential form for $D_{s}(c)$ remains valid after a change from good to near-theta solvent conditions.

As an aside, the data of Deschamps, et al.\cite{deschamps1986a} 
illustrate well the principle that an 
experimental test of a particular theoretical model is sometimes 
not optimal as a 
test for a different theoretical model.  Deschamps and Leger's objective was to 
search for deGennes-type scaling behavior\cite{degennes1979a} 
of polymer self-diffusion in a theta 
solvent.  Scaling behavior is only 
expected in the semi-dilute concentration regime $c > c^{*}$.  In Deschamps, et 
al.'s 
systems, the semi-dilute regime was expected 
to be found only for $c > 50-100$ g/L.  In the 
context of their objective of studying scaling behavior, there was no 
strong reason for Deschamps, et al.\ to measure $D_{s}$ at lower 
concentrations, so they rationally did not do 
so for the 657 kDa system.  However, an accurate stretched-exponential fit 
requires measurements of $D_{s}$ at small as well as large 
concentrations.  Through no fault of the original authors, the 
range of concentrations studied for the 657 kDa polymer 
restrains the utility of the fits that can be made to some of their data.

\begin{figure} 

\includegraphics{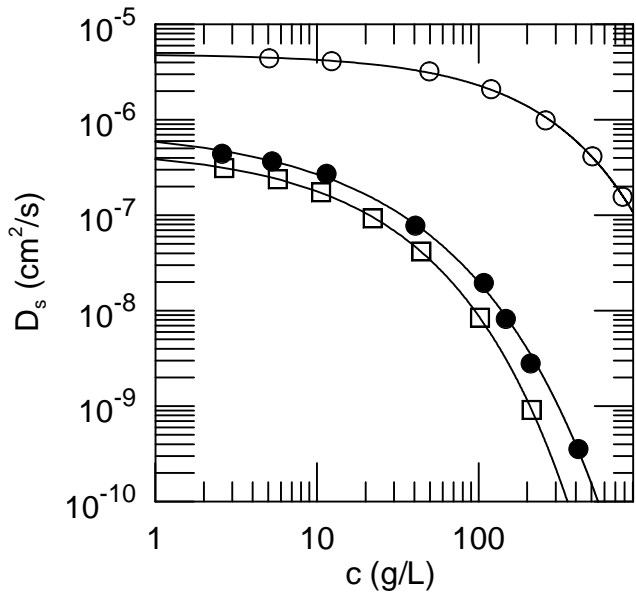} %GI01

\caption{\label{figure8} 
$D_{s}$ of [top to bottom]
15, 530, and 730 kDa polydimethylsiloxane in toluene,
as obtained with PFGNMR and reported by Giebel, et al.\cite{giebel1993a}
based in part on work of Skirda, et al.\cite{skirda1988a}, 
and fits to stretched exponentials.}
\end{figure}

\begin{figure} 

\includegraphics{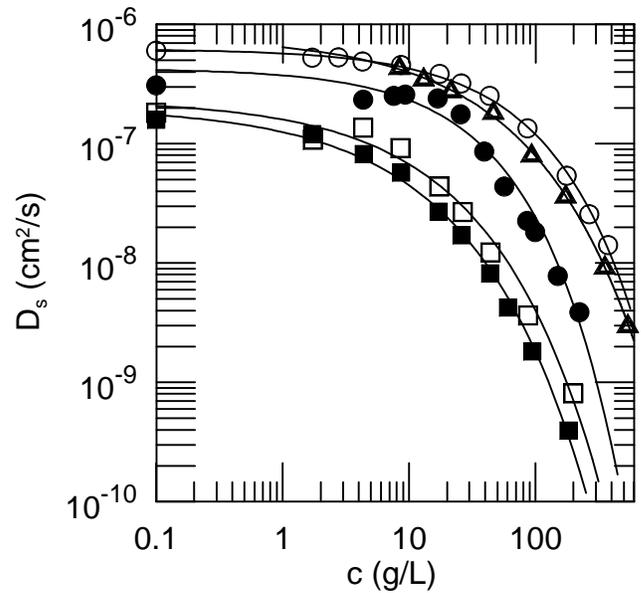} %LE02

\caption{\label{figure9} $D_{s}$ of [from top to bottom] 78, 123, 245, 
599, 
and 745 kDa polystyrene in benzene, and fits to stretched exponentials (fit patameters, Table \ref{table1}, using
data of
Hervet, et al.\cite{hervet1979a} and Leger, et al.\cite{leger1981a}.}
\end{figure}

\begin{figure} 

\includegraphics{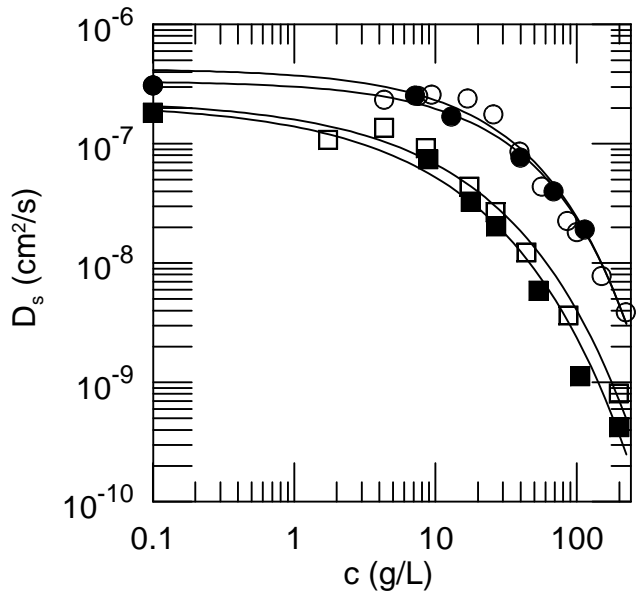} %LE014

\caption{\label{figure10} $D$ of 
245 kDa polystyrene in 245 and 599 kDa polystyrene:benzene
(open, filled circles, respectively),  and
599 kDa polystyrene in 599 and 1800 kDa 
polystyrene:benzene (open, filled squares), 
using data of Leger, et al.\cite{leger1981a} and patameters in Table \ref{table1}.} 
\end{figure} 

Fleischer\cite{fleischer1999a} used PFGNMR to observe self-diffusion of 125 kDa 
polystyrene, $M_{w}/M_{n} \approx 1.02$,
in toluene for concentrations 80-320 g/L.  These concentrations 
were estimated to cover $0.5c_{e} \leq c \leq 2 c_{e}$, where $c_{e}$ is the 
concentration above which entanglements were said to be present.  $c_{e}$ was 
estimated by applying a deGennes-type model to rheological 
data\cite{osaki1985a} .  Fleischer's measurement appear in Fig.\ \ref{figure7}.  As 
seen in the Figure, the measured $D_{s}(c)$ is in good agreement with a 
stretched-exponential form.  

Fleischer's PFGNMR measurements\cite{fleischer1999a} of the incoherent dynamic 
structure factor $S_{inc}(q,t)$ of this system show only a single fast 
relaxation, even under conditions under which QELSS reveals that a slow mode 
dominates the QELSS spectrum.  Fleischer's observation that the QELSS slow mode 
cannot be seen in the PFGNMR data is immediately reminiscent of results of Zero 
and Ware\cite{zero1984a} on fluorescence recovery after photobleaching  in 
poly-L-lysine solutions at low salt concentration.  In each set of results, the 
QELSS spectrum shows distinct slow and fast modes.  The slow mode becomes 
dominant under conditions (higher polymer concentration, lower ionic strength) 
that enhance 'glassy' behavior.  In each case, the appearance and then 
dominance of the slow mode, when 'glassy' behavior is enhanced, has no effect on 
the self-diffusion of individual chains.  This behavior is hard to understand 
if the QELSS slow mode is interpreted as arising from long-lived chain clusters 
with a fixed list of members.  The behavior is, however, understandable in 
terms of a slow mode arising from the appearance of long-lived dynamic 
structures within which individual chains only have short residence times.  
Such dynamic properties of $D_{s}$ and $D_{m}$
for interacting interpenetrating particles 
are also seen in Johnson et al.'s\cite{johnson1998a,klein2000a} model glasses.

Giebel, et al.\cite{giebel1993a} report $D_{s}$ of 15, 530, and 730 kDa 
polydimethylsiloxane in toluene as obtained with PFGNMR, based in some part on 
data of Skirda, et al.\cite{skirda1988a}.  The original measurements cover the 
concentration range 2-900 g/L of polymer.  Fitting 
parameters are in Table I.  Figure \ref{figure8} shows the actual data and their 
fits.   For each polymer molecular weight, 
$D_{s}$ is described well by a stretched exponential in concentration.  

Hadgraft, et al.\cite{hadgraft1979a} used QELSS to measure the diffusion of 
polystyrenes in benzene as a function of molecular weight for $24.8 \leq M \leq 
8870$ kDa at 25 C and very low polymer concentration.  This data is of specific 
interest here in that it supplies values for $D_{s}$ in the low-concentration 
region, as a supplement to data sets in which $D_{s}$ was only obtained at 
relatively elevated concentrations.  Omitting the largest-$M$ point (and thereby
reducing the RMS fractional error in the fit from 18\% to 8.6\%), 
their data follow $D_{o} 
= 4.54 \cdot 10^{-4} M^{-0.588}$.  

Hervet, et al.\cite{hervet1979a} and Leger, et al.\cite{leger1981a} studied 
$D_{s}$ of polystyrene in benzene using FRS to measure 
the relaxation of photoexcitation patterns.  Self-diffusion coefficients were 
obtained for polymers of molecular weight 78.3, 123, 245, 599, and 754 kDa, 
with $M_{w}/M_{n} \approx 1.06-1.12$, for concentrations up to 550 g/L.   
The 
same technique was used to obtain diffusion coefficients of labelled 245 kDa 
probe chains in a 599 kDa polystyrene matrix, and 599 kDa polystyrenes in a 
1800 kDa polystyrene matrix.   Matrix concentrations ranged from 5 to 
400 g/L.  To supplement these measurements, many of
which were made at elevated polymer 
concentrations, we used the dilute-solution 
self-diffusion measurements of Hadgraft\cite{hadgraft1979a} to estimate $D_{s}$ 
at very low $c$.  

\begin{figure} 

\includegraphics{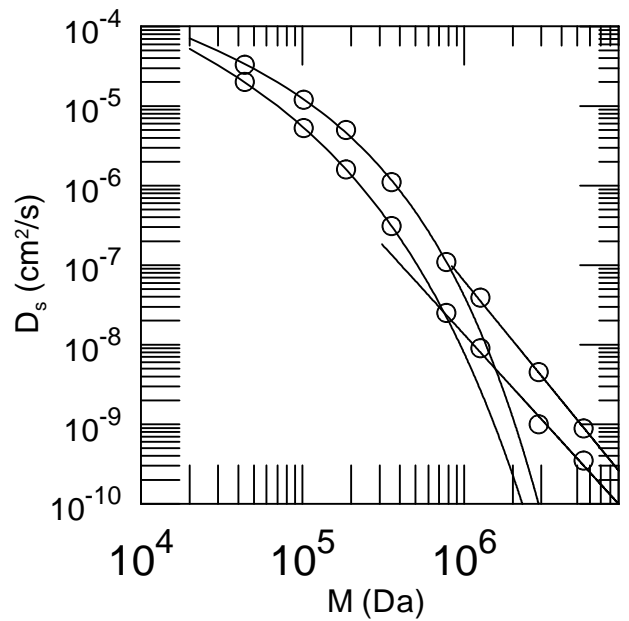} %NE011

\caption{\label{figure11}  $D_{s}$ of 130 and 180 g/L
polystyrene in dibutylphthalate as a function of molecular weight, showing the 
lower-molecular-weight stretched-exponential and the higher-molecular-weight 
power-law molecular-weight dependences of $D_{s}$, 
using data of Nemoto, et al.\cite{nemoto1991a}, Table II.} 
\end{figure} 

\begin{figure} 

\includegraphics{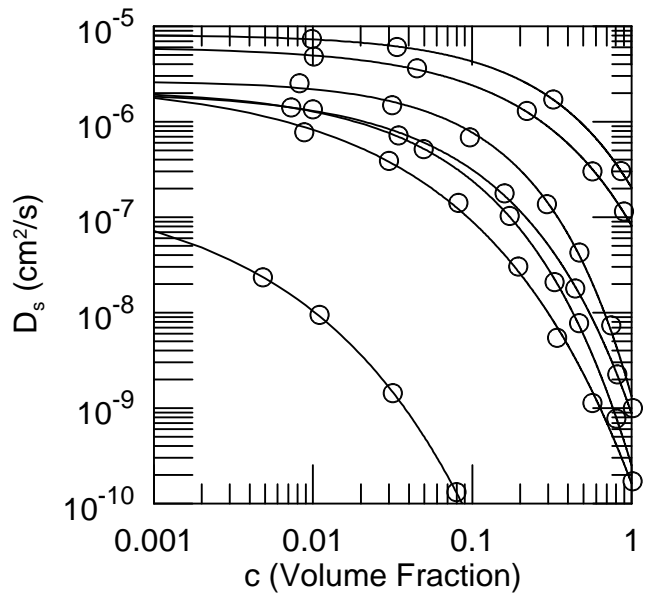} %SK01

\caption{\label{figure12} $D_{s}$ of poly(ethylene oxide), [from top to 
bottom, at $\phi \approx 0.1$] namely 2kDa in CHCl$_{3}$ and in dioxane, 20 kDa 
in benzene and in dioxane, 40 kDa in chloroform and in dioxane, and 3600 kDa in 
dioxane, after Skirda, et al.\cite{skirda1987a} Fig.\ 1a. Lines are fits to stretched 
exponentials in volume fraction.}

\end{figure}

\begin{figure}

\includegraphics{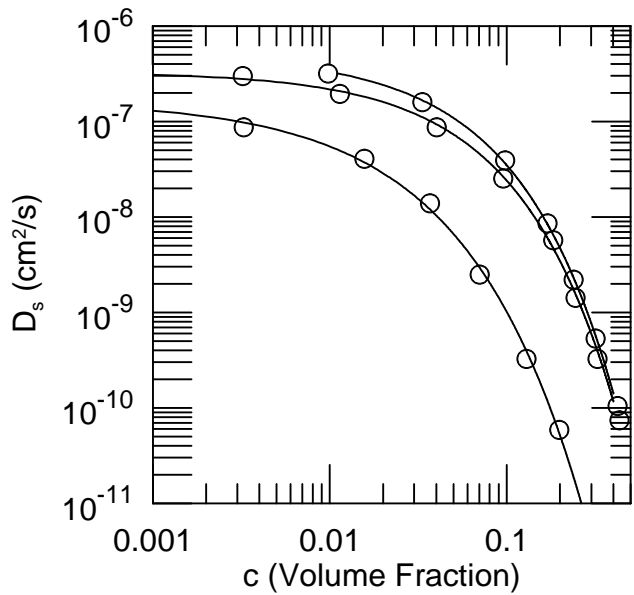} %SK012

\caption{\label{figure13}$D_{s}$ of polystyrene, namely [from top to bottom] 
240kDa in benzene and 1300 kDa in benzene and carbon tetrachloride, 
respectively, after Skirda, et al.\cite{skirda1987a} Fig.\ 1b. Lines are fits 
to stretched exponentials in the volume fraction $\phi$.} 

\end{figure} 

Figure \ref{figure9} displays the 
self-diffusion data of Hervet, et al.\cite{hervet1979a} and 
Leger, et al.\cite{leger1981a} and the fits to stretched exponentials. Except 
for the 245 (filled circles) and 598(open squares) kDa polystyrenes, 
there is good agreement between the reported $D_{s}(c)$ and the fits.  In the 
two anomalous systems, $D_{s}$ at first increases with increasing $c$ and then 
decreases at larger $c$.  The initial increase in $D_{s}$ with increasing $c$ 
appears to be substantially larger than the random error in the measurements as 
inferred from random scatter in the data. The non-monotonic dependence of 
$D_{s}$ on $c$ is unique to this specific set of measurements.  The lack of 
agreement between the stretched exponential form and the measured $D_{s}$ is 
therefore ascribed to some unique feature of this data, and not to a general 
property of polymers in good solvents.

Figure \ref{figure10} presents $D_{p}(c)$ for the two probe-matrix systems and
compares with $D_{s}(c)$ of polystyrenes of similar molecular weights.  
For both probe-matrix systems, $D_{p}(c)$ is described well by a stretched 
exponential, with no indication of the non-monotonic concentration dependence 
seen for $D_{s}(c)$ of the 245 and 598 kDa chains.  The 245 kDa polystyrene 
diffuses approximately equally rapidly through 245 and 599 kDa matrices.  
The 599kDa polystyrene diffuses markedly more slowly through the 1800 kDa 
matrix than through the 599 kDa matrix.

Nemoto, et al.\cite{nemoto1991a} used FRS
to measure the self-diffusion coefficient of 
polystyrene as a function of polymer molecular weight at fixed concentration.
The solvent was dibutylphthalate.  Polymer molecular weights
covered a range $43.9 \leq M_{w} \leq 5480$ kDa with polymer 
concentrations of 130 and 180 g/L.  The polymers were quite monodisperse, with 
$1.01 \leq M_{w}/M_{n} \leq 1.09$, except for the 5480 kDa material, for which 
$M_{w}/M_{n} \approx 1.15$.  

Nemoto, et al.'s results\cite{nemoto1991a} appear in Fig.\ \ref{figure11}.  A 
stretched-exponential 
molecular weight dependence of $D_{s}$ is not observed at all $M$.  
At the two concentrations studied by Nemoto, et al.,
the molecular weight dependence of $D_{s}$ has a transition near $M \approx 
800$kDa.  At lower molecular weights, a stretched exponential in $M$ describes 
$D_{s}(M)$ extremely well.  At larger molecular weights, one finds a power-law 
dependence $D_{s} \sim M^{-\gamma}$ with $\gamma \approx 2.49$ at 130 g/L and 
$\gamma \approx 2.22$ at 180 g/L.  

Nemoto, et al.\cite{nemoto1988a} used FRS and a cone and plate viscometer to 
determine $D_{s}$ and the steady-state shear viscosity $\eta$ of concentrated 
solutions (40 and 50 wt\%) of 44 and 355 kDa polystyrene in dibutylphthalate.  
In terms of the transient lattice 
models, the solutions of the 44kDa polymer are expected to be unentangled, 
while the solutions of the 355 kDa polymer are expected to be entangled.  
They found that $D_{s}/T$ and the fluidity $\eta^{-1}$ have virtually the same 
dependence on temperature, at either concentration, both for the 44 kDa 
polystyrene and for the 355 kDa polystyrene.

\begin{figure} 
\includegraphics{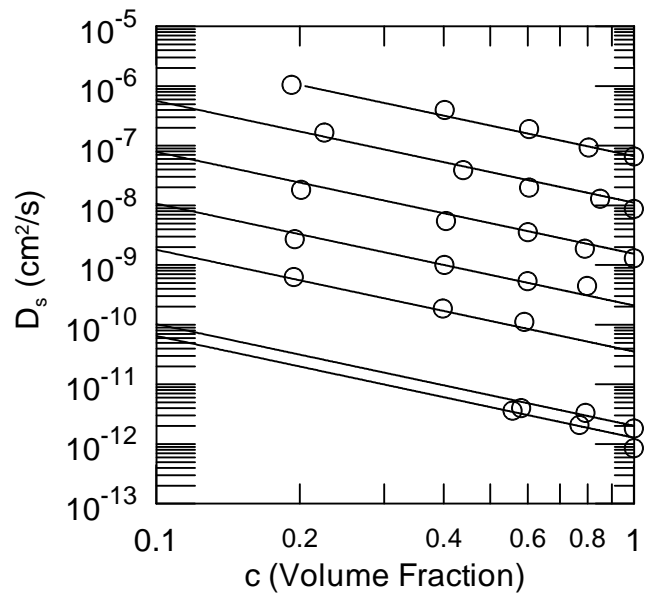} %TA011
\caption{\label{figure14} 
$D_{s}$ of hydrogenated polybutadienes in 
alkanes, with 
polymer molecular 
weights [from top to bottom] 4.9, 10.3, 23.3, 53.2, 111, 364, and 440 kDa,
and the fitted scaling form $D_{s} = 55.0 \phi^{-1.7} M^{-2.42}$, 
showing that three parameters and the scaling form
suffice to describe $D_{s}(c,M)$ over a wide 
range of both parameters.        
Measurements are from
Tao, et al.\cite{tao2000a}+.}
\end{figure} 

\begin{figure}
\includegraphics{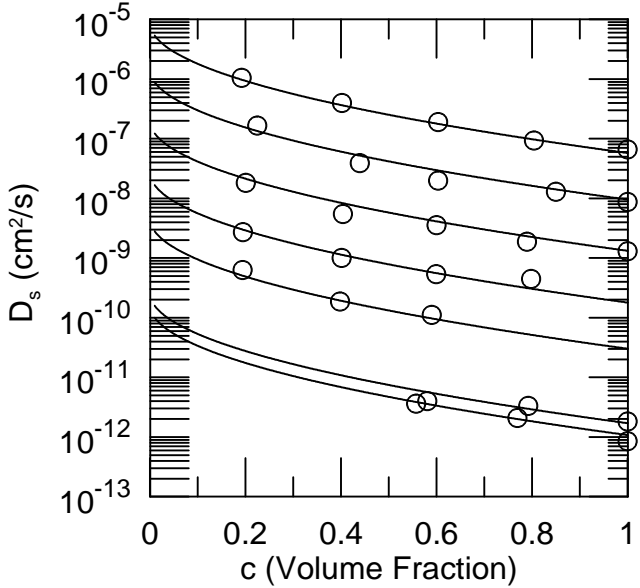} %TA012
\caption{\label{figure15} 
Same as Fig.\ \ref{figure14}, except lines now show the fitted 
form
$D_{s} = 5.54 \cdot 10^3 M^{-2.42} \exp(-5.026 c^{0.5} M^{0.00019})$.
RMS fractional errors in the fits shown in 
Figs.\ \ref{figure14} and \ref{figure15} are equal; a stretched exponential in $c$ and $M$ 
thus suffices to describe this data. }
\end{figure} 

Skirda, et al.\cite{skirda1987a} used PFGNMR to study the self-diffusion of 
polyethylene oxides ($M=$ 2, 20, 40, and 3000 kDa) and polystyrenes ($M_{n} =$ 
240 and 1300 kDa) in chloroform, benzene, dioxane, and carbon tetrachloride 
over a full range of polymer volume fractions $\phi$.  $M_{w}/M_{n}$ was 
$\approx 1.1$ for the polyethylene oxides (except the 3000 kDa polymer, for 
which $M_{w}/M_{n} \approx 2$) and $\approx 1.2$ for the polystyrenes.  Figures 
\ref{figure12} and \ref{figure13} show results for PEO and polystyrene, respectively, and 
the matching stretched-exponential fits, using parameters from Table I.  For 
each polymer:solvent combination, a stretched-exponential form fits the data 
well.

Tao, et al.\cite{tao2000a} measured $D_{s}$ (using PFGNMR 
and forward recoil spectroscopy) and $\eta$ (from the dynamic 
shear moduli) of hydrogenated polybutadienes in alkane solvents.  Polymer 
volume fractions $\phi$ extended from 0.2 up to the melt while polymer 
molecular weights cover two orders of magnitude, from 4.8 to 440kDa, with 
$M_{w}/M_{n}$ of 1.01-1.03 or less.  Tao, et al. fit their self-diffusion data to a scaling 
description $D_{s} \sim \phi^{a} M^{b}$.  When they\cite{tao2000a}
forced $a = -1.8$, a 
one-parameter fit found that the averaged $D_{s} \phi^{1.8}$ is $\sim M^{-
2.41}$.  

Figures \ref{figure14} and \ref{figure15} show Tao, et al.'s data as fit to to power-law 
and stretched-exponential forms, respectively.   In Fig.\ \ref{figure14}, all points 
were fit simultaneously to a scaling equation, finding $D_{s} \sim c^{-1.71} 
M^{2.42}$, with a fractional root-mean square fitting error of 20.2\%.  These 
exponents differ slightly from those of Tao, et al.\cite{tao2000a}:
Tao, et al.\ did sequential 2-parameter fits first to determine $a$ and then to 
determine $b$, while we did a single three-parameter fit to all points.  Figure 
\ref{figure15} shows a fit of all data points to the stretched-exponential form 
$D_{oo} M^{-z} \exp(- \alpha c^{\nu} M^{\gamma})$.  The factor $M^{-z}$ appears here because we are combining data on polymers with multiple molecular weights, and the extrapolations $c \rightarrow 0$ of measurements at different molecular weight should extrapolate to a different $D_{o}$ at each $M$.  Formally, $D_{oo}$ is the extrapolated zero-concentration diffusion coefficient of a highly hypothetical polymer having a molecular weight of unity. On forcing $\nu = 0.5$, we 
obtain $\gamma = 1.9 \cdot 10^{-4}$ and $z=2.42$ with an RMS fractional error 
of 20.2\%.  Treating $\nu$ as a free parameter finds $\nu \approx 0.24$, with 
virtually the same value of $z$, $\gamma \approx 0$, and only a slightly 
improvement (to 19.6\%) in the fit error.  With either value for $\nu$, the 
molecular-weight dependence of $D_{s}$ is almost entirely
determined by the prefix $M^{-z}$.  
The exponential itself has only a negligible dependence on $M$.  

In Tao, et al.\cite{tao2000a}'s systems, scaling-law and stretched-exponential 
forms for $D_{s}$ thus provide equally good descriptions of the concentration 
dependence of $D_{s}$.  The scaling and stretched-exponential forms also agree 
as to the molecular weight dependence of $D_{s}$ at fixed $c$, namely $D_{s} 
\sim M^{-2.4}$.  Tao, et al.\cite{tao2000a} concluded that a scaling-law 
description of their data is correct.  The analysis here corroborates this 
statement, but shows that it is incomplete, in that stretched-exponential forms 
describe equally accurately the measured $D_{s}(c,M)$.

\begin{figure} 

\includegraphics{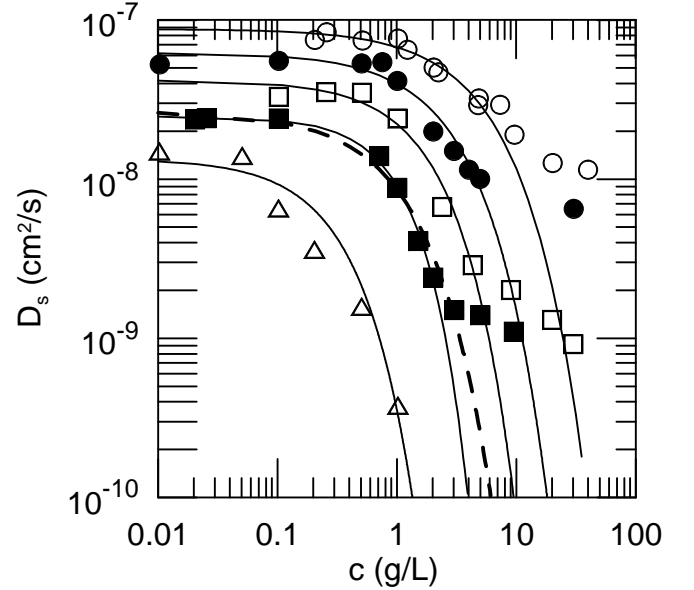} %TI01

\caption{\label{figure16}  $D_{s}$ of xanthan,
molecular weights [from top to bottom] 0.45, 0.99, 1.9, 3.8, and 9.4 MDa,
in water, and fits of the lower-concentration data to stretched exponential 
forms.  For the 3.8Mda polymer, solid and dashed lines represent fits to the 
first 7 and 8 data points, respectively.  Measurements are from
Tinland, et al.\cite{tinland1990a} Fig.\ 3.}

\end{figure}

\begin{figure} 

\includegraphics{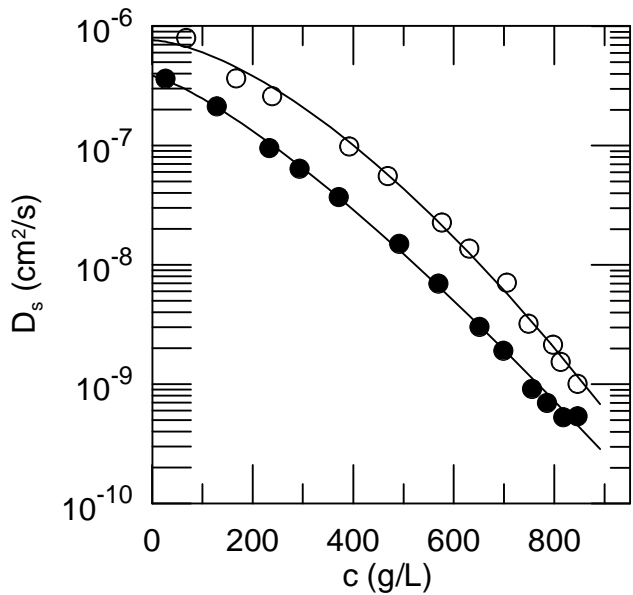} %VMA21

\caption{\label{figure17} $D_{s}$ of $f=2$ (open circles) and $f=8$
(closed circles) polyisoprenes in C$_{6}$F$_{5}$Cl, both with $M_{span}=5$kDa,
and fits to stretched exponentials in $c$.  Data from von Meerwall et
al.\cite{vonmeerwall1982a}, Fig.\ 1.}
\end{figure}

\begin{figure}

\includegraphics{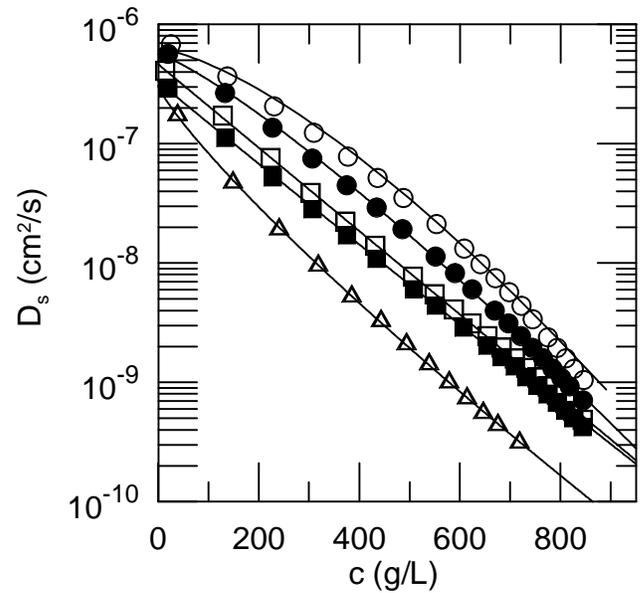} %VMA23

\caption{\label{figure18} $D_{s}$ from von Meerwall, et al.\cite{vonmeerwall1982a} on [from top to bottom] $f=2, 3, 8, 18$ 
polyisoprenes with $M_{span}=5$kDa and $f=8$ polyisoprene with $M_{span}=14$kDa  
in CCl$_{4}$, and fits to stretched exponentials in $c$.}  

\end{figure} 

Qualitatively, the larger part of Tao, et al.'s viscosity 
measurements\cite{tao2000a} 
might have been expected to be 
in the larger-concentration meltlike ($\eta \sim c^{x}$) rather than 
the smaller-concentration solutionlike ($\eta \sim \exp(\alpha c^{\nu})$) regime.  
The transition in the concentration dependence of $\eta$ between these two 
functional forms is not transparently evident in $D_{s}(c)$, 
whose concentration dependence is
consistent with a stretched exponential in $c$ for concentrations up 
to the true melt.  However, contrary to those other systems reviewed in this 
article, from which molecular-weight dependences can be extracted, 
here the molecular weight exponent $\gamma$ of 
$\exp(- \alpha c^{\nu} M^{\gamma})$ is very nearly zero, 
so in Tao, et al.'s systems the concentration dependence of $D_{s}$ is 
very nearly the same at all $M$.

Tinland, et al.\cite{tinland1990a} report on self-diffusion of xanthans of 
molecular weight 0.45-9.4 MDa, with $M_{w}/M_{n}$ in the range 1.2-1.4, 
at concentrations 0.01-40 g/L.  Xanthan forms 
wormlike chains.  The experimental data of Tinland, et al.\ obtained with FRAP
thus differs from 
$D_{s}$ of almost all other polymers, in that $d \ln(D_{s})/d \ln(c)$ does not 
decrease monotonically with increasing $c$.  This phenomenon, which was noted 
by the original authors, is plausibly related to the 
appearance\cite{tinland1990a} of a lyotropic liquid-crystalline phase in this 
material at elevated $c$.  None of the polymer dynamics models discussed above would be expected to remain valid while the system underwent a phase transition, so the behavior observed by Tinland, et al.\ does not contradict any model for polymer dynamics.  
Figure \ref{figure16} shows Tinland, et al.'s\cite{tinland1990a} measurements and fits 
of the lower-concentration data at each molecular weight to a 
stretched-exponential form.  As seen from Table I, 
the somewhat large RMS fractional errors show that 
agreement between the measured data and the 
functional forms is not outstanding.  The  fit 
depends marginally on the number of data points included in the analysis.  For 
the 3800 kDa polymer, we indicate (solid, dashed lines) the fits to the first 7 
or 8 data points.

For all but the largest-$M$ polymer,  there is a  
concentration $c^{**}$ at which $D_{s}(c)$ deviates from its low-concentration 
decline.  In Tinland, et al.'s language\cite{tinland1990a}, 
$c^{**}$ is the concentration of the 
higher-concentration boundary of the semidilute regime.  
From Fig.\ \ref{figure16}, $c^{**}$ appears to 
decrease with increasing polymer molecular weight.

von Meerwall, et al.\cite{vonmeerwall1982a} used PFGNMR to measure $D_{s}$ of 
linear and star polyisoprenes over a near-complete range of concentrations 
(polymer  weight fraction $0.01 \leq x \leq 1$).
The arm number $f$ ranged from 2 to 18.  Molecular weights of single arms were 5
and 18 kDa.  Figure \ref{figure17} compares 
von Meerwall, et al.'s\cite{vonmeerwall1982a} data on $f=2$ and $f=8$ 
polyisoprene stars having 5kDa arms.  Figure \ref{figure18} shows $D_{s}$ for $f= 2, 
3, 8,$ and $18$ star polyisoprenes in CCl$_{4}$.  In the original paper,  
data  were reported as smooth curves, not as measured 
points.  To create Fig.\ \ref{figure18}, the smooth curves were sampled; fits were 
made to the sampled points.  From Fig.\ \ref{figure18}, at all concentrations 
increasing the arm count at fixed span molecular weight reduces $D_{s}$: 
Increasing four-fold the number of arms reduces $D_{s}$ by a factor of 2 to 3.  
At $f=8$, an increase in span molecular weight reduces $D_{s}$.  

In both figures, the solid lines are fits to 
stretched exponentials. Without exception, $D_{s}(c)$ for each $f$ and $M$ is 
described well by a stretched exponential in $c$.  From the fit parameters in 
Table I, for fixed $M_{span}$ and increasing $f$ one finds that $D_{o}$ and 
$\nu$ fall while $\alpha$ increases, a two-fold decrease in $D_{o}$ via 
increasing $f$ being accompanied by a 20-fold increase in $\alpha$.  For the 
smallest molecular weights (10-16 kDa) studied, $\nu > 1$ is observed.

\begin{figure} 

\includegraphics{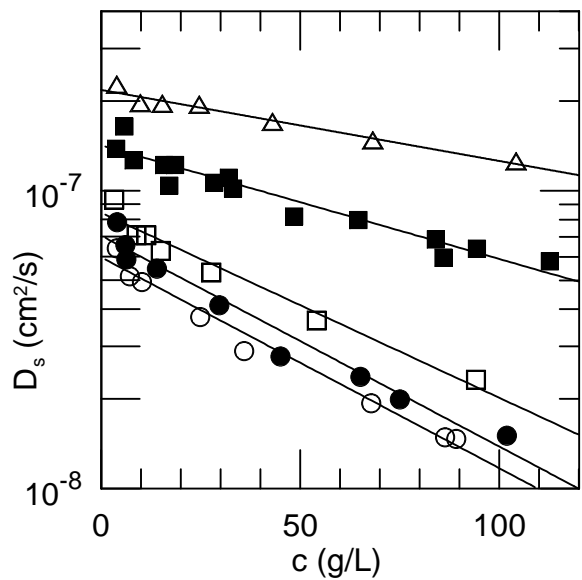} %vm032e.

\caption{\label{figure19} $D_{s}$ of [from top to bottom] 2.3 kDa 
linear, and 
6.5, 16.1, 21, and 26 kDa three-armed polybutadienes in CCl$_{4}$, and fits to 
stretched exponentials, using data of von Meerwall, et 
al.\cite{vonmeerwall1983a}, Fig.\ 2.} 
\end{figure}

\begin{figure} 

\includegraphics{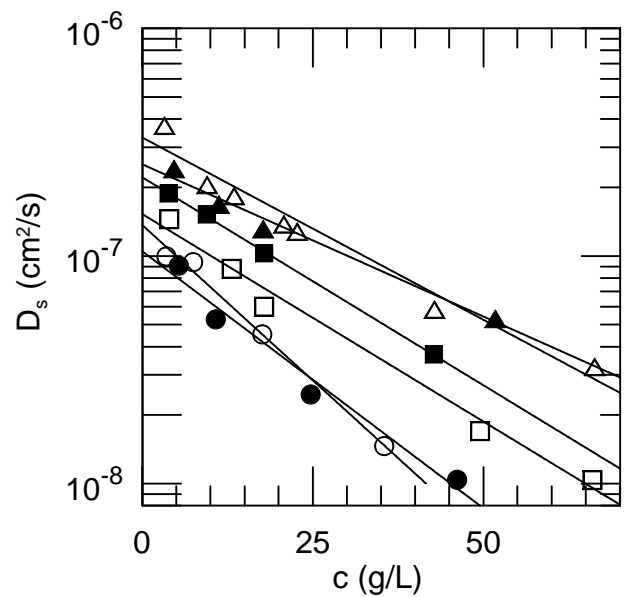} %vm033

\caption{\label{figure20} $D_{s}$ of [from top to bottom] 75, 76, 90, 
161,
227, and 281 kDa polybutadienes in CCl$_{4}$, 
the 75 and 90 kDa polymers being linear and
the others being three-armed stars, and fits to 
pure exponentials, using data of von Meerwall, et 
al.\cite{vonmeerwall1983a}, Fig.\ 3.} 
\end{figure} 
                                                                    
\begin{figure} 

\includegraphics{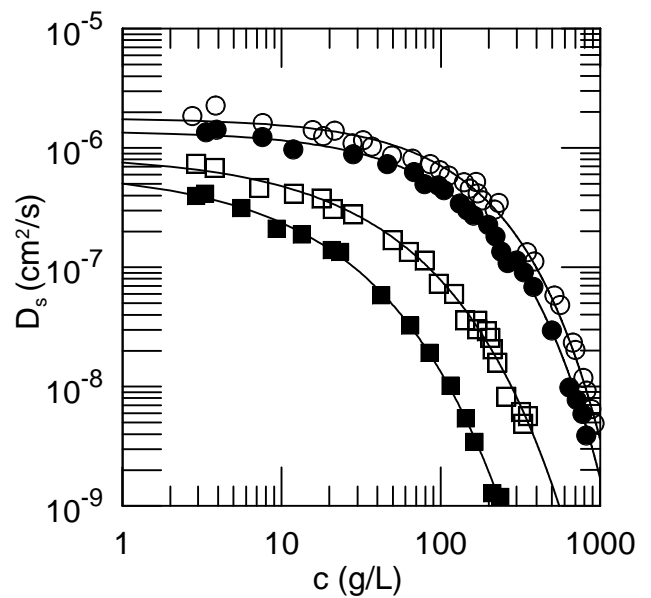} %vm034

\caption{\label{figure21} $D_{s}$ of [from top to bottom] 6.5
8.7, 29, and 76 kDa three-armed polybutadienes in CCl$_{4}$, and fits to 
stretched exponentials, using data of von Meerwall, et 
al.\cite{vonmeerwall1983a}, Fig.\ 4.} 
\end{figure} 

\begin{figure} 

\includegraphics{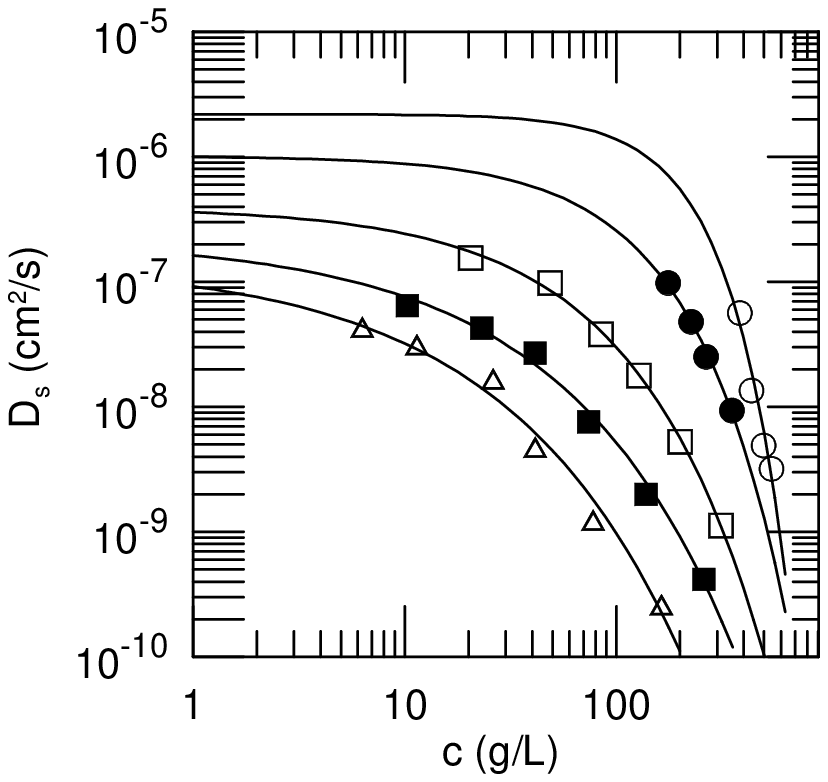} %vm011

\caption{\label{figure22} 
$D_{s}$ of 10, 37, 198, 498, and 1050 
kDa linear polystyrenes in tetrahydrofuran,
and fits to stretched exponentials in $c$.  Data after von Meerwall et
al.\cite{vonmeerwall1985a}, with supplemental low-concentration data based on 
Hadgraft, et al.\cite{hadgraft1979a}.}
\end{figure}

von Meerwall, et al.\cite{vonmeerwall1983a} used PFGNMR to study linear 
and 3-armed star polybutadienes and polystyrenes in CCl$_{4}$.  They report 
low-concentration data on all systems, and extensive concentration dependence 
measurements on some star polymers.  Polymer molecular weights ranged from 2.3 
to 281 kDa with polydispersities $M_{w}/M_{n}$ of 1.03-1.07.  
As seen in Figs.\ \ref{figure19} and \ref{figure20}, for most systems the 
observed concentration range afforded a one-order-of-magnitude variation in 
$D_{s}$.  

Figure \ref{figure21} gives Ref.\ \onlinecite{vonmeerwall1983a}'s measurements on the 6.5, 
8.3, 29, and 76 kDa 3-armed stars, which were made over a far wider 
concentration range that afforded  a 2.5-order-of-magnitude variation in 
$D_{s}$.  Fits of these data to stretched-exponential forms describe well 
$D_{s}(c)$.  As initially noted by von Meerwall, et al.\cite{vonmeerwall1983a}, 
'the slopes [in Fig.\ \ref{figure21}] change continuously' '\ldots deGennes' 
prediction of a concentration scaling regime $D_{s} \sim c^{-1.75}$ ($c^{*} < c 
< c^{**}$) is not borne out by our data \ldots at any molecular weight'.

von Meerwall, et al.\cite{vonmeerwall1985a} used PFGNMR to 
measure self-diffusion of 10, 37.4, 179, 498, and 1050 kDa polystyrene in 
tetrahydrofuran at concentrations 6-700 g/L.  The same technique was also used 
to measure $D_{s}$ of tetrahydrofuran and hexafluorobenzene in the same polymer 
solutions.    Von Meerwall, et al.\ did not report $D_{s}$ for their polymers 
in dilute solution.  $D_{s}$ for these systems in the dilute limit is
therefore inferred here from the molecular weight dependence of $D_{s}$ observed by 
Hadgraft, et al.\cite{hadgraft1979a} for polystyrene in benzene, together with 
the viscosities of tetrahydrofuran and benzene.  von Meerwall, et al.'s measurements, 
and stretched-exponential fits appear 
in Fig.\ \ref{figure22}, using parameters given in Table I.

\begin{figure} 

\includegraphics{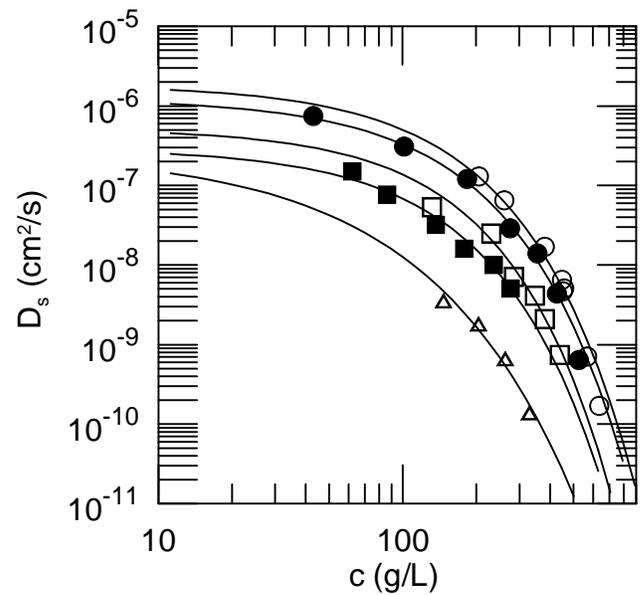} %we01

\caption{\label{figure23} $D_{s}$ of [from top to bottom] 32, 46, 105, 
130, and 
360 kDa polystyrene in tetrahydrofuran, and associated stretched exponential 
fits.  Data are from Wesson, et al.\cite{wesson1984a}, Table II and associated 
Figures, as supplemented by the low-concentration measurements of Hadgraft, et 
al.\cite{hadgraft1979a}.}
\end{figure} 

Wesson, et al.\cite{wesson1984a} used FRS to measure 
self-diffusion of polystyrenes in tetrahydrofuran and benzene.  The 
polystyrenes had $M$ of 32, 46, 105, 130, and 360 kDa, and were observed for 
concentrations in the range 40-500 g/L.   Wesson, et al.'s measurements were 
here supplemented by extreme low-concentration points calculated from 
results of 
Hadgraft, et al.\cite{hadgraft1979a}.  These points were included in the fits 
as having been taken at 0 g/L.  Experimental results and the corresponding 
stretched-exponential fits appear in Fig.\ \ref{figure23}.  Because $D_{s}$ here covers four orders of magnitude, on the scale 
of the figure the fits look very good.  In fact, RMS fractional 
errors are in the 
range 15-21\%, making these among the poorer fits in this section
to a stretched exponential.

\begin{figure} 

\includegraphics{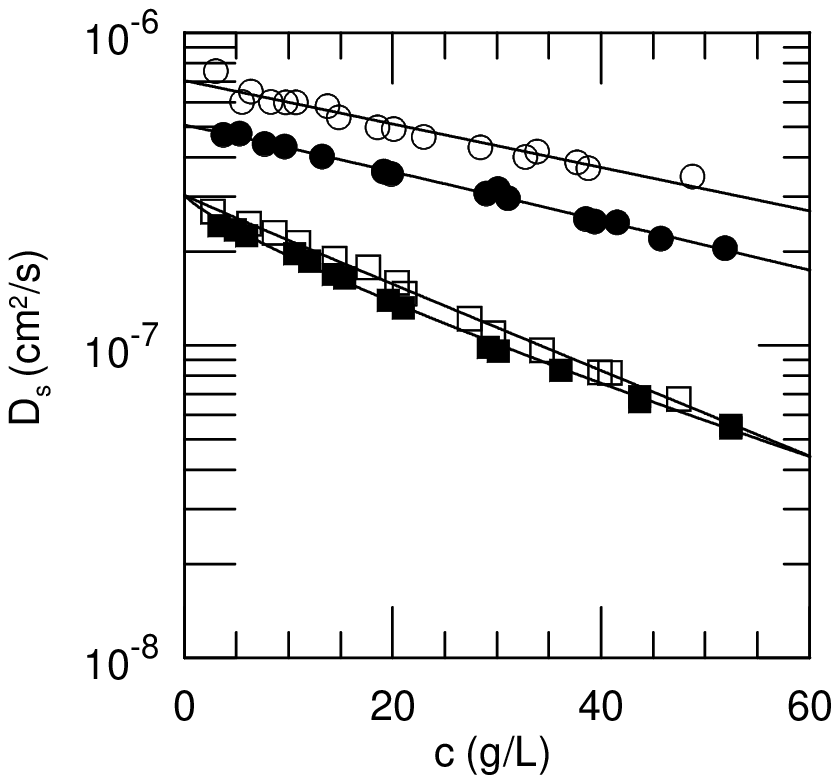} %xue012

\caption{\label{figure24} $D_{s}$ of [from top to bottom] 61, 92, 193, 
and 216 
kDa 18-arm star
polyisoprenes 
in CCl$_{4}$, and fits to stretched exponentials, using data of Xuexin, et 
al.\cite{vonmeerwall1984a}, Fig.\ 3.  } 
\end{figure}

\begin{figure} 

\includegraphics{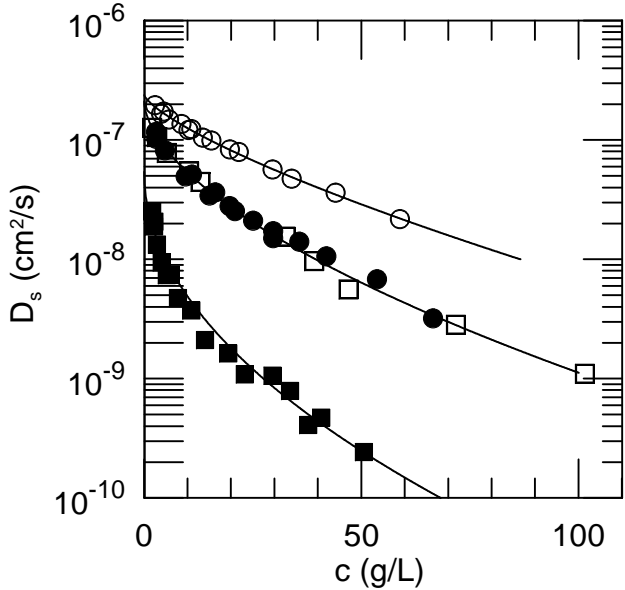} %xue011

\caption{\label{figure25}  $D_{s}$ of 302 kDa linear polyisoprene (filled 
circles)
and [from top to bottom] 340, 800, and 6300 kDa 18-armed star
polyisoprenes (all in CCl$_{4}$)
and fits to stretched exponentials, using data of Xuexin, et 
al.\cite{vonmeerwall1984a}, Fig.\ 2.}
\end{figure}

\begin{figure} 

\includegraphics{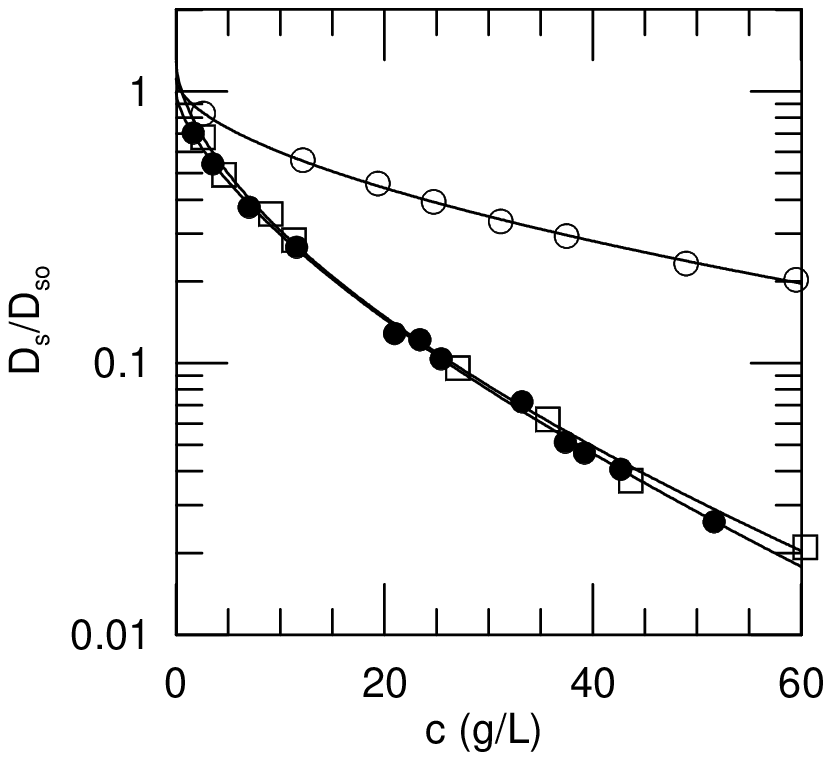} %xue014

\caption{\label{figure26} $D_{s}$ of [from top to bottom] 70.8, 251, and 
302 
kDa linear
polyisoprenes 
in CCl$_{4}$, and fits to stretched exponentials, using data of Xuexin, et 
al.\cite{vonmeerwall1984a}, Fig.\ 4.} 
\end{figure}

Xuexin, et al.\cite{vonmeerwall1984a} used PFGNR to measure $D_{s}$ of 
linear and 18-armed star polyisoprenes in CCl$_{4}$ over a wide range of $c$ 
and a 100-fold range of $M$.  Their results (with concentrations replotted in 
g/L) appear as Figs.\ \ref{figure24}, \ref{figure25}, and \ref{figure26}.  In every case the 
concentration dependence of $D_{s}$ is described well by the stretched 
exponential form, with fitting parameters given in Table I.

Figure \ref{figure24} shows $D_{s}(c)$ for four low-molecular-weight (61, 92, 193, 
216 kDa) 18-arm polyisoprenes.  With increasing $M$ and fixed $f$, $D_{o}$ and 
$\nu$ decrease, while $\alpha$ increases, the increase in $\alpha$ being 
modestly more rapid with increasing $M$ than is the decrease in $D_{o}$.  These 
trends are continued in Fig.\ \ref{figure25}, which shows 
data from ref.\ \onlinecite{vonmeerwall1984a}
for larger-$M$ (344, 800, 6300 kDa) stars.  The displayed data on 302 kDa linear 
polyisoprene are fit by very nearly the same $D_{o}$, ${\alpha}$ and $\nu$ as 
is the data on a much larger (800 kDa) 18-arm star polyisoprene.  Finally, 
Fig.\ \ref{figure26} shows Xuexin, et al.'s data on
$D_{s}(c)$ for three linear polyisoprenes, molecular 
weights of 70.8, 251, and 302 kDa, albeit over a narrower range of $c$ than 
in Figure \ref{figure24}.

\begin{figure*} 

\includegraphics{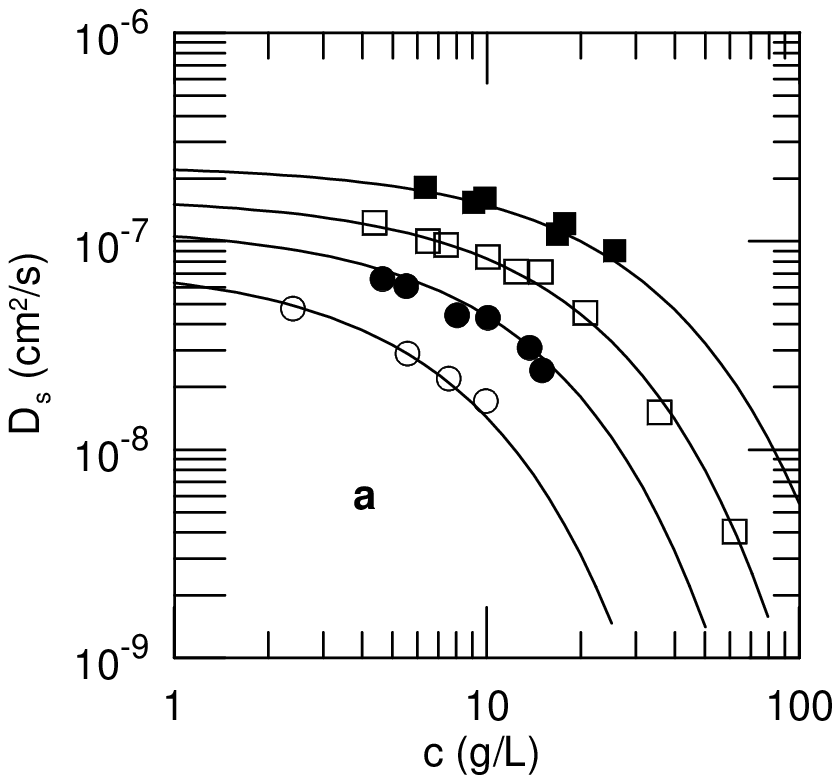} %br02j,ca03j,ca03k,de02j
\includegraphics{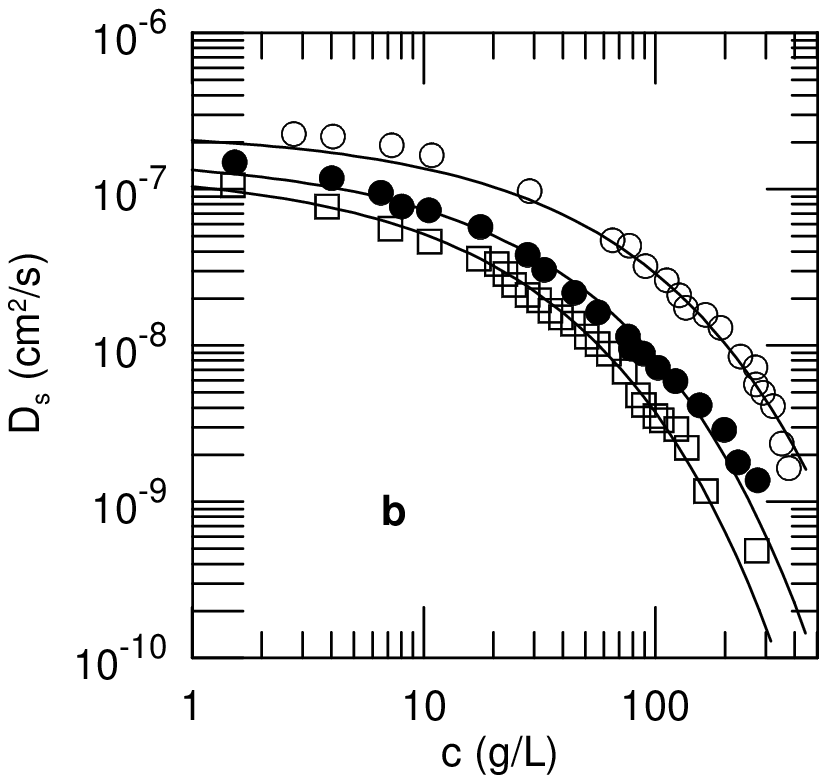} %br02j,ca03j,ca03k,de02j
\includegraphics{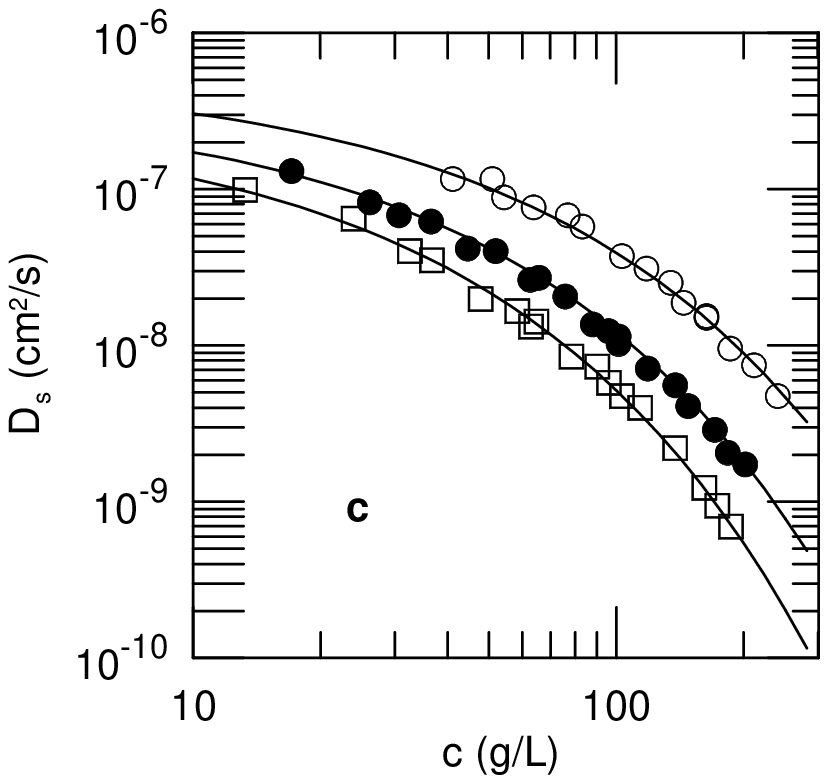} %br02j,ca03j,ca03k,de02j
\includegraphics{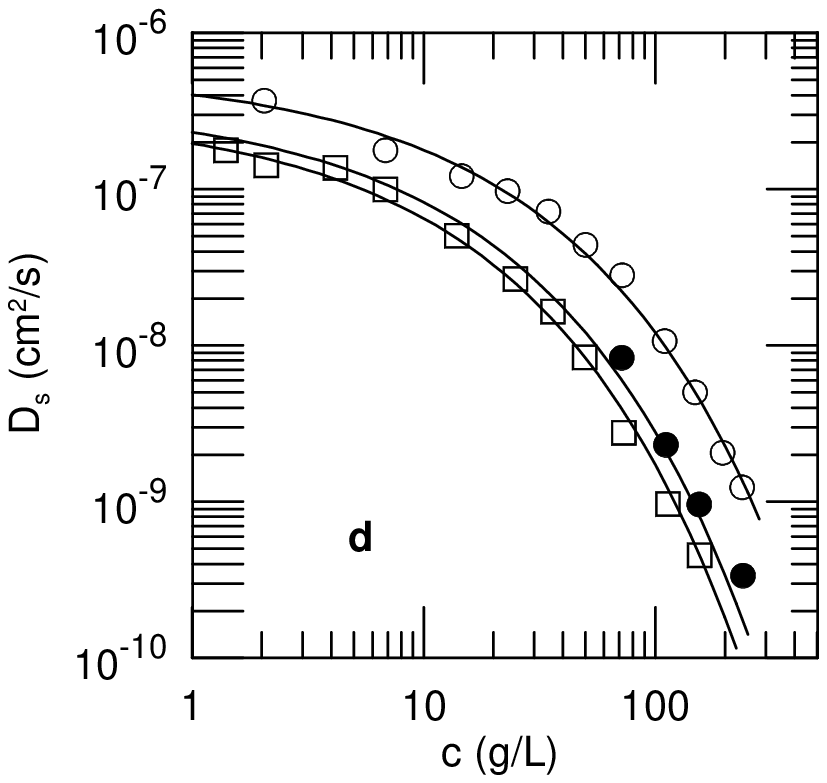} %br02j,ca03j,ca03k,de02j

\caption{\label{figure27} $D_{s}$ as measured by (a) Brown, et 
al.\cite{brown1983a} (cf.\ Fig.\ \ref{figure3}), (b) Callaghan, et al.\cite{callaghan1981a} (cf.\ Fig.\ \ref{figure4}),  (c) Callaghan, et al.\cite{callaghan1981a} (cf.\ Fig.\ \ref{figure5}), and (d) Deschamps, et al.\cite{deschamps1986a} (cf.\ Fig.\ \ref{figure6}), and fits to eq.\ 
\ref{eq:Dsseeq2}, leading to the parameters in Table \ref{table4}.}
 
\end{figure*}

\begin{figure*}
\includegraphics{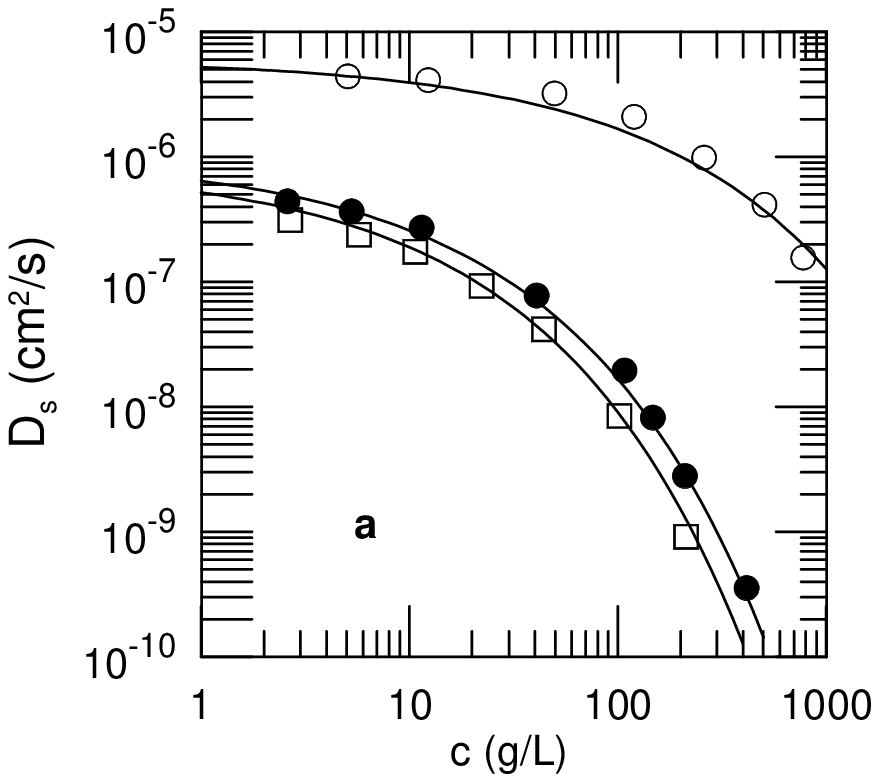} %gi01j,vm34j,xu011j,xu012k
\includegraphics{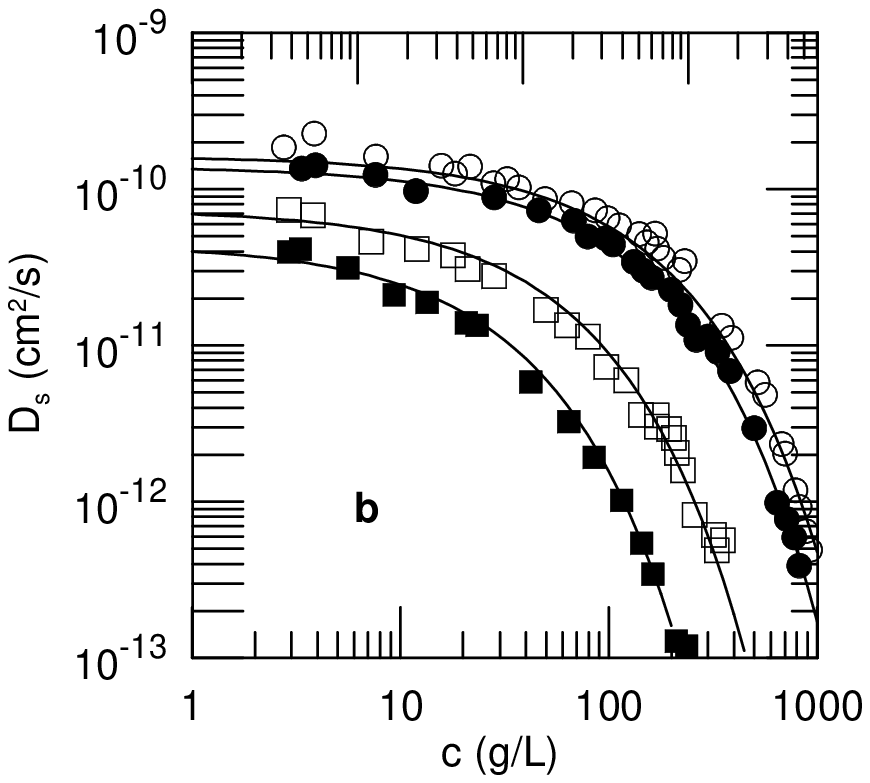} %gi01j,vm34j,xu011j,xu012k
\includegraphics{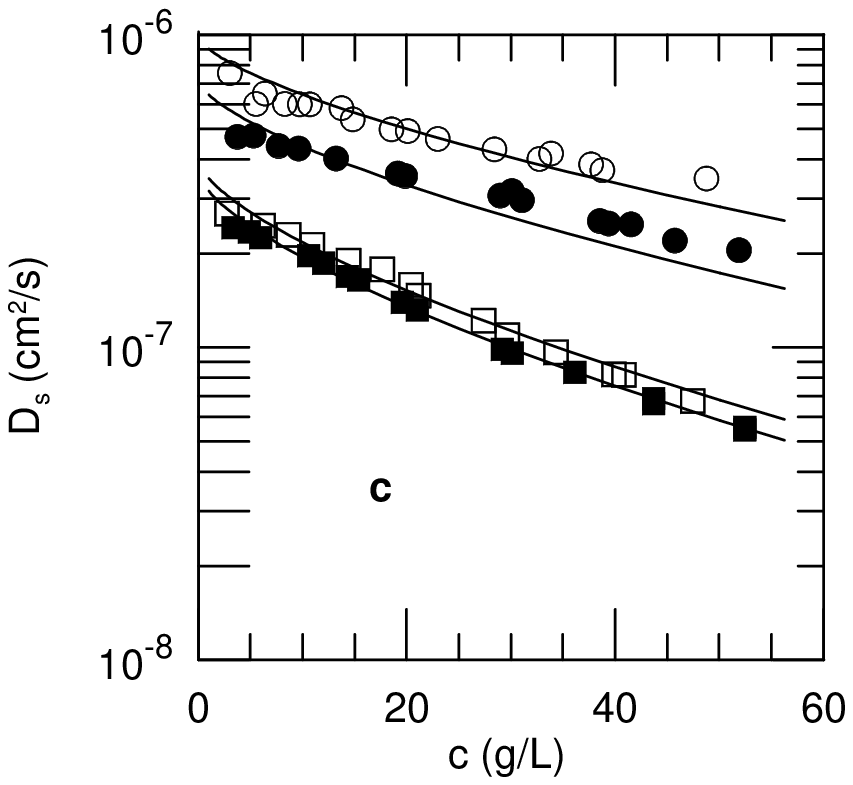} %gi01j,vm34j,xu011j,xu012k
\includegraphics{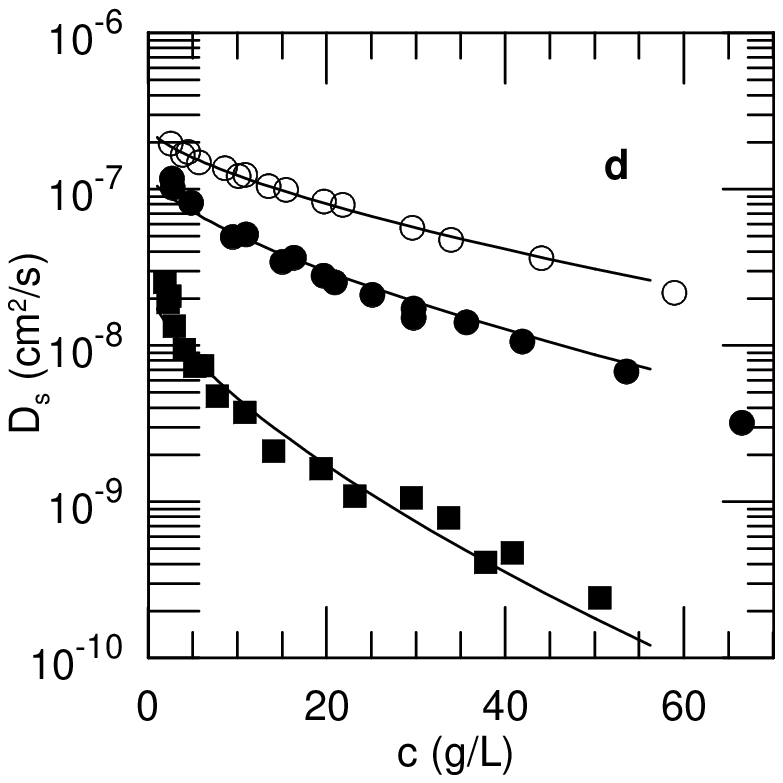} %gi01j,vm34j,xu011j,xu012k

\caption{\label{figure28} $D_{s}$ as measured by (a) Giebel, et 
al.\cite{giebel1993a}, (b) von Meerwall, et al.\cite{vonmeerwall1983a} and (c), 
(d) Xuexin, et al.\cite{vonmeerwall1984a},
and fits 
to eq.\ \ref{eq:Dsseeq2}, leading to the parameters in Table \ref{table4}.  Other details 
as in Figs.\ \ref{figure8}, \ref{figure21}, \ref{figure24}, and \ref{figure25}, respectively.} 
\end{figure*}

\begin{figure}

\includegraphics{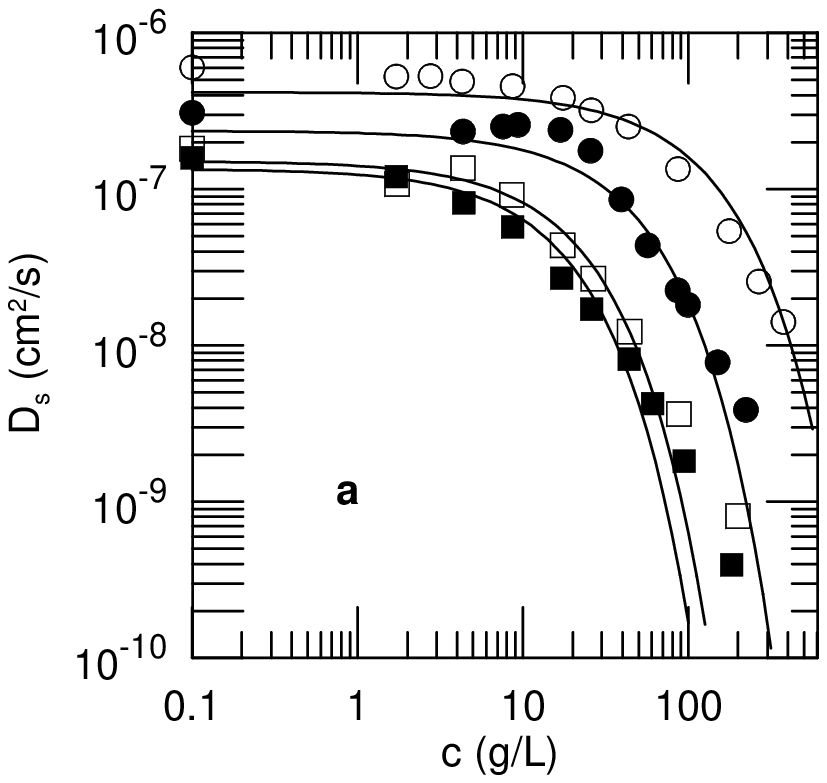} %
\includegraphics{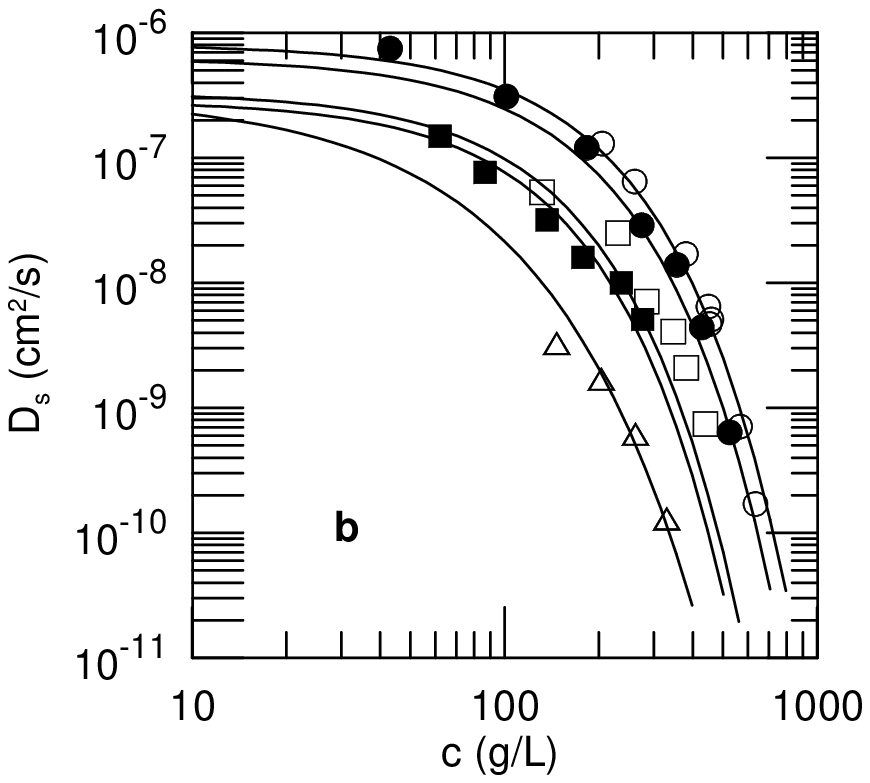} %

\caption{\label{figure29ab}  Joint fits to all data of (a) Leger, et 
al.\cite{leger1981a} and (b) Wesson, et al.\cite{wesson1984a}, showing the 
poor quality of fits of eq.\ \ref{eq:Dsseeq2} to these data sets.  
Fit parameters are 
in Table \ref{table4}; all other plot properties are the same as Figs.\ \ref{figure9}
and \ref{figure23}, 
respectively.}
\end{figure}

Nine of the above papers, namely Brown and Stilbs\cite{brown1983a}, Callaghan 
and Pinder\cite{callaghan1984a}, Deschamps and Leger\cite{deschamps1986a}, 
Giebel, et al.\cite{giebel1993a}, Leger, et al.\cite{leger1981a}, von Meerwall, 
et al.\cite{vonmeerwall1983a}, Wesson, et al.\cite{wesson1984a}, and Xuexin, et 
al.\cite{vonmeerwall1984a} report self-diffusion coefficients 
at a 
series of concentrations and polymer molecular weights
for a series of 
homologous polymers.   
A simultaneous fit of 
each of these sets of measurements 
to a joint function of $c$ and $M$ is then practicable.  
Here fits were made to eq.\ \ref{eq:Dsseeq2},
using the convention that the molecular weight of the sole polymer 
species is $P$.  The fits forced $\delta = 0$ to eliminate a notional
dependence  of $D_{s}$ on the 
molecular weight $M$ of a non-existent second polymer.
Results of the fits appear in Table IV and Figs.\ \ref{figure27}-\ref{figure29ab}.

In a majority of cases, the fits were quite good, with RMS fractional errors in 
the range 6-18\%.  Fits to the measurements of Leger, et al.\cite{leger1981a} 
and Wesson, et al.\cite{wesson1984a} were less satisfactory; these are 
discussed separately below. Xuexin, et al.\cite{vonmeerwall1984a} cover an extremely 
broad range of $M$ in their study of 18-armed stars.  Over this range, $\nu$ 
from fits to data sets covering chains with a single molecular weight
changes substantially, so a fit of all data over a full range of $M$
to eq.\ \ref{eq:Dsseeq2}, with $\nu$ a fixed 
constant, does not work well for very small or very large molecular weights.  On 
limiting the fit to intermediate values for $P$, good agreement between the 
data and the fitted forms are encountered.  Solid curves in Fig.\ \ref{figure28}
were calculated using the parameters to the fit to data in the intermediate-$P$ 
range.

Omitting momentarily the fits to results of Leger, et al.\cite{leger1981a} and Wesson, et 
al.\cite{wesson1984a}:  The exponent $a$ for the molecular weight dependence of 
the bare diffusion coefficient is consistently -0.5.  Except for Browne, et 
al.'s work on polyethylene oxide: water, the concentration exponent $\nu$ is in 
the range 0.5-0.75; Brown, et al.'s data \cite{brown1983a} 
imply $\nu \approx 0.93$.  The molecular weight exponent $\gamma$ is in 
the range 0.32-0.46, again with the exception of fits to data of Brown, et 
al.\cite{brown1983a}, for which $\gamma \approx 0.6$.  Inspection of the 
figures 
indicates that a joint stretched exponential in $c$ and $M$
fits each data set well, with no systematic deviations 
for particular values of $c$ or $M$.

The two data sets that are fit less well by eq.\ \ref{eq:Dsseeq2}
appear in Figure \ref{figure29ab}.
The merged fit to Leger, et al.\cite{leger1981a}'s data on 
polystyrene:CCl$_{4}$ is poor.  However, 
Leger, et al's data shows features -- notably a non-monotonic dependence of 
$D_{s}$ on $c$ -- that appears in data on no other polymer system, including 
other experiments that determined the 
concentration dependence of $D_{s}$ of the same polymer.  We infer that the 
poor fit of eq.\ \ref{eq:Dsseeq2} to Leger, et al.'s data arises from features 
unique to this polymer sample and set of measurements and not to a 
generic behavior of $D_{s}$ for polystyrene solutions.

The merged fit to Wesson, et al.'s\cite{wesson1984a} measurements is also
poor.  However, this set of data is limited to elevated concentrations in
which $D_{s}/D_{s0} < 1$, generally substantially without matching 
low-concentration measurements on the regime in which $D_{s} \approx D_{s0}$.  
By inspection of Fig.\ \ref{figure29ab}b, the merged fit works reasonably well at
small and large polymer molecular weight, but is least satisfactory at
intermediate molecular weights.  In particular, the 105 and 130 kDa
polymers have nearly the same $P$ but at elevated concentrations
substantially different self-diffusion coefficients.

\section{Diffusion of Probe Chains through Matrix Polymer Solutions}

This Section reviews measurements on the diffusion of polymeric probe 
molecules through solutions of a different polymer. These experiments involve
intrinsically ternary solutions in which the molecular 
weight $P$ of the probe polymer and the molecular weight $M$ of the matrix 
polymer are not the same.  In some cases, the probe and matrix 
polymers in the solution have a common monomer, and differ only in their molecular weights.  In other cases, the 
probe and matrix polymers are chemically distinct.  
Studies are again presented alphabetically by first author, 
together with fits of the data sets to 
stretched exponentials in concentration and chain molecular weights.  
Fitting 
parameters appear in Tables \ref{table2} and \ref{table4}.

Brown and Rymden\cite{brown1988a} used quasielastic light scattering to study 
the diffusion of linear polystyrenes and coated silica spheres through 
polymethylmethacrylate in toluene.  
Toluene and PMMA are almost exactly index-matched, so scattering from these 
systems was dominated by scattering from the dilute probe chains.  The matrix 
PMMA's had molecular weights in the range 110 kDa--1.43 MDa.  
Probe polystyrenes had molecular weights of 2.95, 8, and 15 MDa, 
with $M_{w}/M_{n}$ of 1.06, 1.08, and 1.30, respectively.  In the original 
paper, data were reported after normalization by
an unspecified diffusion coefficient obtained in the absence of the matrix 
polymer.  

\begin{figure} 

\includegraphics{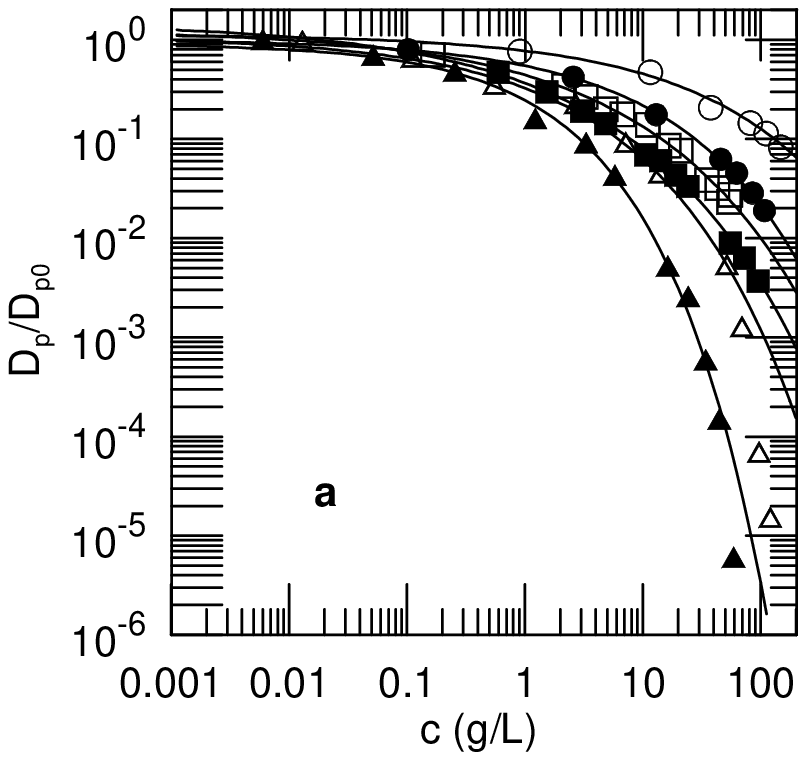} %br04a
\includegraphics{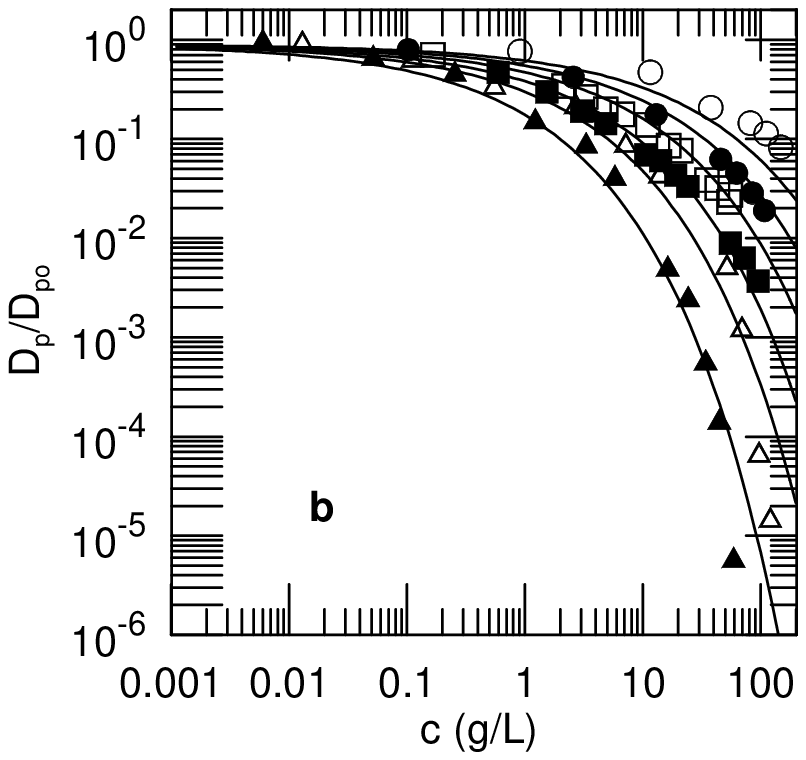} %br04aa

\caption{\label{figure29}   $D_{p}/D_{p0}$ of 8 MDa polystyrene diffusing 
through [top to bottom] 101, 163, 268, 445, 697, and 1426 kDa 
polymethylmethacrylate in toluene\cite{brown1988a}, and  (a) separate fits of the data for each 
matrix polymer to a stretched exponential in $c$, and (b) simultaneous fit of all data to a 
single stretched exponential in $c$ and $M$.}
\end{figure}

Reference \onlinecite{brown1988a} reported how $D_{p}$  depends on matrix 
concentration and molecular weight.  Figure \ref{figure29}a shows $D_{p}/D_{p0}$ for 
the 8MDa polystyrene diffusing through each of six matrix 
polymethylmethacrylates.  Each solid line represents a 
fit to a stretched exponential in 
matrix polymer concentration, using parameters in Table II.   For 
$D_{p}/D_{p0} < 10^{-3}$ or so, a condition attained only with the two largest 
matrix polymers, the measured $D_{p}$ deviates markedly downward from a 
stretched exponential, so that fits to all data on these two systems show 
fractional RMS errors much worse (32--45\% rather than 5--8\%) than fits to the 
same probe polymer with the smaller matrix polymers.  On excluding the final 
few points from the two fits, curves with
far lower RMS fractional errors were generated.  These are the curves shown
in the Figure.  It is difficult to determine from these 
measurements whether there is a systematic change from stretched-exponential to 
some other concentration dependence at very small $D_{p}/D_{p0}$, or whether 
the apparent deviations arise from experimental challenges at very small 
$D_{p}$.  

Figure \ref{figure29}b shows the same data, now fit simultaneously to eqn.\ \ref{eq:Dsseeq2}, the 
joint stretched exponential in $c$ and $M$.  Fitting parameters appear in 
Table IV. Except for the lowest-$M$ matrix 
polymer, the fits are almost as good as 
the individual fits shown in Figure \ref{figure29}a.  For the lowest-$M$ 101 kDa 
matrix polymer, the fitted form predicts too strong a dependence of $D_{p}$ 
upon $c$.

\begin{figure}

\includegraphics{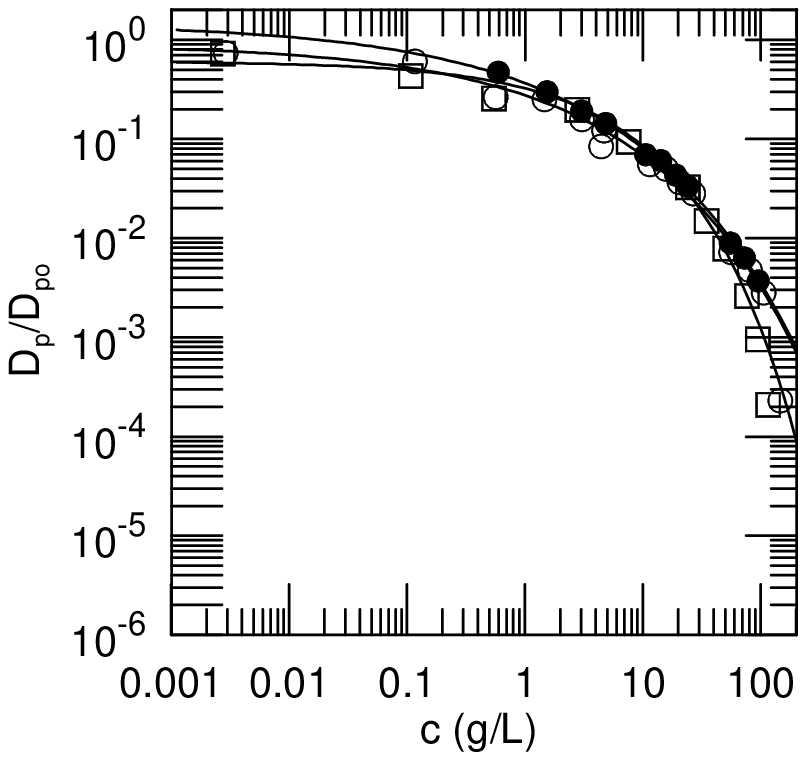} %br04b

\caption{\label{figure31} $D_{p}/D_{p0}$ against $c$
for 3 (open circles), 8 (closed circles), and 15 (squares) 
MDa polystyrenes diffusing through 445 kDa 
polymethylmethacrylate in toluene, using data of Brown and 
Rymden\cite{brown1988a}, and fits to stretched exponentials in matrix 
concentration. }
\end{figure}

Brown and Rymden\cite{brown1988a} also examined how $D_{p}$  depends on probe 
molecular weight.  Figure \ref{figure31} shows $D_{p}$ of the 3, 8, and 15 MDa probe 
polystyrenes, all diffusing through the 445kDa PMMA.  Fits are to stretched 
exponentials in $c$, leading to parameters given in Table \ref{table2}.  The matrix 
chains are all much larger than the probe chains.  As noted by the original 
authors\cite{brown1988a}, the three curves come very close to superposing except perhaps at 
the very highest matrix concentrations examined.  

Brown and Stilbs\cite{brown1983b} used PFGNMR to measure the probe 
diffusion 
coefficient of polyethylene oxide in aqueous solutions of dextran.  
Polyethylene oxides had molecular weights of 73, 278, and 1200 kDa with 
$M_{w}/M_{n}$ of 1.02-1.12; dextrans had molecular weights of 19, 110, and 510 
kDa.  The 1200 kDa PEO represented the lower limit at which $D_{p}$ could be 
determined with then-available technology; the authors limited their detailed 
analysis to the two lower-molecular-weight probes.  

\begin{figure} 

\includegraphics{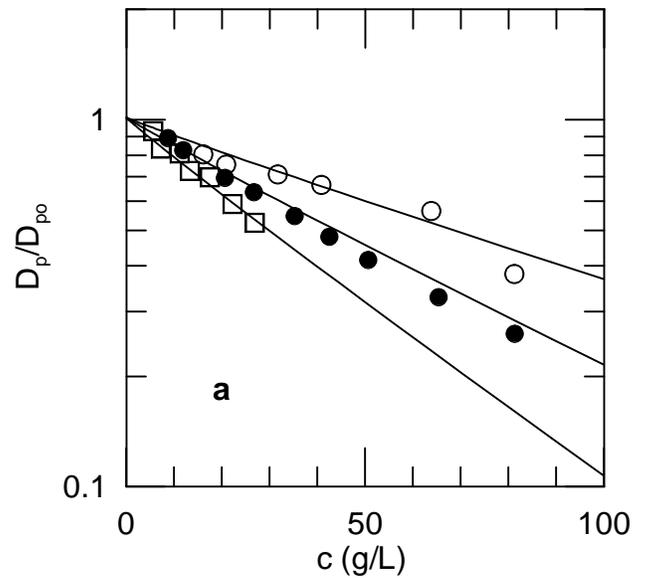} %br034
\includegraphics{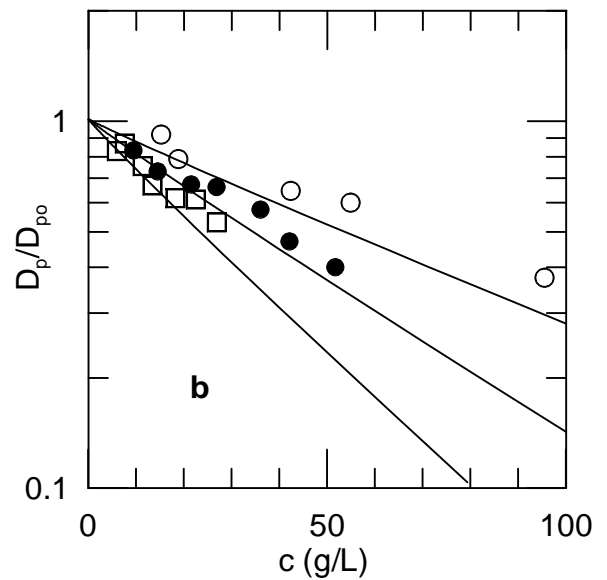} %br035

\caption{\label{figure32}  $D_{p}/D_{o0}$ of  (a) 73 kDa polyethylene oxide and (b) 278 kDa polyethylene oxide
in [top to bottom] 19, 110, and 510 kDa dextrans in aqueous solution,
and simultaneous fit of all measurements 
to a stretched exponential in $c$, $M$, and $P$, 
using data of Brown and Stilbs\cite{brown1983b}.}
\end{figure}

Figures \ref{figure32}a and \ref{figure32}b show $D_{p}$ of the 73 and 278 kDa polyethylene 
oxides, normalized by the measured $D_{o0}$ of the same probes in the absence 
of the matrix polymers.  All measurements were simultaneously fit to 
$D_{po}/D_{o0} \exp(-\alpha c^{\nu} P^{\gamma} M^{\delta})$, yielding the 
smooth curves shown in the Figures and the parameters listed in Table \ref{table3}.  
Agreement between the data and the
fitted curves was very good for the 73 kDa probe.  For 
the 278 kDa probe in the 19 kDa matrix, the fitting function significantly 
underpredicts $D_{p}$.  From the fitting parameters, $D_{p}/D_{o0}$ has a very 
weak dependence on the probe molecular weight (other than that hidden in 
$D_{o0}$), but has a marked dependence on the matrix molecular weight.  The 
same data were also fit, individually for each probe:matrix pair, to a
stretched and a pure
exponential in $c$.  
Without exception, for each probe:matrix pair $D_{s}(c)$  
follows accurately the exponential form, RMS fractional errors being 
in the range 1-4\%.  Because $D_{p}$ varied over a limited range, the 
parameters reported in Table II reflect fits to the pure exponential. 

\begin{figure} 

\includegraphics{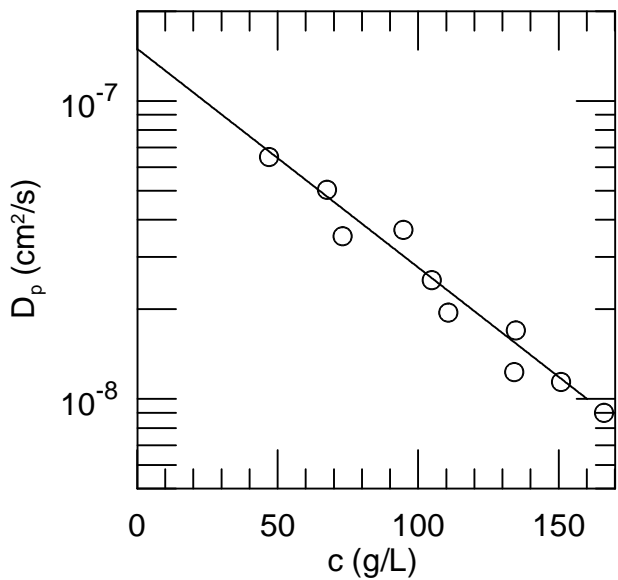} %da011

\caption{\label{figure34} $D_{p}$ of 864 kDa dextran in solutions of 20 kDa 
dextrans 
and fits to stretched exponentials, using data of Daivis, et 
al.\cite{daivis1984a}.} 
\end{figure}

Daivis, et al.\cite{daivis1984a} used quasielastic light scattering 
spectroscopy to measure diffusion of relatively dilute 864 kDa dextran in 
not-necessarily-dilute solutions of 20.4 kDa dextran.  Polymer polydispersities 
were in the range 1.24-1.3.  The concentration of the lower-molecular-weight 
dextran ranged up to 166 g/L.  Analysis of the bimodal QELSS spectra of these 
systems shows that the slower mode corresponds to probe diffusion by the 864 
kDa dextrans.  As seen in Figure \ref{figure34}, the data fit reasonably well to a 
simple exponential concentration dependence, using parameters found in Table 
II.  

In a separate paper, Daivis, et al.\cite{daivis1992a} used QELSS
and PFGNMR to measure the diffusion of a 110 kDa 
polystyrene, $M_{w}/M_{n} = 1.06$, through solutions of 110 kDa 
polyvinylmethylether, $M_{w}/M_{n} \approx 1.3$, in the PVME's isorefractive 
solvent toluene.  Good agreement was found 
between the QELSS measurements of $D_{p}$ and the earlier measurements of 
$D_{p}$ by Martin\cite{martin1986a} on the same system.  

\begin{figure} 

\includegraphics{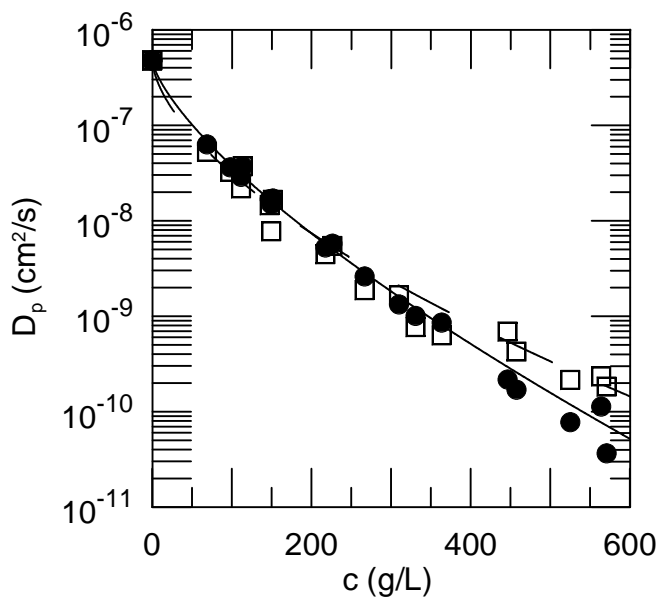} %da012

\caption{\label{figure35}  $D_{p}$ of 110 kDa polystyrene through 110 kDa 
polyvinylmethylether:toluene, based on QELSS (circles) and PFGNMR (squares) 
measurements of Daivis, et al.\cite{daivis1992a}, together with fits (solid, 
dashed lines, respectively) to stretched exponentials in the matrix 
concentration.} 
\end{figure}

Figure \ref{figure35} shows Daivis, et al.'s data\cite{daivis1992a} as obtained using 
both
physical methods.
In the figure, lines represent separate fits of each data set to a stretched 
exponential in matrix polymer concentration, with fitting parameters given in 
Table \ref{table2}.  Stretched exponentials in $c$ describe well each data set.  The 
QELSS data is significantly less scattered than the PFGNMR data, so the 
parameters from the former's fit are probably to be preferred.  

\begin{figure} 

\includegraphics{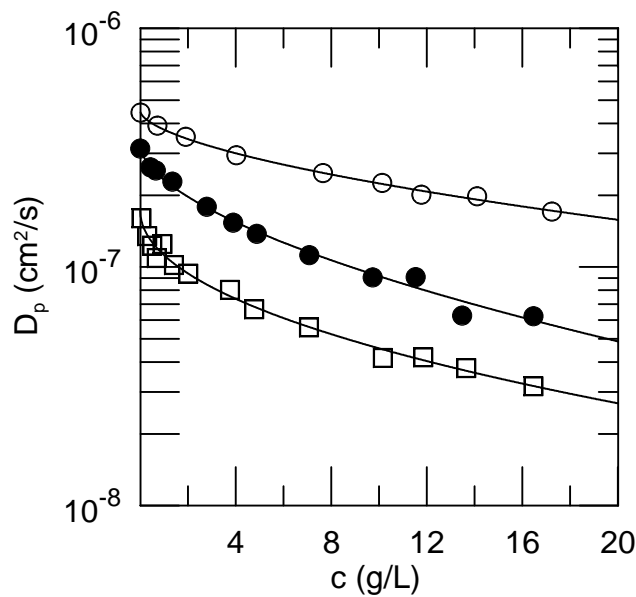} %de01

\caption{\label{figure36}  $D_{p}$ of [from top to bottom] 71, 148, and 487 
kDa 
dextrans in $M_{w}$ 680 kDa hyaluronic acid, 
and fits to stretched exponentials, using data of De Smedt, et 
al.\cite{desmedt1994a}.} 
\end{figure}

De Smedt, et al.\cite{desmedt1994a} used FRAP 
to measure the diffusion of 71, 148, and 487 kDa dextrans 
($M_{w}/M_{n} < 1.35$),
labeled with fluorescein isothiocyanate, through solutions of hyaluronic acid 
($M_{n} = 390$ kDa; $M_{w} = 680$ kDa).  Hyaluronic acid concentrations ranged 
from the dilute up to 18 g/L. $D_{p}$ of the dextrans varied roughly five-fold 
over this concentration range.  De Smedt, et al.'s data\cite{desmedt1994a} 
appear in Fig.\ \ref{figure36}, together with stretched-exponential fits using the 
parameters in Table II.  As seen in the Figure 
and initially reported by the original 
authors, the data fits well to stretched exponential forms.

\begin{figure} 

\includegraphics{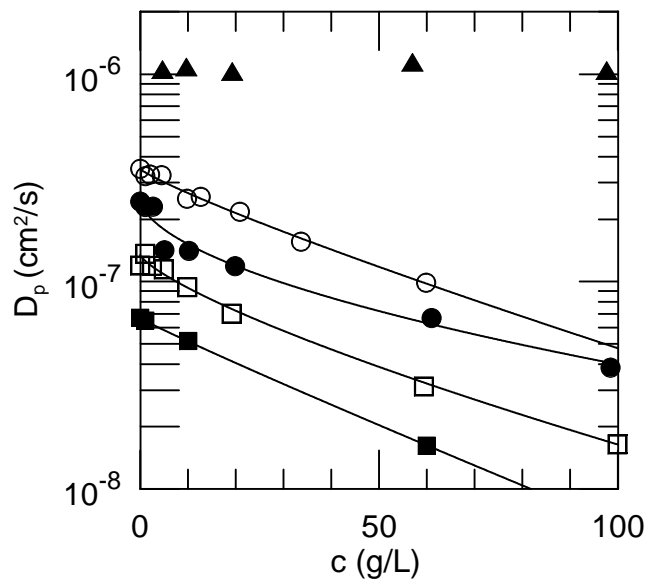} %ha021

\caption{\label{figure37} $D_{p}$ of [from top to bottom] 25, 162, 410, 1110, 
and 4600 kDa 
polystyrenes 
in 1.05 MDa polymethylmethacrylate: CCl$_{4}$ as a function of polymethylmethacrylate 
concentration, and fits to stretched exponentials, using data of Hadgraft, et 
al.\cite{hadgraft1979a}.} 
\end{figure}

\begin{figure} 

\includegraphics{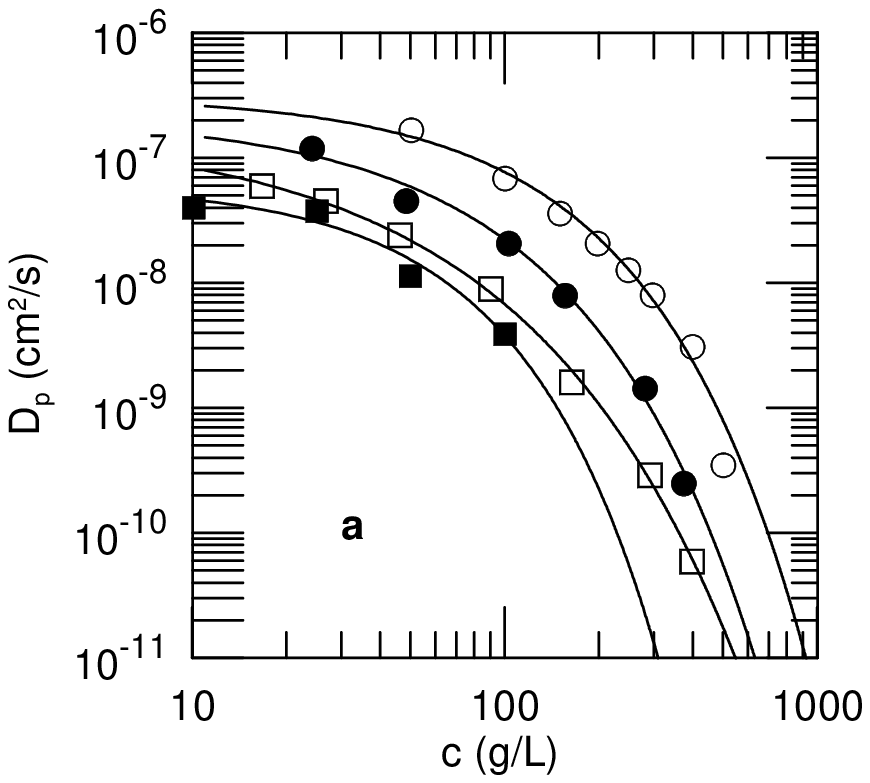} %ha01
\includegraphics{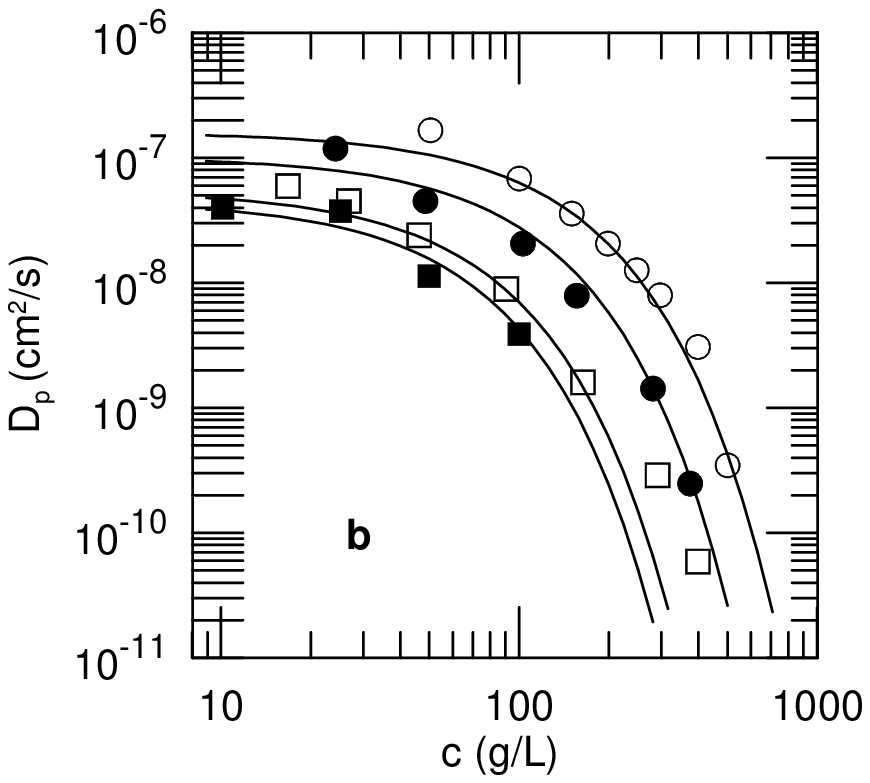} %ha015

\caption{\label{figure38}  $D_{p}$ of [from top to bottom] 50, 179, 1050, 
and 1800 kDa polystyrenes in orthofluorotoluene solutions of 60kDa 
polyvinylmethylether against matrix polymer concentration, 
using data of Hanley, et al.\cite{hanley1985a} and (a) separate 
fits at each $P$ and (b) simultaneous fits at all $P$, with fitting parameters from Tables 
\ref{table2} and \ref{table4}, respectively.}  
\end{figure} 

Hadgraft, et al.\cite{hadgraft1979a} used QELSS to
study the diffusion of polystyrene probe polymers, molecular weights 25, 162, 
410, 1110, and 4600 kDa through solutions of 105 kDa polymethylmethacrylate in 
its isorefractive solvent benzene at 
PMMA concentrations up to 100 g/L.  Polystyrene and PMMA are not compatible, 
implying that the radius of the polystyrene chains may have depended very 
strongly on the matrix polymer concentration.  As seen in Fig.\ \ref{figure37}, 
$D_{p}$ of the 25 kDa polystyrene was substantially independent of PMMA 
concentration.  The data on the 410 kDa polymer is significantly more 
scattered than is data on the other polystyrenes.  The range of variation of 
$D_{p}$ is sufficiently small (roughly a factor of three) that the fits 
to these data are 
less reliable than are the fits to data on
some other systems.  Nonetheless, $D_{p}$ of the 
higher-molecular-weight 
polystyrenes shows a stretched-exponential dependence on PMMA concentration, 
with parameters seen in Table II.  

Hanley, et al.\cite{hanley1985a} used light scattering spectroscopy to examine
the diffusion of polystyrenes through the matrix polymer polyvinylmethylether 
in its isorefractive solvent orthofluorotoluene.  The polystyrenes had 
molecular weights of 50, 179, 1050, and 1800 kDa.  The polyvinylmethylether had 
$M_{w} \approx 60$ kDa with $M_{w}/M_{n} \approx 3$.  Hanley, et al.'s data\cite{hanley1985a} are given in Fig.\ \ref{figure38}a, together with fits of separate stretched-exponentials in $c$ to each data set, yielding 
parameters shown in Table \ref{table2}.  The entire data set was also fit simultaneously 
to the joint stretched exponential of eq.\ \ref{eq:Dsseeq2}, yielding 
parameters seen in Table \ref{table4}, and fitted curves seen in Fig.\ \ref{figure38}b.   The 
lack of measurements at very low matrix concentration substantially broadens 
the range of fitting parameters that yield reasonable descriptions of this 
data.  The fit to the joint stretched exponential is 
substantially less satisfactory than the individual fits at each $P$
to separate stretched exponentials (RMS fractional 
errors of 34\% rather than 9-24\%).  The inadequacy of eq.\ \ref{eq:Dsseeq2}
relates primarily to the 1050 kDa probe chains, for which $D_{p}$ was determined only over a limited range.  As seen in Table \ref{table4}, excluding data on the 1050 kDa probes 
from the simultaneous fit leads to a marked reduction 
in the RMS fractional fit error (to 24\%), modest changes in $\alpha$ and 
$\nu$, but only small changes in the other fitting parameters.  

Kent, et al.\cite{kent1992a} applied static and quasielastic light scattering 
to measure radii of gyration and diffusion coefficients of 233 and 930 kDa 
polystyrenes in 7, 66, 70, 840, and 1300 kDa polymethylmethacrylates in ethyl 
benzoate (for static light scattering) and toluene (for quasielastic light 
scattering).   With one exception (66 kDa PMMA), $M_{w}/M_{n}$ was always $\leq 
1.10$. Different probe:matrix combinations were used for static and 
quasielastic light scattering.  $D$ of the polystyrene was measured as a 
function of polystyrene concentration, and linearly extrapolated to the 
dilute-in-polystyrene limit, thereby obtaining both the probe diffusion 
coefficient $D_{p}$ of the polystyrene and the initial linear dependence of 
$D_{p}$ on polystyrene concentration.  

\begin{figure}

\includegraphics{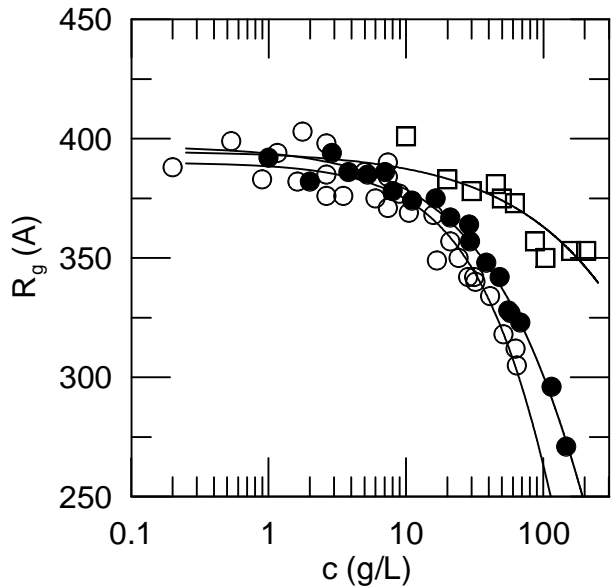} %ke011

\caption{\label{figure40} $R_{g}$ of 930 kDa polystyrene in
[from top to bottom] 7, 70, and 1300 kDa
polymethylmethacrylate: ethylbenzoate as a function of polymethylmethacrylate 
concentration, and fits to stretched exponentials, using data of 
Kent, et 
al.\cite{kent1992a}.} 
\end{figure}

Figure \ref{figure40} shows Kent, et al.'s\cite{kent1992a} measurements of the radius
of gyration of 930 kDa polystyrene in solutions of 7, 70, or 1300 kDa PMMA as a 
function of PMMA concentration, together with fits of $R_{g}$ to a stretched 
exponential
\begin{equation}
  R_{g} =  R_{g0} \exp(- \alpha c^{\nu})
  \label{eq:rg(c)}
\end{equation}
in matrix concentration $c$.  The fits are good throughout, using 
parameters in Table \ref{table5}.  Unlike the 
scaling-law prediction $R_{g} \sim c^{-0.25}$, the stretched exponential 
form shows acceptable behavior down to zero matrix concentration.

\begin{figure} 

\includegraphics{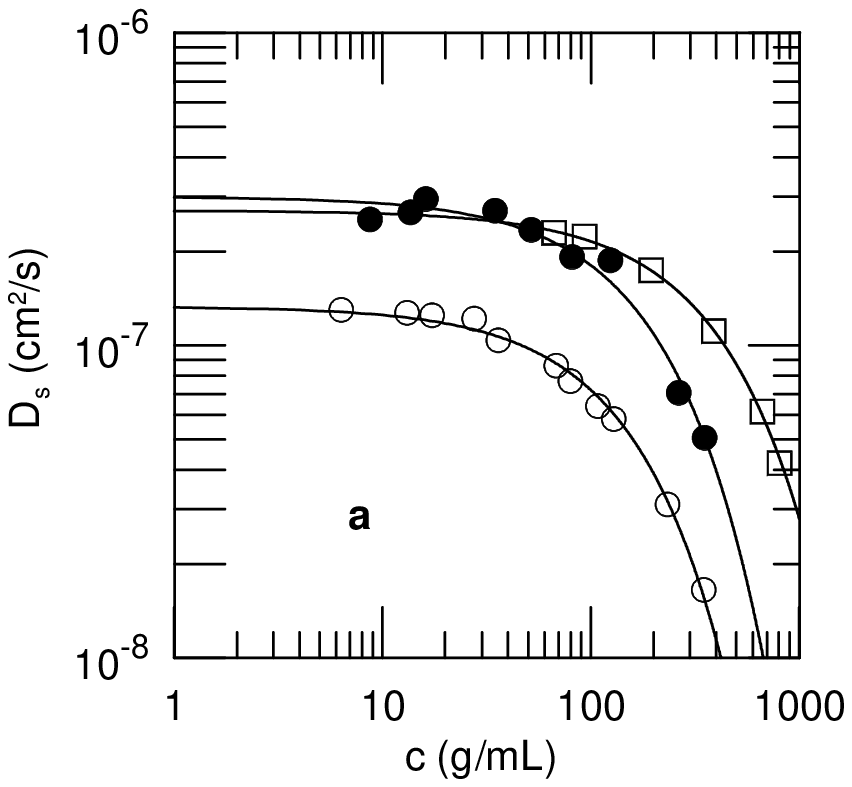} %ke012

\includegraphics{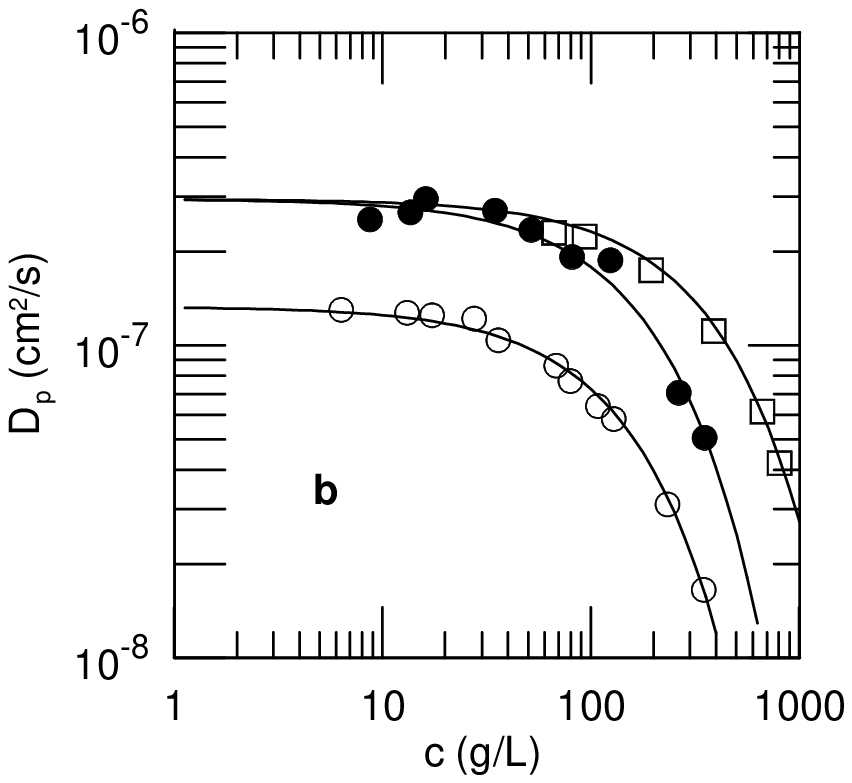} %ke013

\caption{\label{figure41} $D_{p}$ of polystyrene through 
polymethylmethacrylate: toluene with molecular weight combinations $P:M$
[from top to bottom]  233:66, 233:840, and 930:840 kDa
as functions of polymethylmethacrylate 
concentration,  using data of Kent, et 
al.\cite{kent1992a}, and (a) a separate fit to a
stretched exponential for each combination and (b)a single fit to a joint stretched exponential in $c$, $P$, and $M$. }
\end{figure}

Kent, et al.\cite{kent1992a} also measured $D_{p}$ of 233 kDa polystyrene 
in solutions of 66 and 840 kDa PMMA, and 930 kDa polystyrene through 840 kDa PMMA.
Their experimental data is shown in Fig.\ \ref{figure41}a, together with fits to 
exponentials using parameters given in Table \ref{table2}.  Within experimental 
error, the simple exponential fits with $\nu = 1$ forced are as good as the 
stretched-exponential fits to the data: the former are in the 
figure.  Experimentally, the scaling prefactor $\alpha$ depends strongly on 
matrix molecular weight (a 12-fold change in $M$ leads to a two-fold change 
in $\alpha$) but at most weakly on probe molecular weight. Figure \ref{figure41}b
shows a fit of the same data to a joint stretched exponential in $c$, $P$, and 
$M$, based on parameters in Table \ref{table4}.  The joint fit clearly works well for all 
three polymer pairs.

Kim, et al.\cite{kim1986a}  measured the diffusion of dye-labeled polystyrenes 
through matrix solutions of unlabeled polystyrenes in toluene.    The objective 
was to test the prediction of some scaling models that $D_{p}$ becomes 
independent of matrix molecular weight if $M/P \geq 1$.  The probe polystyrenes 
had molecular weights from 10 to 1800 kDa; matrix chains had molecular weights 
from 51 to 8400 kDa.  Ref.\ \onlinecite{kim1986a} also reported the dependence of 
$D_{p}$ on the matrix molecular weight for several probes (51, 390, 900 kDa) at 
multiple matrix concentrations for matrix molecular weights in the range 35--8400 kDa.  Polymer polydispersities were largely $<1.06$, with a maximum of 1.17.  Kim, et al.\cite{kim1986a} also report limited data using methyl red as 
a low-molecular weight probe.  As seen in Fig.\ \ref{figure43}a, the authors found 
that $D_{p}$ becomes substantially independent of $M$ only if $M/P > 3$.  

With respect to the models discussed in Section II, the published derivations 
of the stretched-exponential form refer to polymer chains whose motions are 
adequately approximated by whole-chain translation and rotation.  These 
approximations are only likely to be adequate if the probe and matrix chains 
are of similar size, because otherwise the whole-chain motions of probe or 
matrix would effectively sample some of the internal modes of the other chain 
species, whether matrix or probe.  At fixed $c$, from the hydrodynamic model $D_{p}$ 
would have a stretched-exponential dependence on $M$ if $M/P \approx 1$, but 
might well not have a stretched-exponential matrix molecular weight dependence 
if $M/P \gg 1$ or $M/P \ll 1$.  Kim, et al.'s\cite{kim1986a} results thus are 
not necessarily  inconsistent with exponential-type models of polymer dynamics.

Kim, et al.'s data\cite{kim1986a} for the domain $M/P < 3$, as seen
in Figs.\ \ref{figure43}, 
were fit to the form 
$D_{o} P^{-a} \exp(-\alpha c^{\nu} P^{\gamma} M^{\delta})$, yielding parameters 
in Table \ref{table4}.
As seen in Fig.\ 
\ref{figure43}, $D_{p}$ is indeed described well by the joint stretched 
exponential over a wide range of $c$, $P$, and $M$, 
except in the regime 
$M/P > 3$ in which the form is not necessarily expected to apply.  

\begin{figure} 

\includegraphics{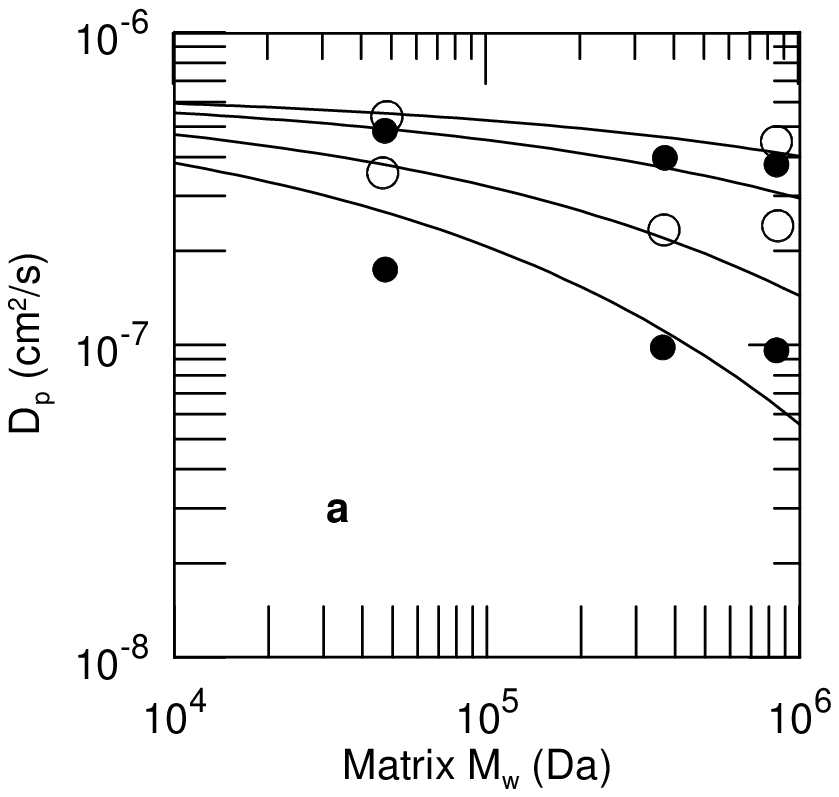} %ki013
\end{figure}
\begin{figure}
\includegraphics{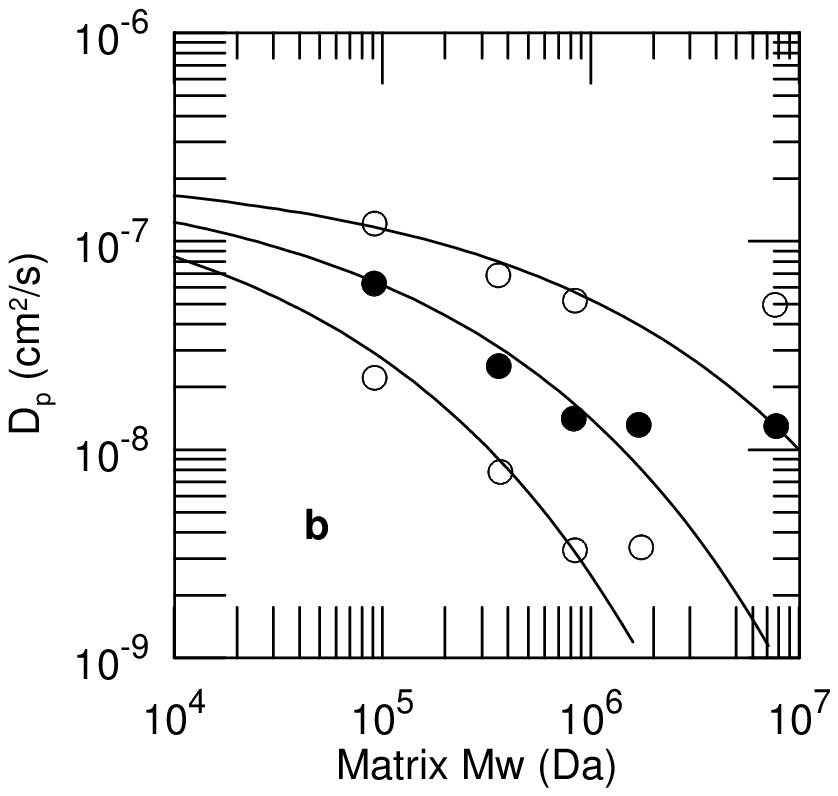} %ki014
\includegraphics{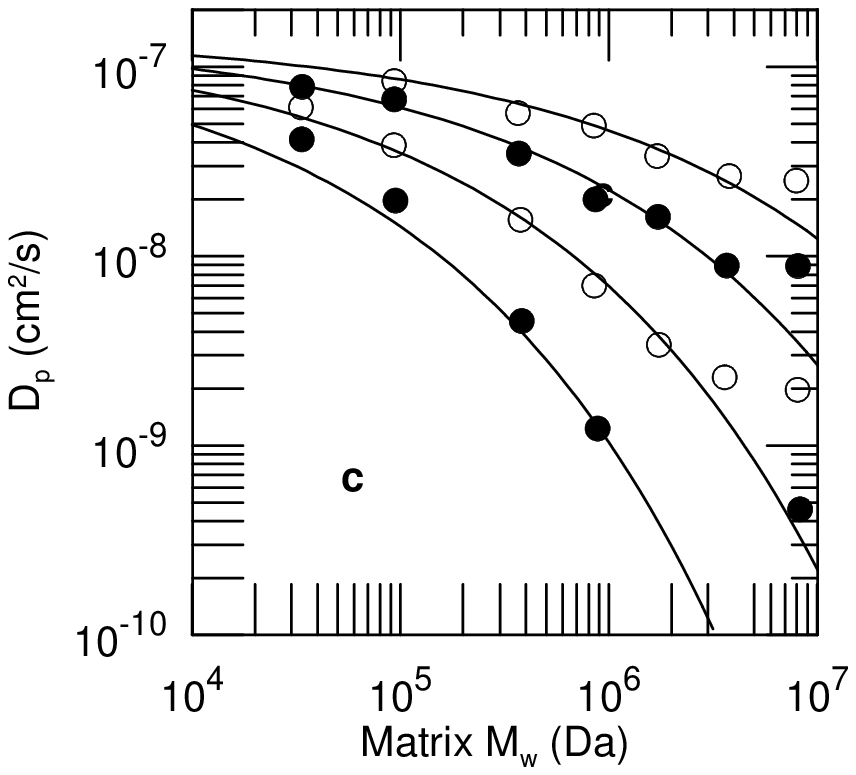} %ki015

\caption{\label{figure43} $D_{s}$ of polystyrene in matrix polystyrene:toluene 
solutions as functions of matrix molecular weight
at various matrix concentrations, based 
on data of Kim, et al.\cite{kim1986a}, Fig.\ 3.  Probe molecular weights were (a) 51 kDa, (b) 
390 kDa, and (c) 900 kDa. Matrix concentrations [top to bottom] were (a) 10, 
20, 50, and 100 g/L; (b) 20, 50, and 100 g/L; and (c) 10, 20, 40, and 80 g/L.
A single stretched exponential 
$D_{o} P^{-a} \exp(-\alpha c^{\nu} P^{\delta} M^{\gamma})$ with constant 
parameters was fit (solid 
lines) to all data in the figures having $M/P < 3$.}  
\end{figure}

\begin{figure} 

\includegraphics{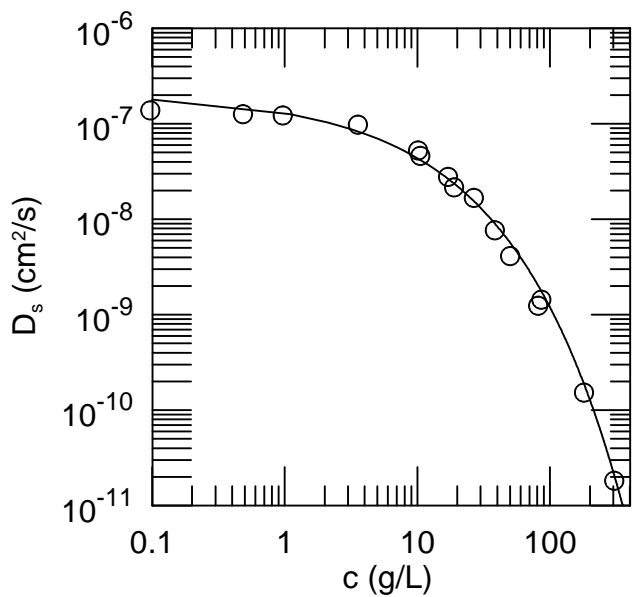} %ki011

\caption{\label{figure44} $D_{s}$ of 900 kDa
in toluene,  
based on data of Kim, et al.\cite{kim1986a}, Fig.\ 1,
and a fit to a stretched exponential.}
\end{figure}

\begin{figure} 

\includegraphics{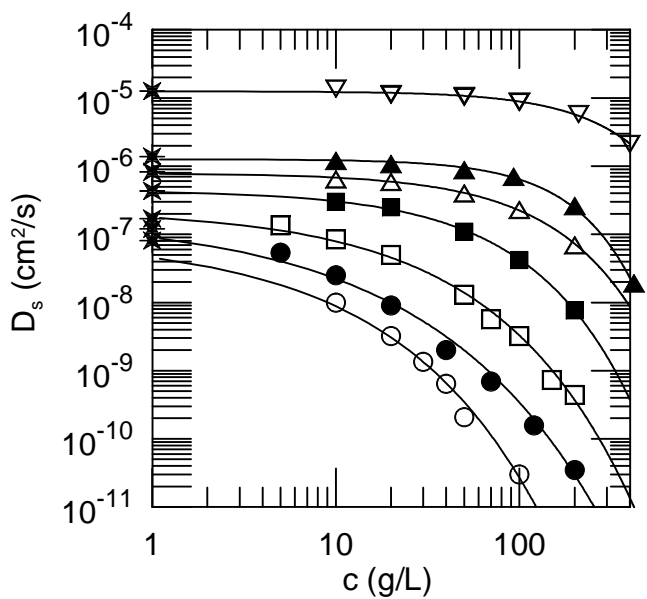} %ki012

\caption{\label{figure45} Probe diffusion of (top to bottom)
methyl red, and 10, 35, 100, 390, 900, and 1800 kDa $D_{s}$ 
polystyrenes through high-molecular-weight ($M/P > 3$)
polystyrenes in toluene, based on data of Kim, et al.\cite{kim1986a}, Table \ref{table2}, 
and fits to stretched exponentials.  The left axis shows zero-matrix-concentration data.}
\end{figure}

Kim, et al.\ also studied the concentration dependence of $D_{s}$ and
$D_{p}$ at large 
$M/P$.  Figure \ref{figure44} shows Kim, et al.'s\cite{kim1986a} determination of 
self-diffusion of 900 kDa polystyrene in toluene, and our stretched-exponential 
fit to the data, yielding the parameters in Table I.  Figure \ref{figure45} shows 
measurements of probe diffusion, using as probes methyl red and 10, 35, 100, 
390, 900, and 1800 kDa polystyrenes, all through matrix polystyrenes in 
toluene.  The molecular weight of the matrix polystyrene ranges from 51 to 8400 
kDa, varying from data point to data point, with $M/P >3$ and generally $M/P > 
6$.  This ratio of $M/P$ was chosen by Kim, et al.\ based on their 
interpretation of their data that showed  $D_{p}$ to be independent of $M$ for 
$M/P > 3$. As seen in Figs.\ \ref{figure44} and \ref{figure45}, stretched-exponential 
functional forms using parameters in Table II describe the matrix-polymer 
concentration dependences of $D_{s}(c)$ and $D_{p}(c)$ well over a wide range 
of polymer concentrations and probe molecular weights, even though the ranges 
of $P$ and $M$ are very wide.  The stretched exponential in $c$ continues to 
describe well the concentration dependence of $D_{p}$ in the large $M/P$ range 
which eq.\ \ref{eq:Dsseeq2} does not represent well the $P$ and $M$ dependences 
of $D_{s}$.  

Lodge and collaborators have reported an extensive series of studies of 
probe diffusion in polymer solutions, using quasielastic light scattering to 
measure $D_{p}$ of a dilute probe polymer, generally polystyrene, through the 
isorefractive matrix polymer:solvent pair polyvinylmethylether: 
orthofluorotoluene.  Variables studied include the probe and matrix molecular 
weights, the matrix concentration, and the topology (linear and star) of the 
probe polymers.

\begin{figure}

\includegraphics{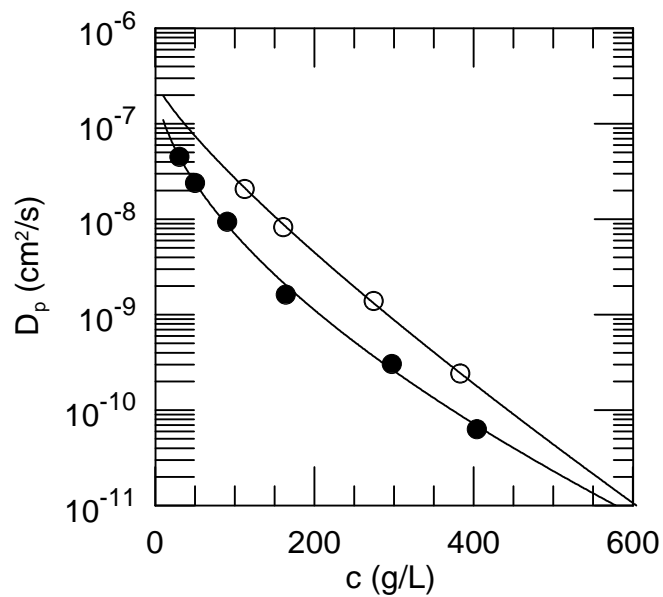} %lo01

\caption{\label{figure46}  $D_{p}$ of 179 kDa and 1.05 MDa polystyrene through 
50 kDa polyvinylmethylether in orthofluorotoluene, 
based on data of Lodge\cite{lodge1983a}, and fits to stretched exponentials in 
$c$.}
\end{figure}

An early letter\cite{lodge1983a} of Lodge reports $D_{p}$ of 179 kDa and 1.05 
MDa polystyrenes through a 50kDa polyvinylmethylether.  The data, and 
corresponding stretched-exponential fits, appear in Fig.\ \ref{figure46}.  Fit
parameters are in Table \ref{table2}.  A stretched exponential in concentration
describes well both data sets.  

Lodge and Wheeler\cite{lodge1986a} compared the diffusion of linear and 3-armed
star polystyrenes through a high molecular weight polyvinylmethylether.
Polystyrene molecular weights were 422 and 1050 kDa for the linear chains and 
379 and 1190 kDa for the stars; the polystyrenes were identified as being 
'relatively monodisperse'.  The PVME had an $M_{w}$ of 1.3 MDa, with an 
estimated $M_{w}/M_{N} \approx 1.3$.  $D_{p}$ was obtained over $1 \leq c \leq 
100$ g/L in PVME concentration.

\begin{figure} 

\includegraphics{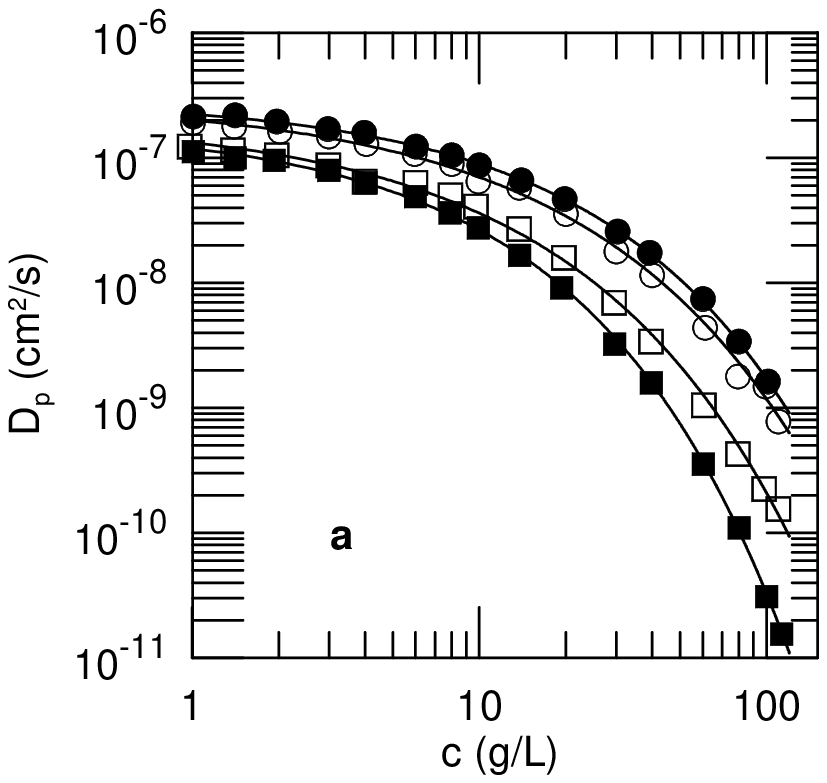} %lo21

\includegraphics{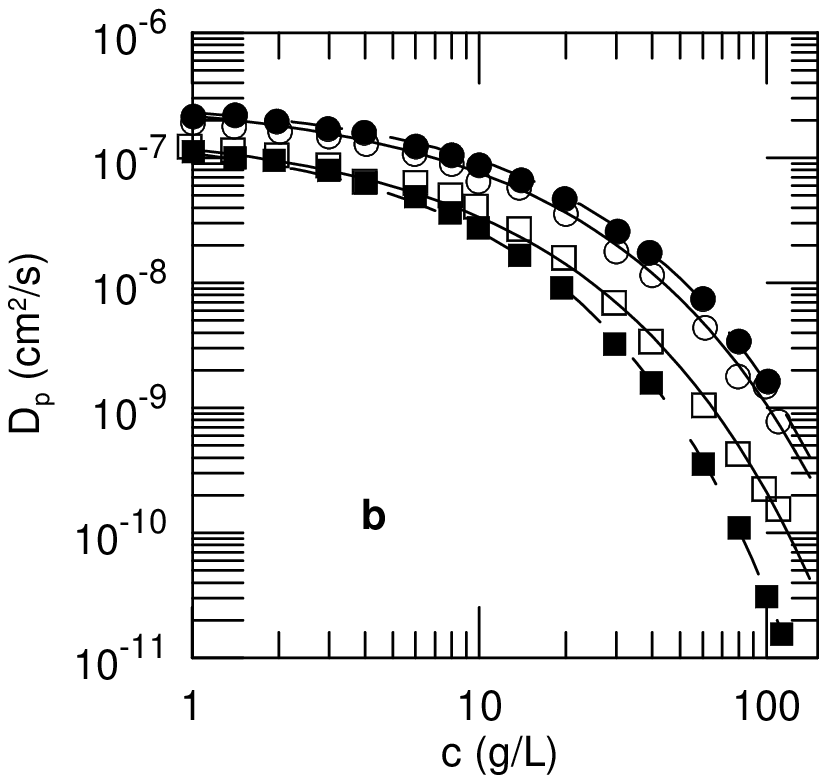} %lo25

\caption{\label{figure47} $D_{p}$ of polystyrenes [from top to bottom: 379 kDa 
$f=3$ star, 422 kDa linear chain, 1050 kDa linear chain, and 1190 kDa $f=3$ 
star] through 
1300 kDa polyvinylmethylether in orthofluorotoluene, 
based on data of Lodge\cite{lodge1986a} and fits of a
stretched exponential to (a) data on each probe separately and (b) all data on linear chains (solid lines)
and separately all data on 3-armed star polymers (dashed lines).}
\end{figure}

Figure \ref{figure47}a shows Lodge and Wheeler's measurements\cite{lodge1986a}, 
together with fits of data on each probe polymer to a stretched exponential in 
$c$.  $D_{p}$ varies over nearly four orders of magnitude.  Over the full range, 
agreement of the data with the functional form is very good, with RMS 
fractional errors of 2.6-10\% and parameters as seen in Table \ref{table2}
Figure \ref{figure47}b shows the same data, with the linear chains and star polymers 
separately fit to a stretched exponential in $c$ and $P$.  
As seen in Table \ref{table4}, the fractional errors in these fits are very nearly as 
good as the fits to individual probe species.  The $P^{-a}$ scaling 
of the zero-concentration diffusion coefficient and the $P^{\gamma}$ scaling of 
the scaling prefactor $\alpha$ account for the dependence of 
$D_{p}$ of linear and star polymers on probe molecular weight.

\begin{figure}

\includegraphics{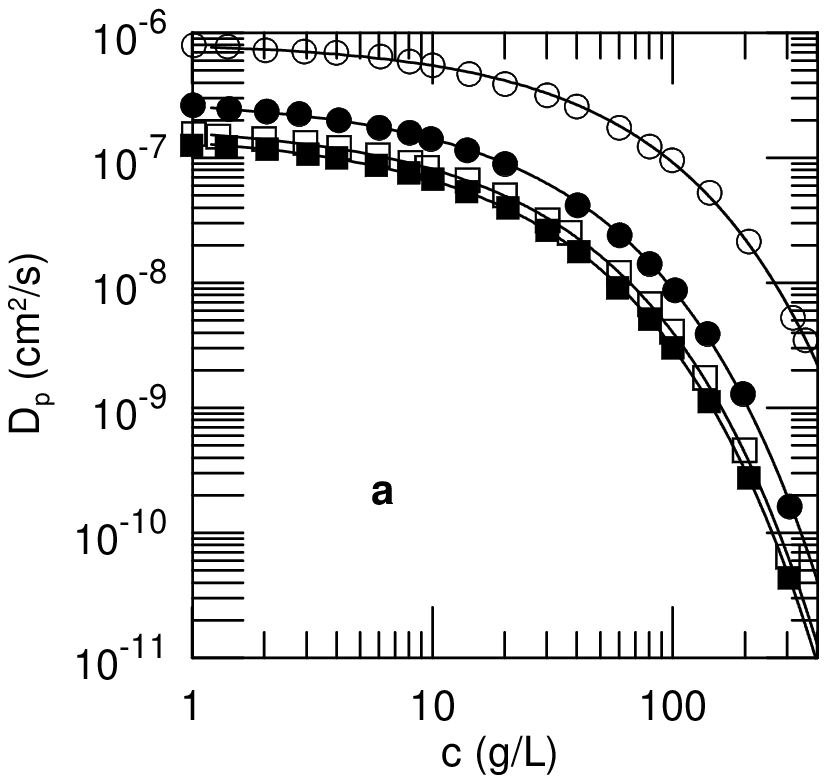} %lo031

\includegraphics{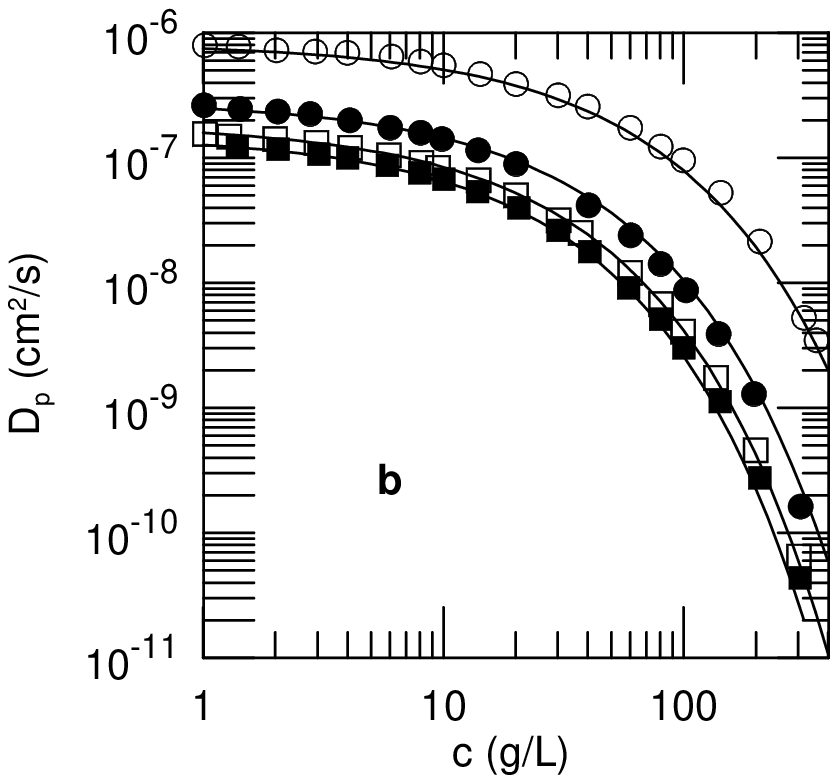} %lo035

\caption{\label{figure49} $D_{p}$  of 12-arm star
polystyrenes (from top to bottom, 
$M_{w}$ of 55, 467, 1110, and 1690 kDa)
diffusing through solutions of 140 kDa polyvinylmethylether:ortho-fluorotoluene
based on 
data of Lodge, et al.\cite{lodge1987a}, their Table 1.
Solid lines are fits of (a) a separate stretched-exponential for each probe, with parameters in \ref{table2} and 
(b) fits of a single stretched exponential in $c$ and $P$ to all data with
parameters in Table \ref{table4}.}
\end{figure}

Lodge and Markland\cite{lodge1987a} used light scattering spectroscopy to 
measure the single-particle diffusion coefficients of tracer 12-armed star 
polystyrenes through solutions of 140 kDa polyvinylmethylether, $M_{w}/M_{n} 
\approx 1.6$,  in the isorefractive solvent orthofluorotoluene.    The 
polystyrenes had $M_{w}$ of 55, 467, 1110, and 1690 kDa, with $M_{w}/M_{n} 
\leq 
1.10$.  Lodge and Markland estimate for the matrix that $c^{*} \approx 20$ g/L 
and $c_{e} \approx 100$ g/L.  Figure \ref{figure49} shows Lodge and Markland's data.  In the two graphs,
the solid lines represent, respectively, fits to individual stretched exponentials in $c$ and to fits to a 
joint stretched exponential in $c$ and $P$.  Fitting parameters 
appear in Tables \ref{table2} and \ref{table4}.  Parameters in Table \ref{table2} differ modestly 
from Table \ref{table2} of Ref.\ \onlinecite{lodge1987a}.  The  
sets of fits are excellent, with RMS fractional 
individual fits here are excellent, with errors of 2-4\%; the RMS fractional 
error for the joint fit was 12\%.  The most notable deviation for the joint exponential is for
the smallest probe at large concentration, where the fitting function 
underestimates $D_{p}$.  

Lodge, Markland, and Wheeler\cite{lodge1989a} used light scattering 
spectroscopy to measure the diffusion of 3-armed and 12-armed star polystyrenes 
through solutions of polyvinylmethylether in its isorefractive solvent 
orthofluorotoluene.   The 3-armed stars had $M_{w}$ of 379 and 1190 kDa; the 
12-armed stars had $M_{w}$ of 55, 467, 1110, and 1690 kDa.  Polyvinylmethyl 
ethers used as matrices had $M_{W}$ of 140, 630, and 1300 
kDa.  
Polystyrenes all had $M_{w}/M_{n} < 1.1$; the matrix polymers had 
$M_{w}/M_{n} 
\approx 1.6$. Light scattering measurements were also made of the radii of 
gyration of linear, 3-arm and 12-arm stars with molecular weights above 1MDa in 
the presence of 250 kDa polyvinylmethylether at concentrations as large as 50 
g/L.  Lodge, et al.\cite{lodge1989a} reported that they modified their 
reported $D_{p}$ to take account of the concentration dependence of a local 
friction factor, using a process described by Wheeler and Lodge\cite{wheeler1989a} 
(see below).  This modification factor was removed from Lodge, et 
al's\cite{lodge1989a} data, before making the analysis here, in order that this data be made more strictly comparable with the remainder of the literature.

\begin{figure}

\includegraphics{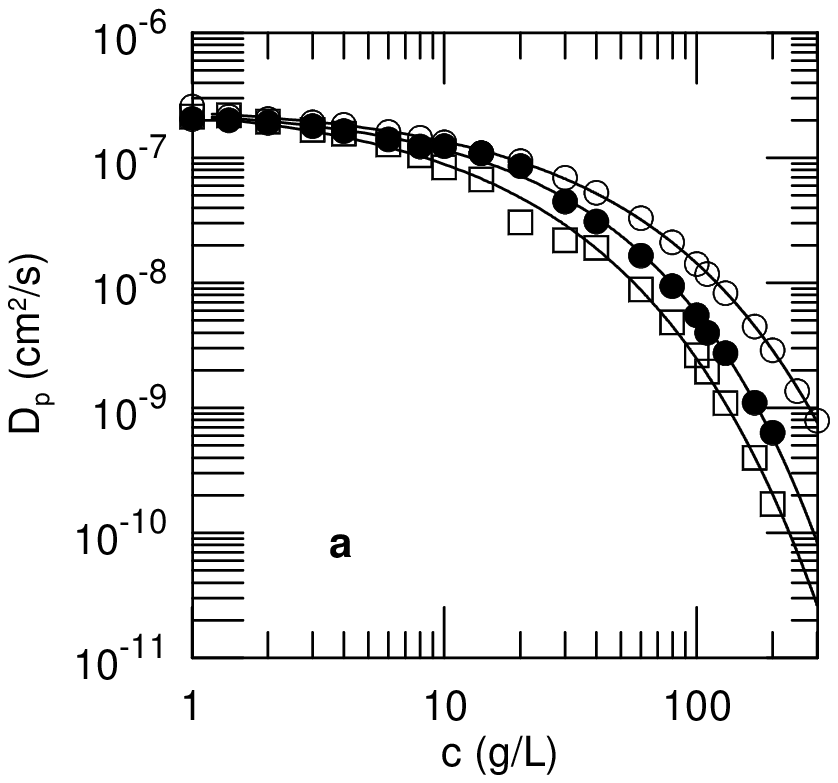} %Lo042
\includegraphics{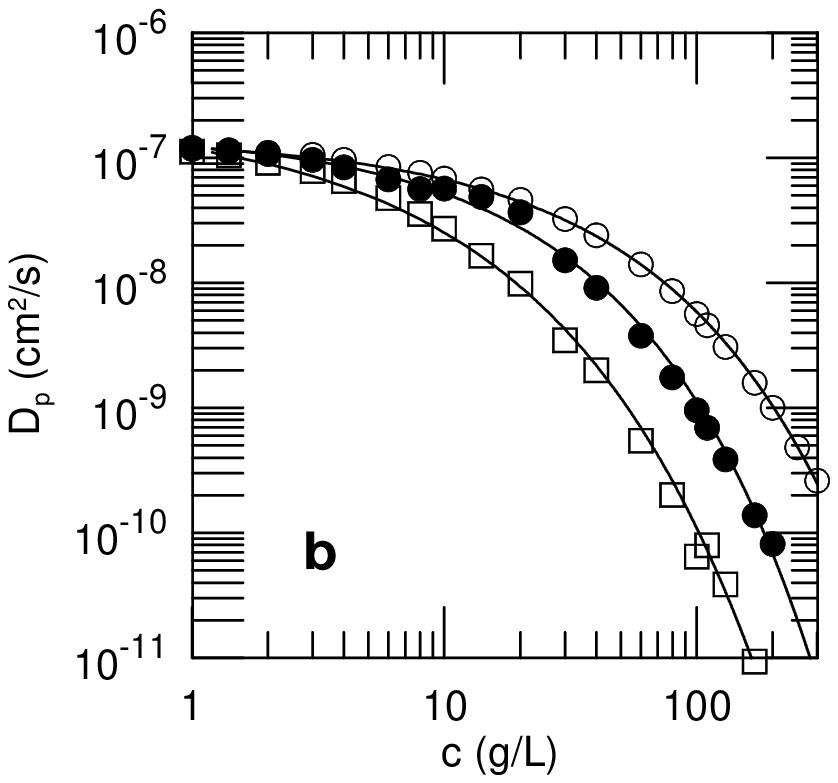} %Lo042

\caption{\label{figure51} $D_{p}$ of (a) 379 kDa and (b) 1.19 MDa 3-arm star
polystyrenes in (from top to bottom, 
140, 379, and 1300 kDa) 
polyvinylmethylethers in orthofluorotoluene,
based on 
data of Lodge, et al.\cite{lodge1989a}.
Solid lines are 
stretched-exponential fits with parameters in Table \ref{table2}.}
\end{figure}

\begin{figure}

\includegraphics{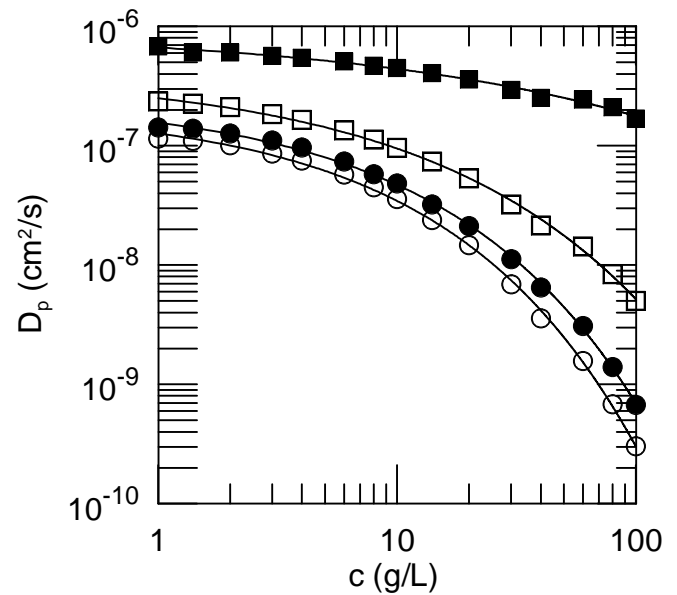} %Lo041

\caption{\label{figure52}  $D_{p}$  of 12-arm star
polystyrenes (from top to bottom, 
$M_{w}$ of 55, 467, 1110, and 1690 kDa) 
diffusing through solutions of 1300 kDa polyvinylmethylether:ortho-fluorotoluene
based on 
data of Lodge, et al.\cite{lodge1989a}.
Solid lines are 
stretched-exponential fits with parameters in Table II.}
\end{figure}

Figures \ref{figure51} and \ref{figure52} show Lodge, et al.'s\cite{lodge1989a}measurements 
of $D_{p}$ for 3-armed and 12-armed stars.  As is apparent from the figures, 
for every $M:P$ combination a stretched exponential gives an excellent 
description of the the concentration dependence of $D_{p}$, with RMS fractional errors in the range 4-18\%
(cf.\ Table \ref{table2}).

\begin{figure}

\includegraphics{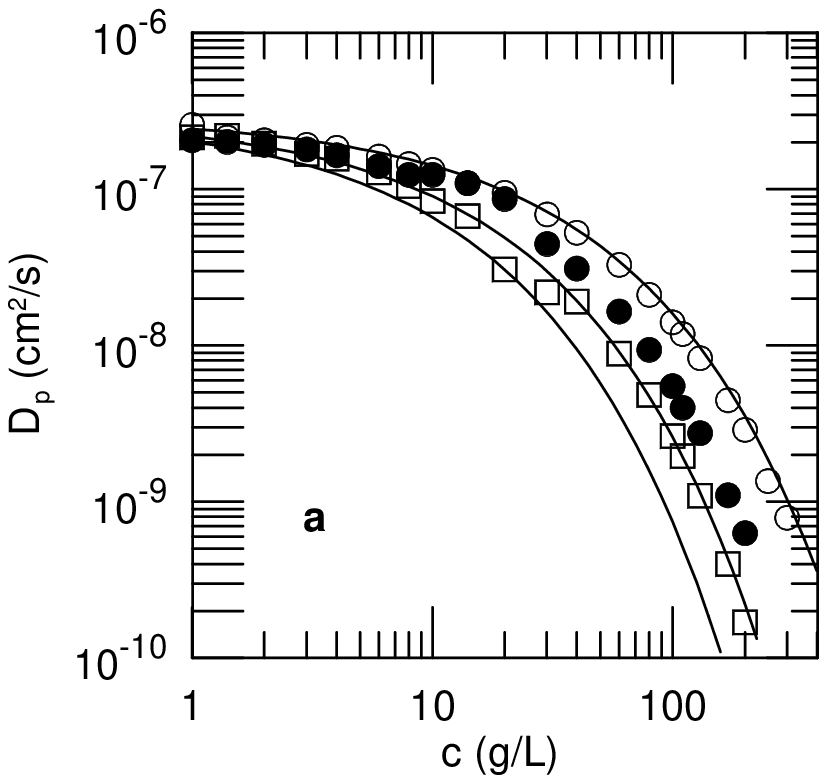} %Lo04t1a,
\includegraphics{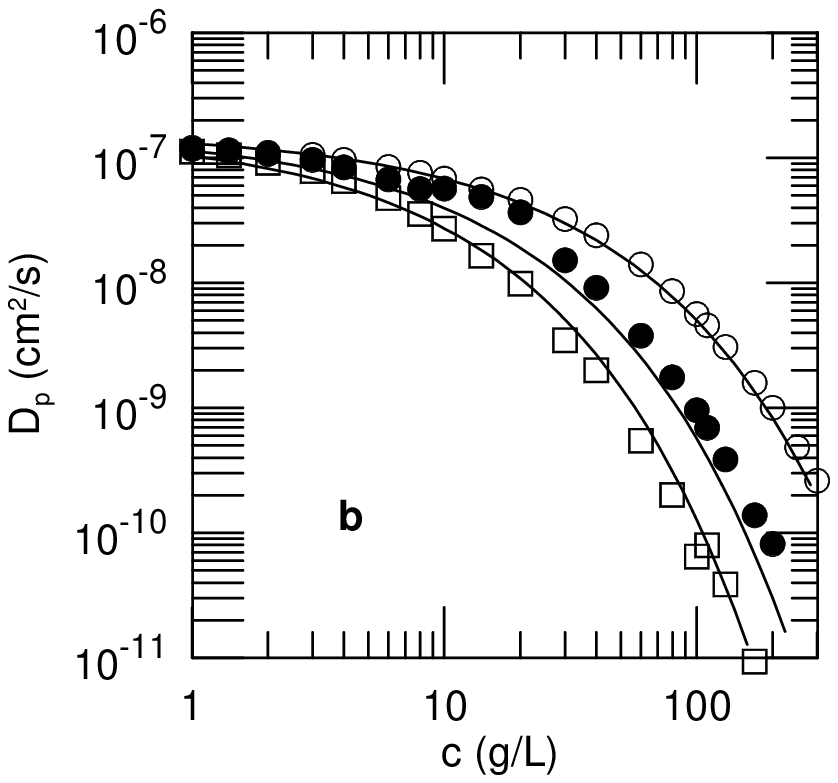} %Lo04t1b

\caption{\label{figure53}  $D_{p}$  of (a) 379 and (b) 1190 kDa
3-arm star polystyrenes in (from top to bottom, 140, 630, and 1300 kDa) 
polyvinylmethylether:ortho-fluorotoluene as jointly fit to eq.\ 
\ref{eq:Dsseeq2}.}
\end{figure}

The entirety of Lodge and collaborators' data\cite{lodge1987a,lodge1989a} on 
12-armed and 3-armed stars were also fit (separately for each arm number)
to equation \ref{eq:Dsseeq2}, 
the joint stretched exponential in $c$, $P$, and $M$.  
Fitting parameters appear in \ref{table4}. 
Figure \ref{figure53} shows the fit of $D_{p}$ for the 3-armed stars 
to the joint stretched exponential.  
As in the few other cases in 
which there are substantial differences between the fitted curve and 
measurements, the stretched-exponential form overestimates the concentration 
dependence of $D_{p}$.  

\begin{figure} 

\includegraphics{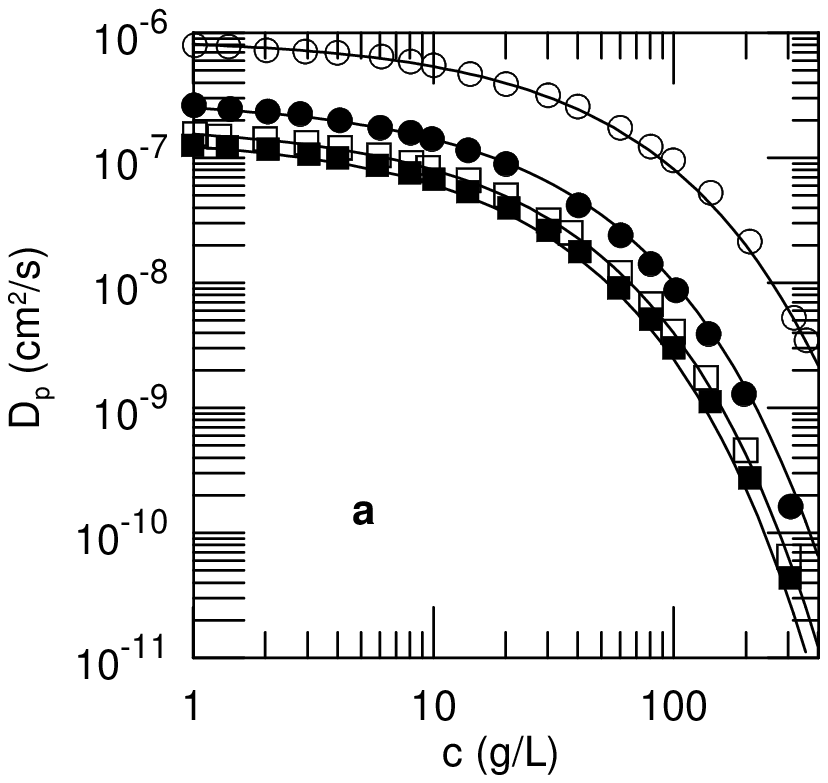} %Lo041u1a
\includegraphics{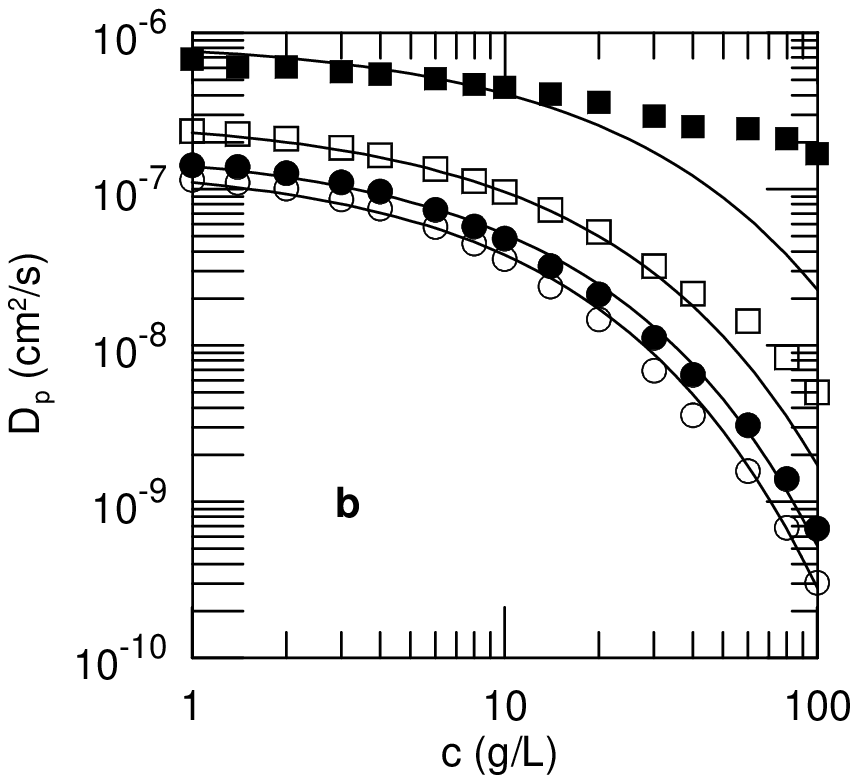} %Lo041u1b

\caption{\label{figure54} $D_{p}$  of 12-arm star
polystyrenes (from top to   bottom, 
$M_{w}$ of 55, 467, 1110, and 1690 kDa) 
in (a) 140 kDa and (b) 1.3 MDa 
polyvinylmethylether:ortho-fluorotoluene\cite{lodge1989a},
and the fit to the joint stretched exponential in $c$, $P$, and $M$.}
\end{figure}
        
Figure \ref{figure54} shows the outcomes of the joint fit to $D_{p}$ of the 12-armed 
stars.    Except for the smallest (55kDa) star in the 1.3MDa matrix polymer,  
eq.\ \ref{eq:Dsseeq2} describes very  well the entire dependence of $D_{p}$ on 
all three variables.  The fit for $f=12$ is 
markedly more outstanding than is the fit to the $f=3$ stars, particularly 
with the 379 kDa pVME as the matrix polymer. 
The displayed curves represent fits to all data points 
except for the 55kDa probe in the 1.3 MDa matrix polymer for $c \geq 10$ g/L.  
Including these 55kDa probe points in the fit raises the fractional RMS error 
from 14.6 to 19.5\%, changes the three scaling exponents by 0.01 each, and 
elsewise has almost no effect on the fitted curves.  

Figures \ref{figure53} and \ref{figure54} show that eq.\ \ref{eq:Dsseeq2} and a single 
set of fitting parameters account well for the dependence of $D_{p}$ on $c$, 
$P$, and $M$ for star polymers of given arm number in linear matrices.  All 
three independent variables ranged over extensive domains: more than two orders 
of magnitude in $c$, a factor of 30 in $P$, and an order of magnitude in $M$.   
The only failure in the fit occurred for the smallest probe in the largest 
matrix polymer ($P/M \approx 25$) at elevated matrix concentrations.  

Martin\cite{martin1984a,martin1986a} examined polystyrenes diffusing through 
polyvinylmethylether (as the matrix polymer) in toluene, with which the matrix 
is isorefractive.  Polystyrenes had molecular weights of 50, 100, 420, and 900 
kDa with $M_{w}/M_{n} \leq 1.1$. The polyvinylmethylether had from 
intrinsic 
viscosity measurements a molecular weight ca.\ 110 kDa and a 'fairly 
polydisperse' molecular weight distribution.  Diffusion coefficients of the 
probe polymers were obtained with QELSS; Martin\cite{martin1984a,martin1986a} 
also determined the viscosities of the polymer solutions.  

\begin{figure} 

\includegraphics{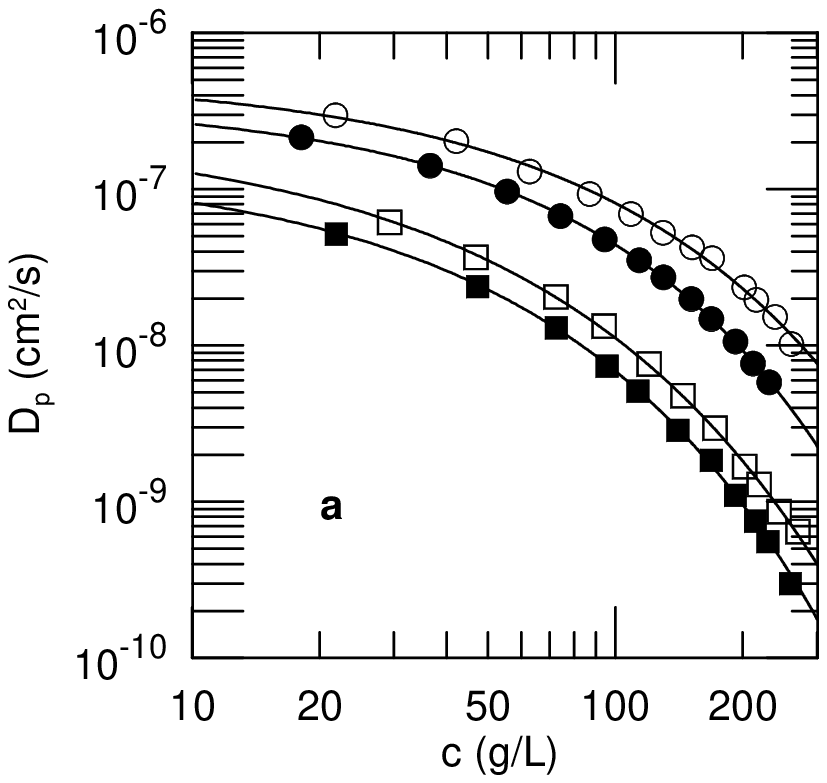} %jm021.eps
\includegraphics{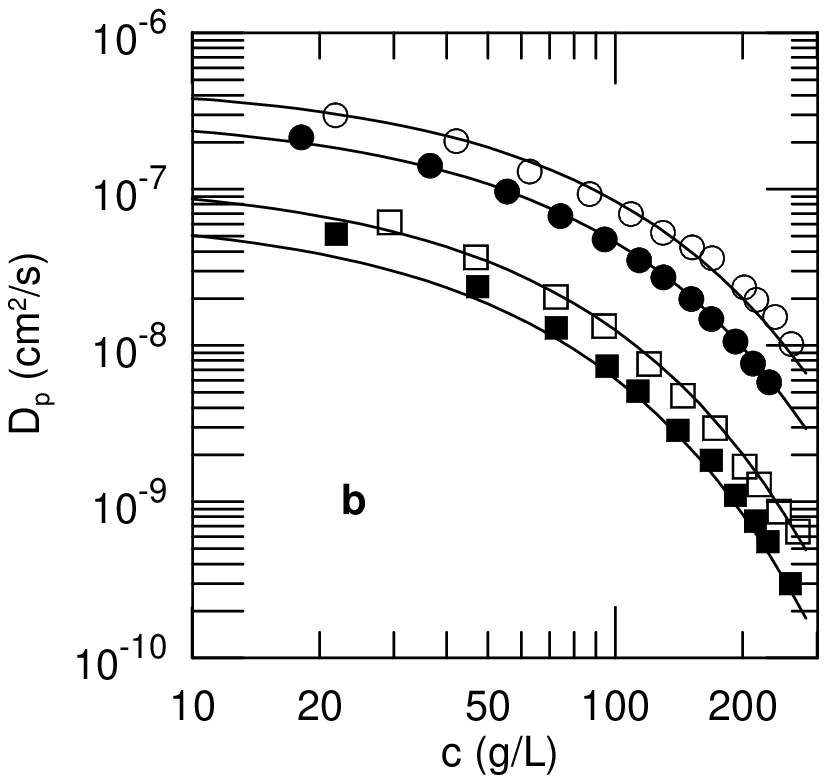} %jm025.eps

\caption{\label{figure55}
Probe diffusion of (top to bottom)
50, 100, 420, and 900 kDa polystyrenes through 110 kDa polyvinylmethyl ether in 
toluene, based on data in Martin\cite{martin1984a,martin1986a}.  Solid lines 
are (a) fits of data for each probe and (b) fits of data for all probes to stretched exponential forms.}
\end{figure}

Figure \ref{figure55} shows $D_{p}$ for each of the four probe 
polymers, as functions of matrix concentration.  In Fig.\ \ref{figure55}a, $D_{p}$ 
for each probe polymer was separately fit to a stretched exponential in $c$, 
yielding the parameters given in Table \ref{table2}
and the four solid lines seen in the Figure.  Stretched exponential forms do an 
excellent job of describing $D_{p}(c)$.  In Fig.\ \ref{figure55}b, all data on 
the four probes was fit simultaneously to a single stretched exponential in $c$ 
and probe molecular weight $P$, obtaining fitting parameters given in Table \ref{table4}.  Martin only reports results for one $M$, making it impossible to evaluate the $M$-dependence of $D_{p}$ form his results. 
Plots of this stretched exponential, as functions of $c$ at fixed $P$, give 
the solid lines of Fig.\ \ref{figure55}b.  Agreement between the fitting function and 
experiment is good in the second Figure, though less good than in the first.

\begin{figure} 

\includegraphics{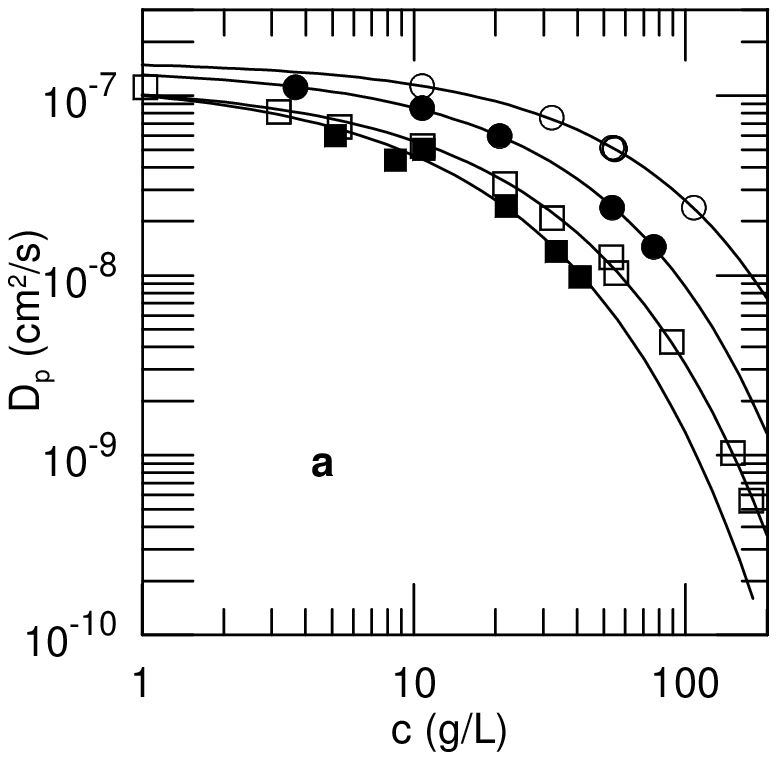} %ne051.eps
\includegraphics{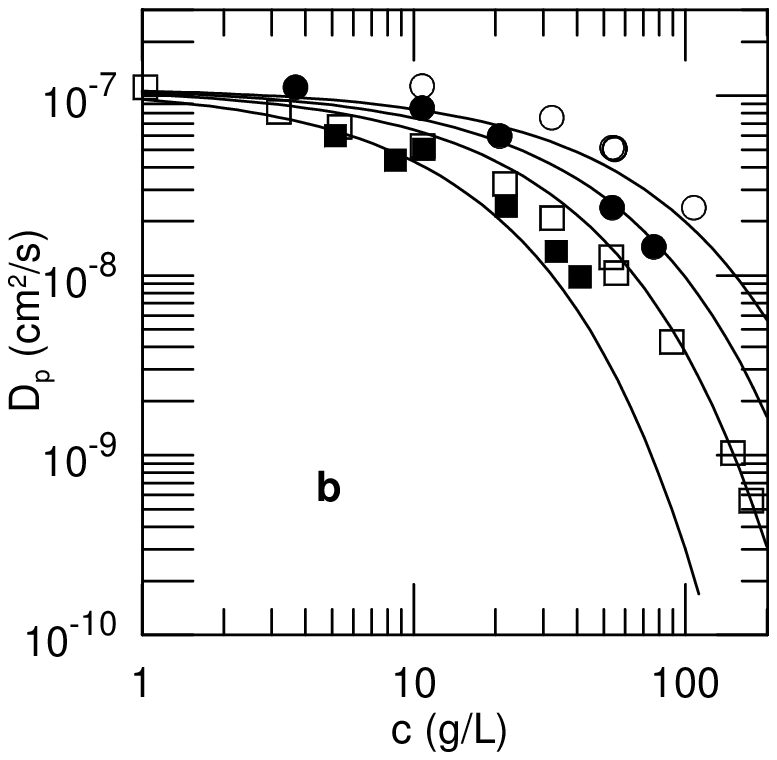} %ne052.eps

\caption{\label{figure58}
 $D_{p}$ of 342 kDa 
polymethylmethacrylate in polystyrene: thiophenol (data of Nemoto, et 
al.\cite{nemoto1985a}) against polystyrene $c$
concentration for polystyrene $M$ of [top to bottom]
44, 186, 775, and 8420 kDa, and fits (a) for each $M$ to a separate
stretched exponential in $c$, and (b) to $D_{o} P^{-a} \exp(-\alpha c^{\nu} P^{\delta} 
M^{\gamma})$ with $D_{o}$, $\alpha$, and $\gamma$ as fitting parameters.} 
\end{figure}

Martin\cite{martin1986a} also measured the viscosity of his matrix polymer 
solutions.  At lower polymer concentrations, especially for the 
lower-molecular-weight probe polymers, Martin\cite{martin1986a} found that 
$D_{p}\eta$ is nearly constant.  At elevated concentrations, especially for the 
larger probe chains, Stokes-Einstein behavior ceases to obtain: $D_{p} \eta$ 
increases with increasing $c$, so that at large $c$ and $P$ the polymer chains 
diffuse faster than might have been expected from the macroscopic solution 
viscosity.   

In addition to the self-diffusion studies noted in the previous Section, 
Nemoto, et 
al.\cite{nemoto1985a} used ultracentrifugation and quasielastic light 
scattering to measure sedimentation and probe diffusion coefficients of dilute 
polymethylmethacrylates (as probe polymers) in isorefractive polystyrene: 
thiophenol solutions.  Polystyrene molecular weights were 43.9, 186, 775, and 
8420 kDa; the PMMA had a molecular weight of 342 kDa.  The reported 
$M_{w}/M_{n}$ were in the range 1.10-1.17.  

Figure \ref{figure58}a shows Nemoto, et al.'s\cite{nemoto1985a} measurements of 
$D_{p}$ of PMMA in solutions of each of the four polystyrenes.  For each 
molecular weight of the matrix, $D_{p}(c)$ is described to good accuracy by a 
stretched exponential in matrix concentration.  Fitting parameters 
are given in Table \ref{table2}.  Figure \ref{figure58}b shows the same data with all 
measurements simultaneously fit to a stretched exponential in $c$ and $M$,
yielding 
parameters in Table \ref{table4}.  Over nearly 
200-fold variations in these variables, the single stretched exponential in $c$ 
and $M$ describes reasonably well the behavior of 
$D_{p}$, with a 20\% RMS fractional error.

Nemoto, et al.\cite{nemoto1985a} report that $D_{p}/D_{o}$ depends more 
strongly on $c$ and $P$ than does $s/s_{o}$.  At large $c$, especially at large 
$M$, $D_{p}/D_{o}$ was found to be
significantly less than $s/s_{o}$. Nemoto, et al.\ concluded 
that at elevated matrix concentrations and polymer molecular weights the 
sedimentation and self-diffusion behaviors of PMMA in polystyrene solutions are 
quite distinct.  

For two samples with the same $D_{p}/D_{o}$ but very different matrix 
molecular weights (44, 8420 kDa), Nemoto, et al.\ also  
measured the shear viscosity $\eta$, 
finding that $\eta$ differed 'by more than two orders of magnitude' between 
the two samples.  Nemoto, et al.\ thus showed 
that $D_{p}$ is not governed by the shear viscosity of 
the matrix solution.  (The original paper 
did not specify which solution was the more 
viscous. Note that the comparison is being made at fixed $D_{p}/D_{o}$,
{\em not} at fixed $c$, so the correspondence is not self-evident.)

Nemoto and 
collaborators\cite{nemoto1989a,nemoto1990a} 
also used forced Rayleigh scattering to 
study the diffusion of probe polystyrenes through polystyrene:dibutylphthalate 
solutions.  A first study\cite{nemoto1989a} focused on self-diffusion and 
tracer diffusion of labeled polystyrene through 40 wt\% solutions of very long 
chains ($M/P > 5$) and very short chains ($M/P < 0.2$).  Thirteen polystyrenes 
having $2.8 \leq M_{w} \leq 8420$ kDa and $M_{w}/M_{n} <1.07$ (except for 
chains larger than 1MDa, for which $M_{w}/M_{n}$ was in the range 1.09-1.17) 
were used in the studies.  

Nemoto, et al.'s data\cite{nemoto1989a} 
appears as Fig.\ \ref{figure61}.  For $D_{s}$ and for $D_{tr}$ 
of short probe chains in
solutions of long matrix molecules, 
a best-fit to the molecular weight dependence 
of $D$ gives parameters seen in Table \ref{table4},
the $P$-dependence of the prefactor being forced rather than obtained from the 
fit.  Because all data is at the 
same concentration, a concentration dependence was not obtained.  Similarly, if 
we fit the data of Nemoto, et al.\cite{nemoto1989a} on $D_{p}$ of long 
probe chains 
in short matrix chains to $D_{o} P^{-a} \exp(- \alpha P^{\gamma})$, we find 
that $D_{p}$ gains its molecular weight dependence almost entirely through the 
factor $P^{-a}$.  The data are fit well with $a = 0.52$, in which case the 
best-fit gives $\gamma \approx 0.03$.  

\begin{figure} 

\includegraphics{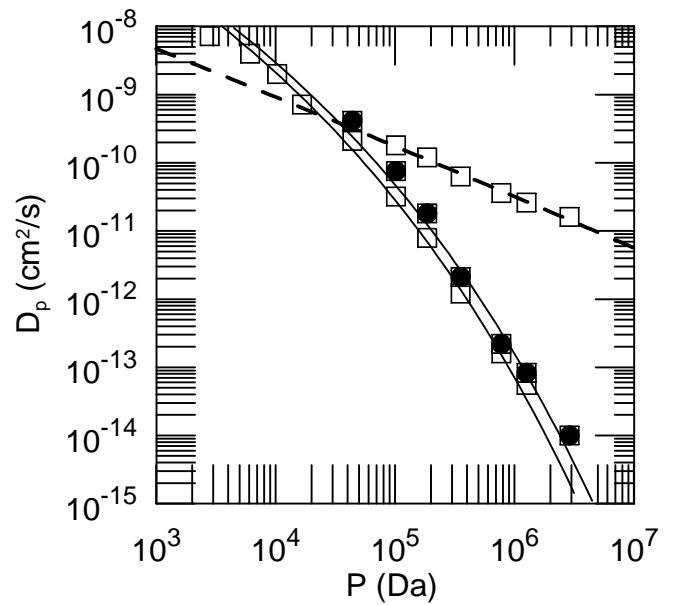} %ne061.eps

\caption{\label{figure60} $D_{s}$(filled points) and $D_{p}$(open points) 
from Nemoto, et al.\cite{nemoto1989a}, Tables 2 and 3.  Dashed line marks 
systems with $M \ll P$; solid lines represent a joint fit to $D_{s}$ and to 
systems with $M/P > 5$.}
\end{figure}

In a separate paper, Nemoto, et al.\cite{nemoto1990a} used forced Rayleigh 
scattering to measure $D_{p}$ of probe 
polystyrenes in dibutylphthalate solutions of high-molecular-weight matrix 
polymers at 13 and 18 \% matrix concentration.  
The probe polystyrenes were in the molecular weight range $6.1 \leq 
M_{w} \leq 2890$ kDa, with polydispersities $M_{w}/M_{n} \leq 1.17$, and 
generally $\leq 1.09$.   Table IV  shows the fits to this data.  RMS fractional 
errors were 12-13\% for single concentrations, but ca. 30\% for the fit to both 
concentrations.  It should be stressed that $D_{p}$ covers more than five 
orders of magnitude, so the errors are not large relative to the total range of 
$D_{p}$.   We find a 
significant dependence of $D_{p}$ on matrix as well as probe molecular 
weight.  With only a few points for any particular $P$ or $M$, 
it is difficult to present the data as a simple figure.

Numasawa, et al.\cite{numasawa1986a} used QELSS to study the diffusion of 
tracer polystyrene chains through index-matched solutions of matrix polymethylmethacrylates 
in benzene.  Polystyrene molecular weights were in 
the range 185-8420 kDa, with $M_{w}/M_{n}$ in the range 1.04--1.17.  
Polymethylmethacrylates had molecular weights 850--4050 kDa, with $M_{w}/M_{n} 
\leq 1.08$, except for the 850 kDa polymer, for which $M_{w}/M_{n}$ 
was 1.35. 
PMMA matrix concentrations were a half-dozen values in the range 0-36 g/L.  The 
tracer polystyrenes were dilute in all solutions.  In addition to measuring the 
probe diffusion coefficients, Numasawa, et al.\cite{numasawa1986a} also report 
the zero-shear viscosity of the matrix polymers and (on the basis of static 
light scattering) determinations 
of the radii of gyration of the probe polymers as a 
function of matrix concentration.  

Numasawa, et al's results\cite{numasawa1986a} appears as Fig.\ \ref{figure61}.  Figure \ref{figure61}a shows $D_{p}$ of the 420 kDa 
and the 8.42 MDa  
polystyrene in solutions of two polymethylmethacrylates, as a function of 
matrix concentration.  Figure \ref{figure61}b shows $D_{p}$ of five polystyrenes in 
each of four polymethylmethacrylates, all at a matrix concentration near 37 
g/L.  One observes that $D_{p}$ decreases monotonically with increasing probe 
molecular weight and with increasing matrix concentration and molecular weight.  

All eight solid lines in both figures represent a simultaneous 
fit of a stretched exponential in $c$, $P$, and $M$ to all of the 
data in both Figures and to one additional data point.  Fit parameters appear 
in Table \ref{table4}.  It is important to emphasize that the same fitting parameters 
were used to generate all eight curves in both Figures.  From 
Fig.\ \ref{figure61}a, the stretched exponential captures well the $c$-dependence of 
$D_{p}$, and the variation of that dependence with $P$ and $M$.  From Figure 
\ref{figure61}b, at fixed $c$, the stretched exponential captures reasonably well the 
dependence of $D_{p}$ on $M$ at fixed $P$, but at fixed $M$  and $c$ does less 
well at capturing the dependence of $D_{p}$ on $P$.

\begin{figure} 

\includegraphics{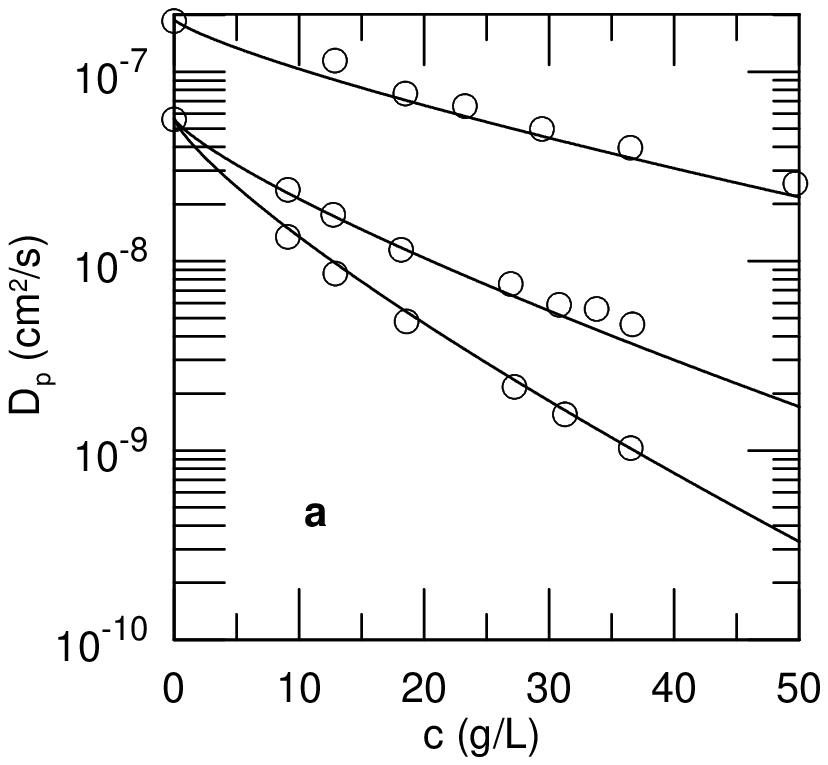} %nu013.eps
\includegraphics{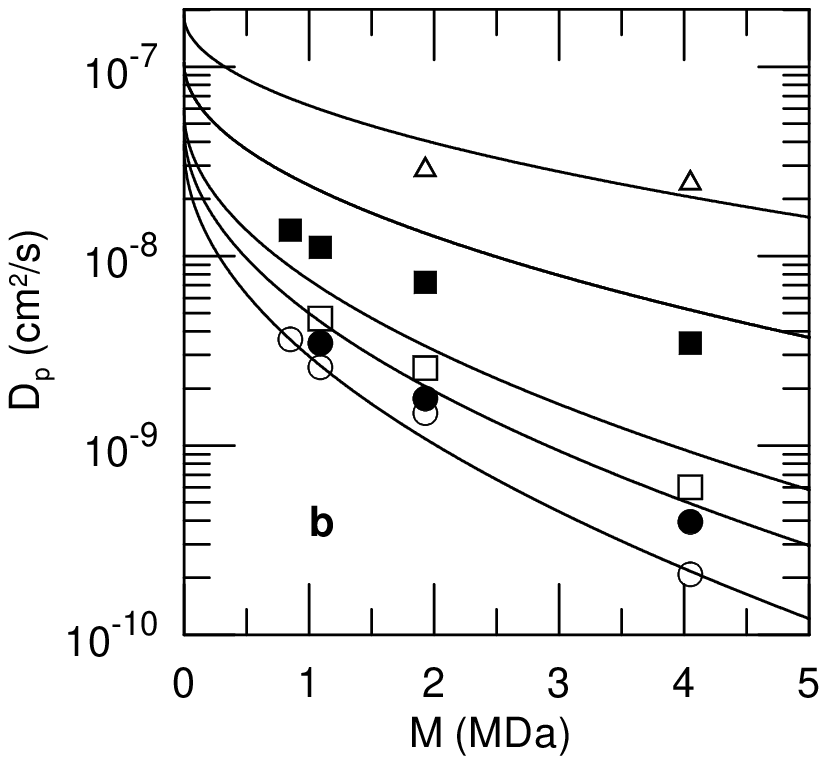} %nu012.eps

\caption{\label{figure61} $D_{p}$ of polystyrenes in polymethylmethacrylate: 
benzene  (a) as a function of matrix concentration, with $P$:$M$  of
[top to bottom] 420 kDa:4.05 MDa, 8.42 MDa: 1.95 MDa, and  8.42 MDa: 4.05 MDa,
and (b) as a function of molecular weight, for [top to bottom] 0.42, 1.26, 3.84, 5.48, 
and 8.42 MDa polystyrene probes
in 36.7 g/L polymethylmethacrylate.  The lines all show the same best-fit stretched 
exponential in $c$, $P$, and $M$.}
\end{figure}

\begin{figure} 

\includegraphics{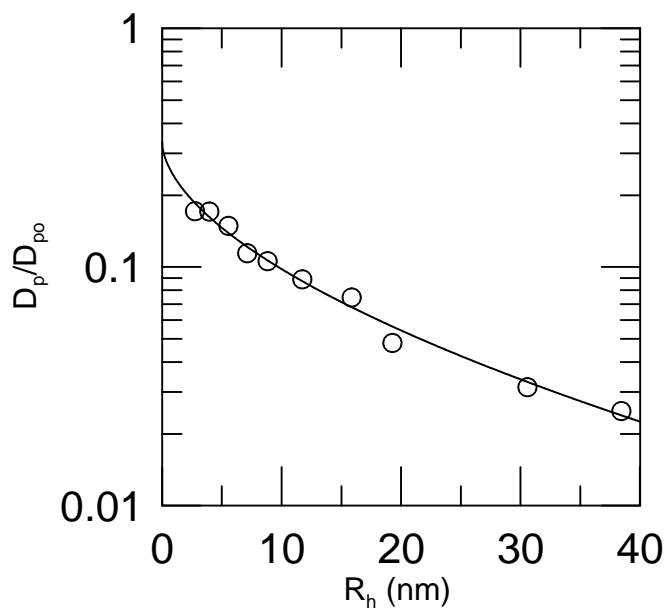} %ny011.eps

\caption{\label{figure63} $D_{p}/D_{p0}$ 
from Nyden, et al.\cite{nyden1999a} for polyethylene oxides with a range of 
molecular weights diffusing in 1\% solutions of 100kDa 
ethylhydroxyethylcellulose, plotted as a function of the hydrodynamic radii of 
the probes in pure water.}
\end{figure}

Nyden, et al.\cite{nyden1999a} used PFGNMR to determine probe diffusion of 
monodisperse ($M_{w}/M_{n} < 1.1$) polyethylene oxides (molecular weights 
10-963 
kDa) diffusing through aqueous solutions of 100kDa ethylhydroxyethylcellulose.  
The authors studied 1\% and 6\% solutions and a 1\% chemically cross-linked gel.  
Figure \ref{figure63} shows $D_{p}/D_{o}$ of the polyethylene oxides in 1\% of the 
matrix polymer as a function of their hydrodynamic radii.  
The solid line in the Figure
represents a stretched exponential 
\begin{equation} 
  D_{p}/D_{p0} = D_{1} \exp(- \alpha R^{\beta}) 
  \label{eq:DpD0R}
\end{equation}
in probe hydrodynamic radius $R$, where $\alpha$ and $\delta$ are a scaling 
prefactor and exponent, $D_{p0}$ is the diffusion coefficient of the probe 
polymer in pure solvent, and $D_{1}$ is the probe diffusion coefficient for a 
nominal $R = 0$ polymer chain.  The best-fit 
parameters were $D_{1} = 0.33$, $\alpha 
= 0.33$, and $\beta = 0.57$; the RMS fractional error in the fit was 7\%.   
Nyden, et al.\cite{nyden1999a} note $R = K P^{a}$ with $a = 0.53$ for probes 
having molecular weight $P$, so equation \ref{eq:DpD0R} is equivalent to a 
stretched exponential in $P^{\gamma}$ with $\gamma \approx 0.30$.  This value 
of $\gamma$ is consistent with values for $\gamma$ 
found for other systems, as seen in 
Table \ref{table4}.

\begin{figure} 

\includegraphics{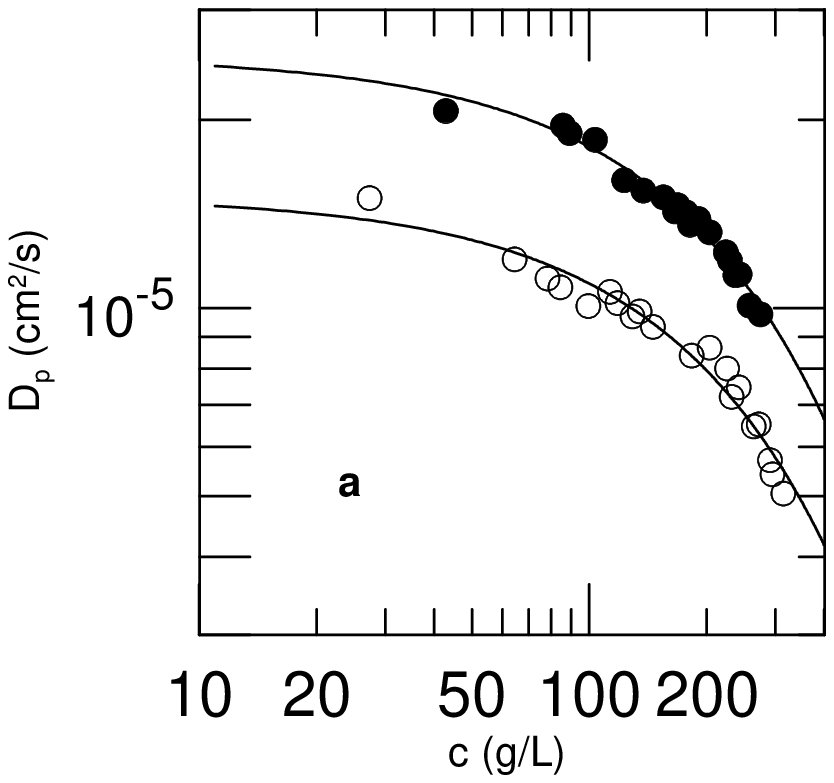} %pi021.eps

\includegraphics{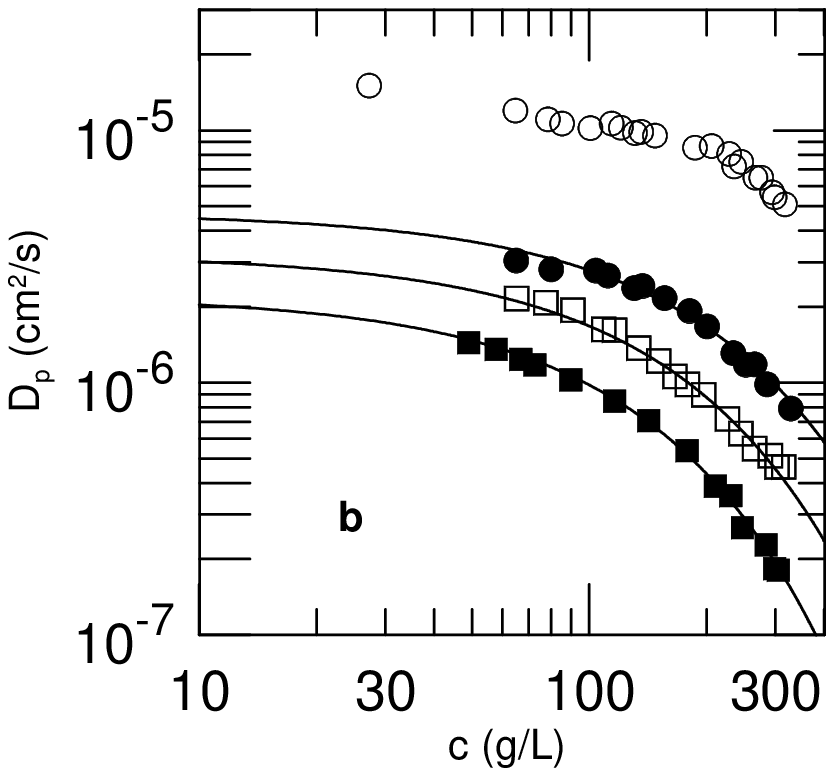} %pi022.eps

\caption{\label{figure64}  $D_{p}$, using data of 
Pinder\cite{pinder1990a}, of (a) styrene monomer through solutions of 
high-molecular-weight polystyrenes in cyclohexane (theta solvent, filled points)
and CCl$_{4}$ (non-theta solvent, open points), and pure-exponential fits, and
(b) [top to bottom] styrene monomer and 580, 
1200, and 2470 Da polystyrene polymers in CCl$_{4}$ and a high-molecular weight
polystyrene matrix polymer with (lines) fits of the polymeric 
probes data to $D_{p} = D_{o} P^{-a} \exp(-\alpha c P^{\gamma})$.
}
\end{figure}

Nyden, et al.\cite{nyden1999a} also examined probe diffusion in 6\% solutions 
of their matrix polymer.  $d D_{p}/dM$ does not change monotonically with 
increasing $M$.  However, at the larger $M$ at which the anomalous behavior 
occurs, the PFGNMR echo decays are no longer simple exponentials, implying that 
diffusive behavior has become more complex.

Pinder\cite{pinder1990a} reported from PFGNMR measurements the tracer diffusion 
coefficient of styrene and low-molecular-weight probe polystyrenes ($P \leq 
2470$ Da) through solutions of deuterated polystyrenes ($10.7 \leq M \leq 
430$ kDa) in theta (cyclohexane) and non-theta (CCl$_{4}$) solvents.  $D_{p}$ 
was measured as a function of matrix polymer concentration for matrix 
concentrations up to 300 g/L. The matrix polymer polydispersities were 
$M_{w}/M_{n} \leq 1.14$, with in most cases $M_{w}/M_{n} < 1.08$.

Figure \ref{figure64}a shows the tracer diffusion coefficient of styrene in 
polystyrene solutions.  The matrix molecular weights were 68 and 200 kDa in the 
non-theta solvent and 68, 87, 200, and 430 kDa in the theta solvent.  Within 
the scatter in the data---there are not a large number of data points for any
particular matrix $M$---$D_{p}$ does not appear to depend on $M$.  $D_{p}(c)$ 
was fit both to pure and stretched exponentials in $c$.  As seen in Table \ref{table2}, 
there appears to be very little improvement in the fit on allowing $\nu \neq 
1$, so the Figure shows the fits to the pure exponentials in $c$.

Figure \ref{figure64}b  shows the tracer diffusion coefficient of styrene and three 
low-molecular-weight polystyrenes  ($P \leq 2470$ Da) though a range of 
high-molecular-weight polystyrenes ($10.7 \leq M \leq 430 $kDa).  For each probe, 
matrix molecular weights were varied by factors of 3-5 with no apparent 
significant effect on $D_{p}$.  In all cases, $M/P$ was in the range 15-20 or 
larger.  $D_{p}(c)$ for each probe was fit to a separate stretched exponential, 
giving excellent results reported in Table II.  The merged results for all four 
probes were also fit jointly to a stretched exponential in $c$ and $P$.  
The computed $D_{p}$ for styrene monomer underestimates $D_{p}(c)$ by a nearly 
constant multiplicative factor.  Because styrene is little larger than a 
solvent molecule, a second fit including only the polymer probes was made.  
Parameters for both fits are in Table IV.  The Figure shows the second fit, to 
the three polymeric probes.  The second fit has a modestly better RMS 
fractional error, and finds in the limit of zero matrix concentration that 
$D_{p} \sim P^{-0.52}$.  If the styrene monomer is included in the fits, in the 
same $c \rightarrow 0$ limit $D_{p} \sim P^{-0.68}$ is obtained.  

\begin{figure} 

\includegraphics{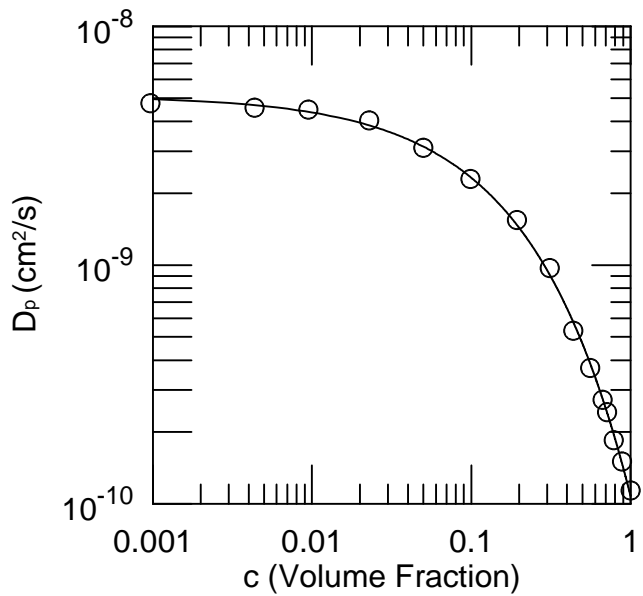} %smo11.eps

\caption{\label{figure66} $D_{p}$ of 33.6 kDa labeled polypropylene oxide 
chains through solutions of 32 kDa PPO chains in a 1 kDa PPO melt, 
using results 
of Smith, et al.\cite{smith1986a}, Figure 1.  The solid line represents a 
stretched-exponential fit.}
\end{figure}

\begin{figure} 

\includegraphics{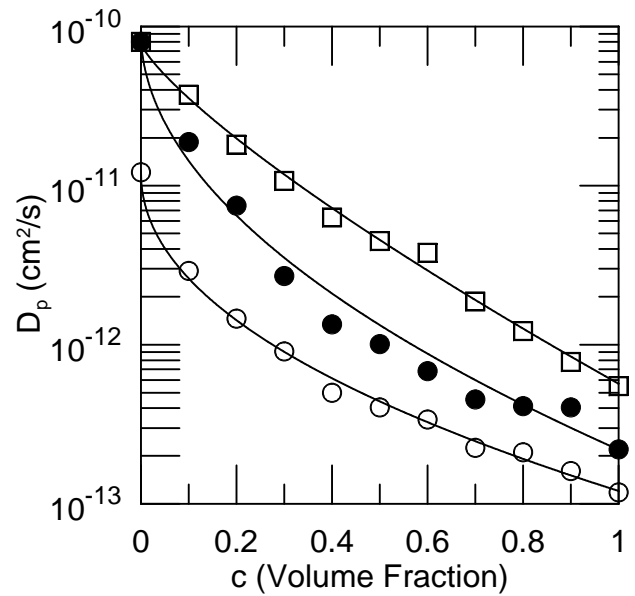} %te011.eps

\caption{\label{figure67} $D_{p}$  of 255 kDa deuteropolystyrene in solutions 
of [top to bottom] 93, 250, and 20 000 kDa polystyrene in molten
10 kDa polystyrene, at 150, 150, and 175 C, respectively, 
from Tead and Kramer\cite{tead1988a} Figures 4 and 6.  Solid lines are 
stretched-exponential fits using parameters given in Table II.
For clarity, the 150 C data has been multiplied 
by a factor of 100.}
\end{figure}

Smith, et al.\cite{smith1986a} used fluorescence recovery after pattern 
photobleaching to measure the diffusion of labeled 33.6 kDa
polypropylene oxide (PPO) chains through solutions of unlabeled 32kDa PPO
chains dissolved in a melt of 1 kDa PPO chains.  The probe and solvent were 
relatively monodisperse ($M_{w}/M_{n} = 1.1$) while the matrix polymer was 
relatively polydisperse ($M_{w}/M_{n} = 1.6$).  The matrix concentration was 
varied all the way from dilute solution up to the matrix melt.  Smith, et al.'s 
data\cite{smith1986a} are shown in Fig.\ \ref{figure66}.  The solid line in the 
Figure is a stretched-exponential fit using parameters in Table \ref{table2}.  A single 
stretched exponential with constant parameters describes the concentration 
dependence of $D_{p}$ all the way from dilute solution up to the matrix melt.  

Tead and Kramer\cite{tead1988a} studied diffusion of 255 kDa deuterated 
polystyrene through solutions of large-molecular-weight
(93, 255, and 20 000 kDa) polystyrenes dissolved in a low-molecular-weight 
(10 kDa) polystyrene melt. The matrix polymer volume fraction covered the full range $0 \leq \phi \leq 1$.
The temperature was 150 C for the 93 and 255 kDa matrix polymers and 175 
C for the 20 000 kDa matrix polymer.  Probe diffusion coefficients were obtained using forward recoil spectroscopy
to measure concentration profiles of the probe molecules as a function of time.
As seen in Fig.\ \ref{figure67}, $D_{p}$ is represented well by 
a stretched exponential in the matrix polymer concentration $c$, no matter whether the matrix molecular 
weight is less than, equal to, or far larger than the probe molecular weight.
The agreement with the stretched-exponential form is good from low 
concentrations of a short (93 kDa) chain out to high concentrations (melt) of a 
very large (20 000 kDa) chain.

\begin{figure} 

\includegraphics{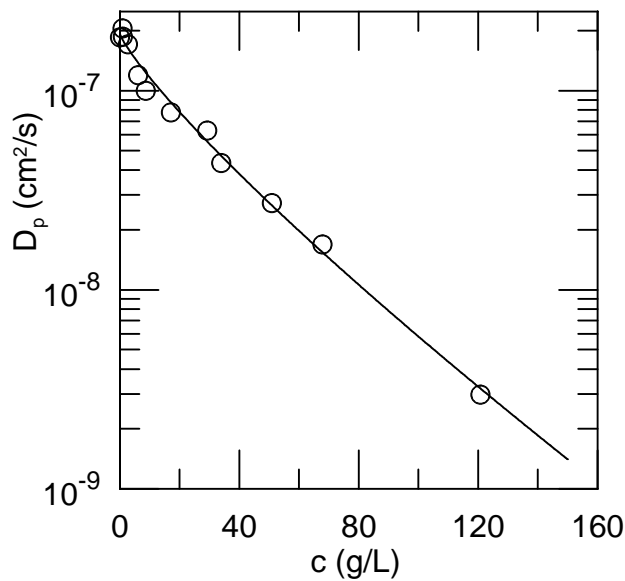} %ti021.eps

\caption{\label{figure68} $D_{p}$  of 433 kDa dextran
diffusing through aqueous solutions of 310 kDa polyvinylpyrrolidone, based on 
data of Tinland and Borsali\cite{tinland1994a}, Table 2 and Figure 3.
Solid line is a 
stretched-exponential fit from parameters in Table \ref{table2}.}
\end{figure}

Tinland and Borsali\cite{tinland1994a} used fluorescence recovery after 
photobleaching and quasielastic light scattering to make independent 
measurements of the probe diffusion coefficient of 433 kDa dextran through 
solutions of 310 kDa polyvinylpyrrolidone in water.  PVP concentrations ranged 
from 0 to 120 g/L.  The polydispersity $M_{w}/M_{n}$ is 1.5 for the matrix 
polymer but ca.\ 1.9-1.95 for the probe chains.  Except perhaps at the very 
highest concentrations studied, values of $D_{p}$ from the two techniques do 
not agree.  In the following analysis, we use $D_{p}$ as obtained from FRAP, 
since the values from this technique do not rely so heavily on detailed 
model assumptions of the relationship between the scattering spectrum and 
the underlying diffusion coefficients.  Tinland and Borsali's data appear in Fig.\ \ref{figure68}, with data points from the original paper\cite{tinland1994a} and a smooth curve showing the 
stretched exponential in $c$ generated from the fitting parameters in Table \ref{table2}.  Agreement between 
the measurements and the functional form is good at all polymer concentrations.

Wheeler, et al.\cite{wheeler1987a} studied tracer diffusion of linear 
polystyrenes having molecular weights 65, 179, 422, and 1050 kDa (with 
$M_{w}/M_{n} \leq 1.1$) through a 1.3 MDa polyvinylmethylether matrix polymer, 
$M_{w}/M_{n} \approx 1.6$, in orthofluorotoluene.  $D_{p}$ was determined using 
quasielastic light scattering, which was possible because the polystyrenes were present 
at trace concentration while the matrix polymer and solvent are isorefractive.
Matrix concentrations covered the range 1--100 g/L.  For this matrix polymer, 
$1/[\eta]=2.2$ g/L, so much but not all of the data is in the range $c > 
c^{*}$.

\begin{figure} 

\includegraphics{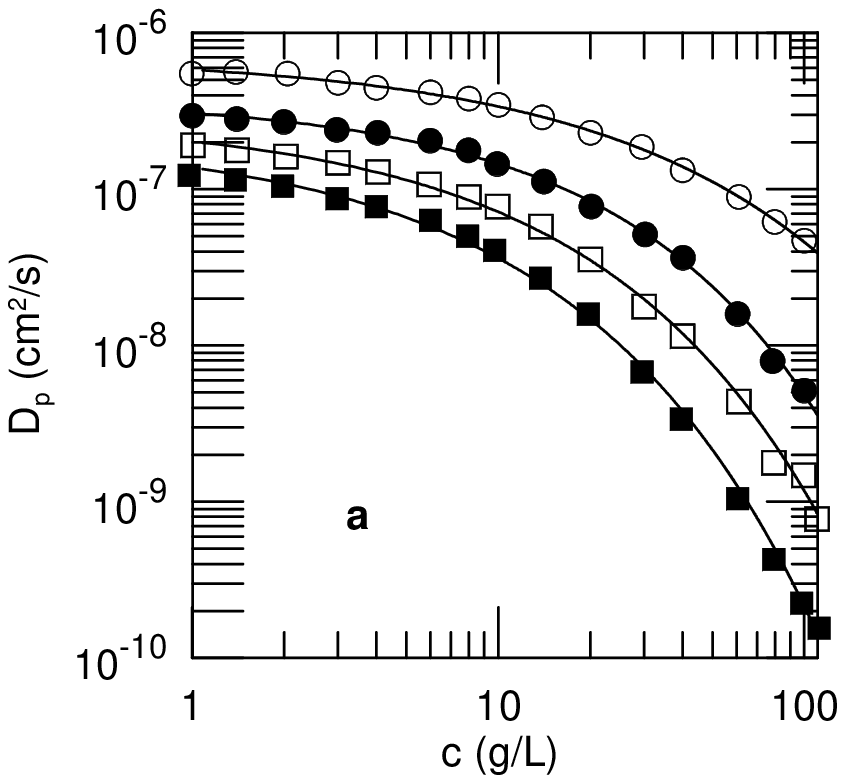} %wh011.eps

\includegraphics{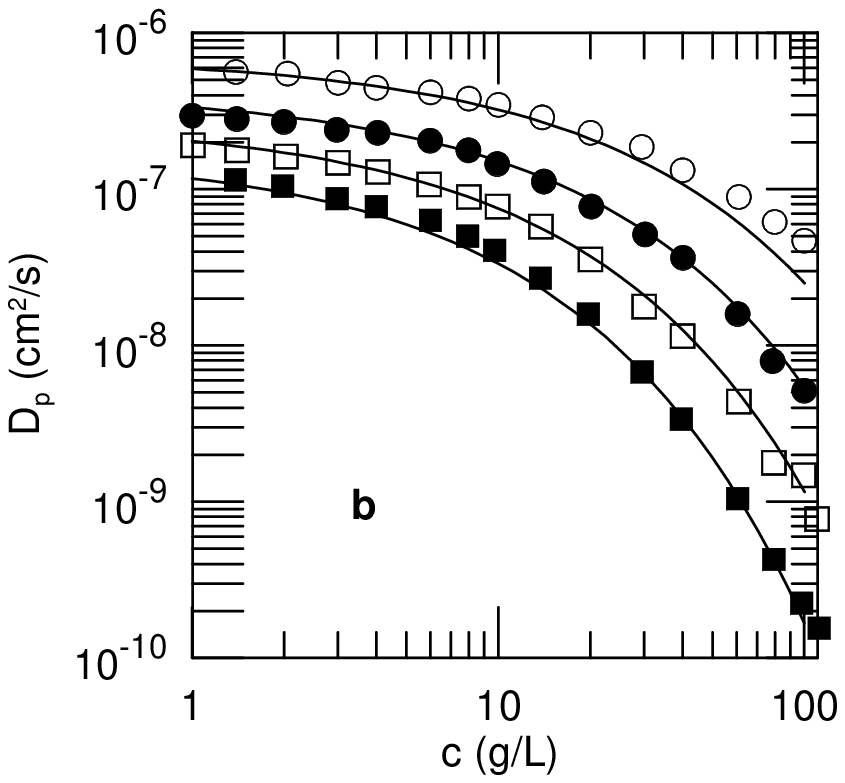} %wh015.eps

\caption{\label{figure69}  $D_{p}$  of polystyrenes (from top to bottom, 
$M_{w}$ of 65, 179, 422, and 1050 kDa)
diffusing through solutions of 1.3 MDa polyvinylmethylether:ortho-fluorotoluene
based on data of Wheeler, et al.\cite{wheeler1987a}, with solid lines showing fits to (a)
separate stretched exponentials at each $P$, and (b)
jointly to a stretched exponential in $c$ and $P$.}
\end{figure}

The data and corresponding stretched-exponential fits are in Figs.\ \ref{figure69}a  
and \ref{figure69}b, with fit parameters in Tables \ref{table2} and \ref{table4}, respectively.  Figure 
\ref{figure69}a shows fits made separately for each probe polymer. As in the other systems discussed 
above, the concentration dependence of $D_{p}$ for each probe is described 
extremely well 
by a stretched exponential in polymer concentration. 
Figure \ref{figure69}b 
shows the outcome of a simultaneous fit of a single stretched exponential in 
$c$ and $P$ to all data on all four probes.  At the scale of the Figures, the 
curves in the two Figures for the 179 and 422 kDa probes are coincident, while 
the curves for the 1050 kDa probes are very close.
However, for the smallest probe and large matrix concentrations the
joint stretched exponential substantially underestimates $D_{p}$.

Wheeler and Lodge\cite{wheeler1989a} used QELSS to measure the diffusion of 
linear polystyrenes through polyvinylmethylether:orthofluorotoluene.  The 
polystyrenes had molecular weights of 65, 179, 422 and 1050 kDa, with 
polydispersities $M_{w}/M_{n} < 1.1$.  The polyvinylmethylether 
samples had molecular weights of 140, 630, and 1300 kDa; concentrations of 
these matrix polymers ranged up to 300 g/L.  Reference \onlinecite{wheeler1989a}
represents a major extension of Ref.\ \onlinecite{wheeler1987a} in the 
range of matrix concentrations, number of matrix molecular weights, and number of concentrations studied.
For the three matrix polymers, 
$c^{*}$ was estimated at 11, 5.7, and 3.3 g/L, respectively (based on $c^{*} 
= 1.5/[\eta]$), while $c_{e}$ was estimated at 50, 12, and 6 g/L, respectively.  
The matrix polymers had $M_{w}/M_{n} \approx 1.6$.   

Wheeler and Lodge also used PFGNMR to measure $D_{s0}$ of orthofluorotoluene 
diffusing through the 1300 kDa polyvinylmethylether at matrix concentrations 
up to 300 g/L.  $D_{s0}$ fell by 62\%  over this concentration range.  To good 
approximation, $D_{s0}(c)$ of the orthofluorotoluene is fit by a simple 
exponential $D_{s0} =  2.15 \cdot 10^{-5} \exp(- 2.98 \cdot 10^{-3} c^{1})$.   
Wheeler and Lodge\cite{wheeler1989a} and also Lodge, et al.\cite{lodge1989a} 
used these data to modify $D_{p}$ of their polystyrene polymer probes to remove the 
concentration dependence of a nominal local friction $\zeta = k_{B} T/D_{s0}$.  This
local friction modification was here removed from the probe diffusion data of 
Refs.\ \onlinecite{wheeler1989a} and \onlinecite{lodge1989a} before analyzing them further.  

\begin{figure} 

\includegraphics{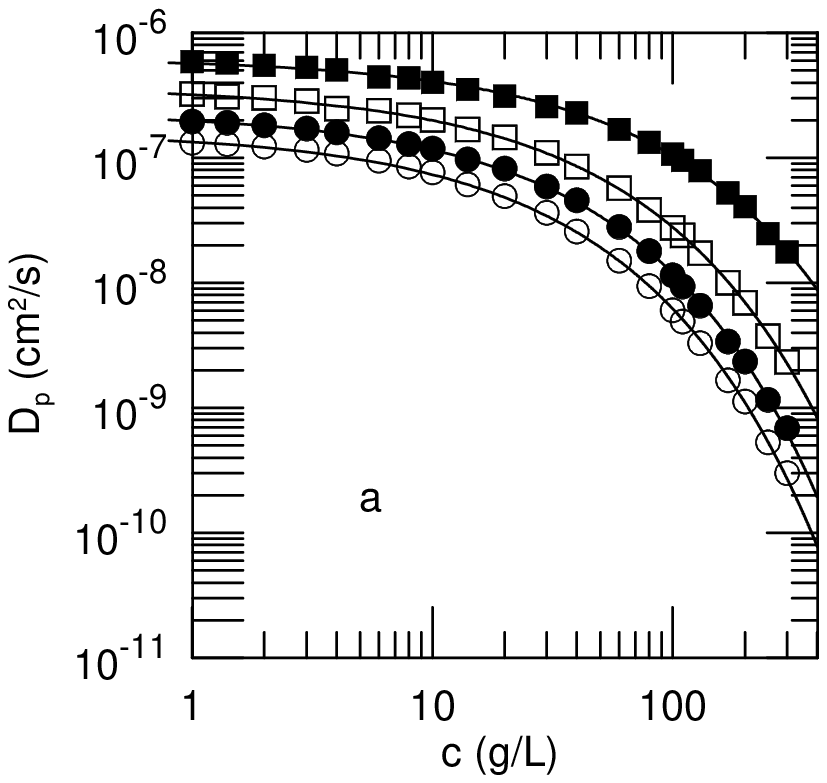} %wh021a.eps
\includegraphics{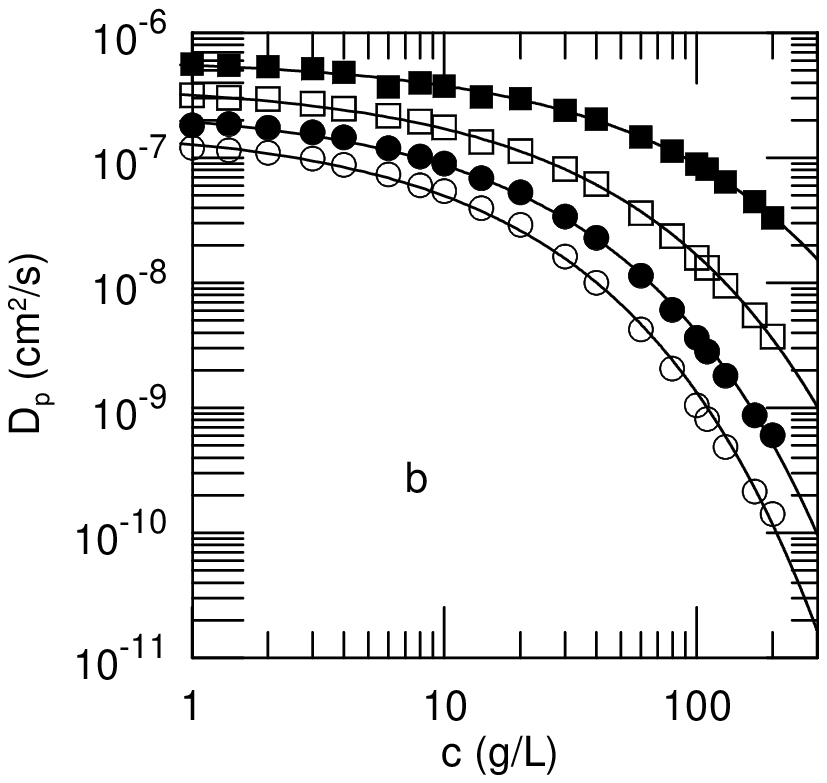} %wh021b.eps

\caption{\label{figure71}
$D_{p}$  of polystyrenes (from top to bottom, 
$M_{w}$ of 65, 179, 422, and 1050 kDa)
diffusing through solutions of (a) 140 kDa 
and (b) 630 kDa polyvinylmethylether:orthofluorotoluene
based on 
data of Wheeler, et al.\cite{wheeler1989a}, and (lines)  
separate stretched-exponential fits for each $P:M$ combination.
}
\end{figure}

Figures \ref{figure71} show Wheeler, et al's\cite{wheeler1989a} data on 
their four probe polystyrenes in the 140 and 630 kDa polyvinylmethylethers.  
Solid lines represent fits to stretched exponentials in $c$ for individual 
probe:matrix pairs.  The fits are uniformly extremely good, with RMS fractional 
errors in the range 2-12\% and fitting parameters as seen in Table \ref{table2}

We also fit all of
Wheeler, et al.'s data\cite{wheeler1987a,wheeler1989a} 
to a joint stretched exponential in $c$, 
$P$, and $M$, as seen in Figs.\ \ref{figure72}.  The solid lines are 
the best-fit $D_{p}(c)$, plotted for fixed $P$ and $M$, 
all lines being computed from a single 
set of 
parameters given
in Table \ref{table4}.   In Fig.\ \ref{figure72}c, measurements from Ref.\ 
\onlinecite{wheeler1989a} were supplemented with the data of Ref.\ 
\onlinecite{wheeler1987a}; the latter reference reported $D_{p}$ for
all probes in 1300 kDa polyvinylmethylether 
solutions having $c \leq 100$ g/L.   

Over an order of magnitude in matrix molecular weight, a factor of 15 in probe 
molecular weight, and a factor of 300 in matrix concentration, the joint 
stretched exponential of eq \ref{eq:Dsseeq2} represents reasonably well the 
joint dependence of $D_{p}$ on $c$, $P$, and $M$.  The RMS fractional error in 
the fit is 25\%.  As noted above in the analysis for Ref.\ 
\onlinecite{wheeler1987a}'s data, for the smallest probe in the 
630 and 1300 kDa matrix 
polymers (Fig.\ \ref{figure72}), the stretched exponential form noticeably 
underpredicts $D_{p}$.  A similar issue arises for the 1050kDa probe in the 140 
kDa matrix polymer(fig.\ \ref{figure72}a):  at large $c$ the predicted $D_{p}$ is too small, because 
the predicted curve  does not bend quite sharply enough between small and large 
$c$.  

\section{Other Experimental Studies}

In addition to the work reviewed above, a variety of other studies of polymer 
tracer diffusion appear in the literature.  These are papers whose experimental 
foci are not the same as those of the papers examined in the previous sections, 
and which are not amenable to analysis on the lines employed above. 
For example, a number of papers report scattering spectra of polymer: polymer: 
solvent mixtures in which neither polymeric species is dilute, and therefore in which the spectral relaxation times do not correspond to the self-diffusion coefficient.

Aven and Cohen\cite{aven1990a} measured the diffusion of dilute polystyrenes 
through 15 vol\% solutions of polydimethylsiloxane in tetrahydrofuran.  
The tracer diffusion coefficients of the polystyrenes, the molecular weight 
dependence, and the initial linear slope of the concentration dependences were 
obtained with light scattering spectroscopy.  

Borsali, et al.\cite{borsali1987a} studied QELSS spectra of 970 kDa 
polystyrene: 950 kDa polymethylmethacrylate: toluene, PMMA and toluene forming 
an isorefractive pair.  Semiquantitative experimental tests were made of the 
theoretical work of Benmouna, et al.\cite{benmouna1987a}.  When neither polymer 
was dilute, the observed spectrum was biexponential.  The mode amplitude ratio 
and relaxation times were within a factor of two of predictions of the Benmouna 
model\cite{benmouna1987a}.  The mutual diffusion coefficient of polystyrene in 
toluene and the cooperative diffusion coefficient of the mixture, both measured 
at the same total polymer concentration, are equal to within 6\%, also in 
agreement with the theory.  

Borsali, et al.\cite{borsali1989a} extended their work on this ternary system 
with measurements at several large (weight fraction $> 0.8$) polystyrene 
concentrations and a range of total polymer concentrations, finding two 
relaxational modes in QELSS spectra.  The diffusion coefficent associated with 
the fast mode increased with increasing polymer concentration.  The diffusion 
coefficient associated with the slow mode decreased markedly with increasing 
polymer concentration. Results were consistent with the Benmouna, et 
al.\cite{benmouna1987a} model.  

Borsali, et al.\cite{borsali1989b} also studied mixtures of polystyrene and 
polydimethylsiloxane in tetrahydrofuran (which is isorefractive with 
polydimethylsiloxane) and in toluene, which is a zero average contrast solvent 
for these polymers and conditions.  Measured spectral forms (one or two 
relaxations), relaxation times, and their concentration dependences were 
consistent with the Benmouna model\cite{benmouna1987a}.  

In an extremely important series of experiments, Chang, et al.\cite{chang1988a} 
report probe diffusion by polystyrene through polyvinylmethylether: toluene, 
using both QELSS and FRS to determine diffusion coefficients.  The two 
techniques differ markedly in the length scales they examine, scattering vectors 
$q$ for the two techniques being $10^{10} \leq q^{2} \leq 10^{11}$ cm$^{-2}$ 
for QELSS but only  $10^{7} \leq q^{2} \leq 10^{8}$ cm$^{-2}$ for FRS.  The 
relaxation rates $\Gamma$ were within experimental error proportional to 
$q^{2}$ over the full range  $10^{7} \leq q^{2} \leq 10^{11}$ cm$^{-2}$, 
confirming that QELSS does measure a simple translational diffusion 
coefficient, even though the probe displacements sampled by QELSS are smaller  
than the diameter of the probe's correlation hole in the solution. 
Any claim that probe motion is more rapid, inside the 
probe's correlation hole, than it would be over larger distances, must therefore 
explain how Chang, et al.'s data are consistent with the claim.  Chang, 
et al.\cite{chang1988a} also 
demonstrate that the initial dependence of the measured 
diffusion coefficient on the probe concentration depends strongly on matrix 
concentration, the initial slope being substantially positive in the absence of 
matrix polymer and becoming significantly negative as polystyrene concentration 
is increased.

B. Chu, et al.\cite{chu1986a,chu1987a} report on the diffusion of polymethacrylate probes 
through a polystyrene matrix.  The solvent was a mixture of toluene and 
$\alpha$-chloronaphthalene.  By adjusting the solvent composition and the 
temperature, it was possible to make a virtually exact match of the indices of 
refraction of the mixed solvent and the matrix, so that scattering arose only 
from the probe.  By varying the solvent composition, the motions of the matrix 
could separately be examined.   A weak fast mode and a dominant slow mode were 
apparent, even at low angles, for probe chains much larger than the matrix 
chains.  The slow mode had a weaker concentration dependence than does the 
solvent viscosity.

\begin{figure} 

\includegraphics{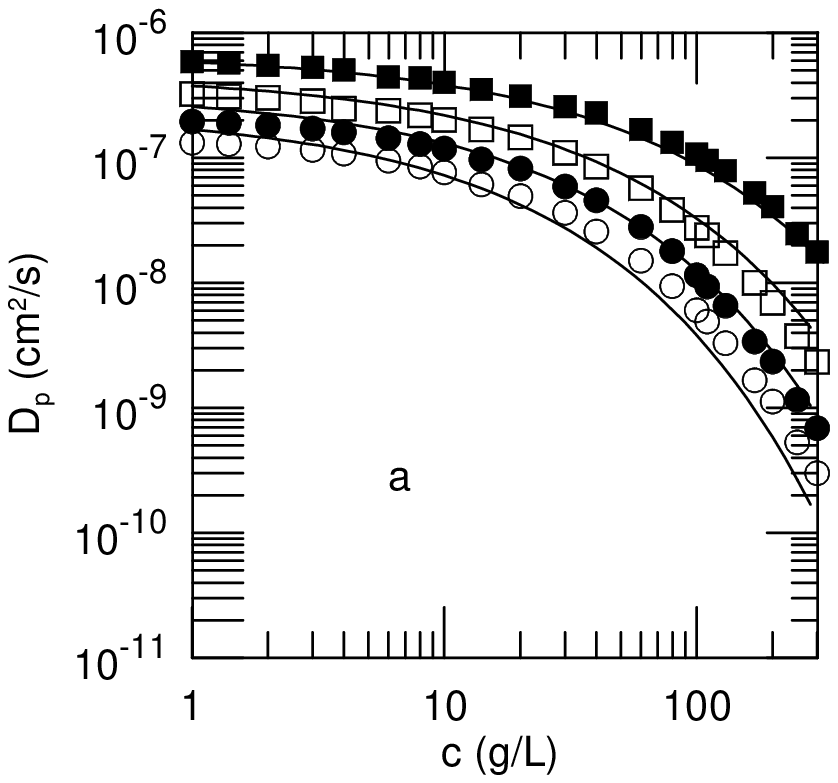} %wh02a.eps

\end{figure}

\begin{figure}
\includegraphics{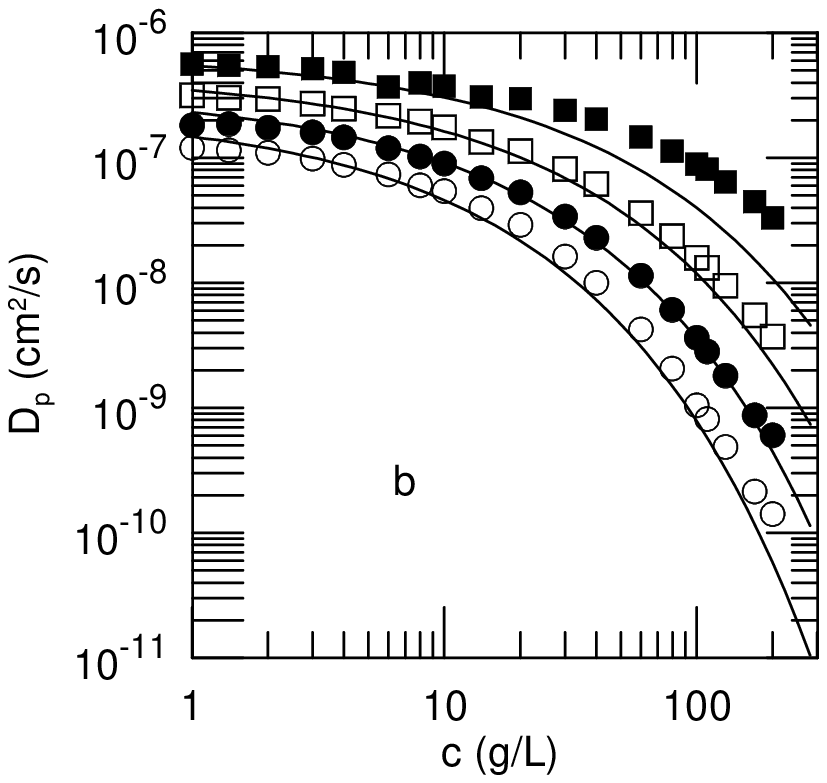} %wh02b.eps
\includegraphics{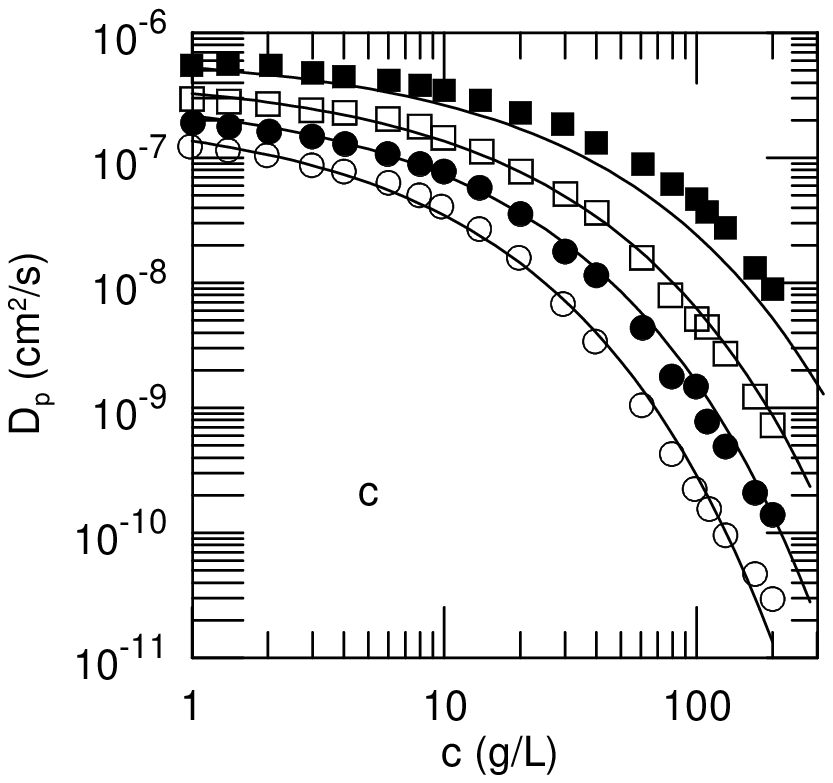} %wh02b.eps

\caption{\label{figure72}
Merged data from Figs.\ \ref{figure69} and \ref{figure71},  
with (a) 140 kDa, (b) 630 kDa, and (c) 1300 kDa matrix polymers, probe symbols 
as in the previous figures, and 
(solid line) fit of all data to a single stretched exponential in $c$, $P$, and $M$.
Parameters are in Table \ref{table4}. 
}
\end{figure}

S. Chu and collaborators\cite{smith1995a} studied the motion of 
fluorescently labelled DNA molecules in non-dilute solution.  They propose that 
they have confirmed the reptation model, based on their observations that
(i) stretched DNA chains relax as though they were confined to tubes, (ii) $D_{s}$ 
scales as $M^{-2}$ for nondilute probe chains, and (iii) a scaling law $D_{s} \sim 
c^{-1.75}$ is consistent with their data.  

However, referring to these points seriatim: (i) Even within the reptation 
model a stretched chain relaxes from stretching
primarily via its higher-order modes, revealing 
only tube confinement on a short time scale.  
(ii) The scaling result $D_{s} \sim M^{-2}$ 
is a common feature of many models, is in no sense a signature of 
reptation\cite{skolnick1990a}, and in any event is inconsistent with the 
modern\cite{lodge1999a} analysis of data on synthetic polymers, (iii) Ref.\ 
\onlinecite{smith1995a}'s measurements of $D_{s}(c)$ are too scattered to demonstrate 
$c^{-1.75}$ behavior.  Furthermore, 
if these data had demonstrated concentration scaling, they would be 
inconsistent with essentially all other 
data in the literature on the concentration dependences of 
$D_{s}$ and $D_{p}$ of polymers in solution, as reviewed here, in that in almost all other systems $D_{s}$ and $D_{p}$ have stretched-exponential, not power-law, concentration dependences.

Finally, as was not emphasized in the original paper, the data of Chu and collaborators\cite{smith1995a} 
conclusively reject the reptation model for their system.  
According to the reptation model, polymer chains only 
perform free diffusion after they escape from their tubes, which occurs at times 
longer than $\tau_{d}$.  At times $t < \tau_{d}$, polymer chains remain confined to their tubes and perform  
confined diffusion, so for $t < \tau_{d}$
a chain's mean-square displacements satisfy $\langle 
x^{2} \rangle \sim t^{x}$ where $x$ is 0.5 or less.  Ref.\ 
\onlinecite{smith1995a} reports that their chains under their conditions remain 
within the hypothesized tube for a time $\tau_{d} \approx 1.2$ or $\tau_{d} > 
2$ minutes, depending on which of several methods is used to estimate 
$\tau_{d}$.  However, ref.\ \onlinecite{smith1995a} measured directly $\langle 
x^{2}(t) \rangle$.  Figure 2 of ref.\ \onlinecite{smith1995a}
shows $\langle x^{2}(t) \rangle \sim t^{1}$  
for times as short as $\tau_{d}/7$ or less.  That is, ref.\ \onlinecite{smith1995a} 
found that their chains were performing free, Brownian, non-constrained diffusion
even even on time scales sufficiently short that under the reptation model the chains would have been confined to their tubes.  This last finding is completely incompatible with the reptation model, which requires 
that chains perform constrained curvilinear diffusion along their tubes so 
long as they remain inside their tubes.  The data of Chu and 
collaborators\cite{smith1995a} thus indicates that a key aspect of 
the reptation picture, namely tube confinement, is incorrect, at least in Chu, 
et al.'s system.  

Corrotto, et al.\cite{corrotto1996a} performed static light scattering and 
QELSS measurements on bidisperse nondilute mixtures of polystyrene in toluene, 
extracting fast and slow mode relaxations.  

Cosgrove and Griffiths\cite{cosgrove1993a} used PFGNMR to study the 
diffusion 
of protonated polystyrenes through solutions of deuterated polystyrenes, 
varying the matrix concentration, and the probe and matrix molecular weights.  
Data was obtained over limited ranges of $D_{p}$ on the dependence of $D_{p}$ 
on each of these variables.  $D_{p}$ generally declined with increasing matrix 
concentration and molecular weight, and also declined with
increasing probe molecular weight, except for very 
large probes in relatively dilute solutions of small polymers, in which 
$D_{p}$ was nearly independent of matrix molecular weight.

Cotts\cite{cotts1983a}  report QELSS measurements on polystyrene: 
polyvinylmethylether: toluene, proposing that radii of gyration, hydrodynamic 
radii, scaling exponents, and modes of diffusive motion can be measured 
systematically for the dilute visible chains.   This and other\cite{lodge1983a} 
early work on polystyrene: polyvinylmethylether: toluene 
flowered into the systematic
studies of Lodge, Wheeler, and 
collaborators\cite{lodge1986a,lodge1987a,lodge1989a,wheeler1987a,wheeler1989a}
on this system .

Daivis and Pinder\cite{daivis1993a} report QELSS studies of polystyrene: 
polyvinylmethylether mixtures dissolved in toluene or carbon tetrachloride.  
The first ternary mixture is nearly unique, in that the 
polymers are nearly compatible, toluene is a good solvent for both 
polymers, and toluene and PVME are isorefractive.  The second ternary mixture 
differs from the first in that CCl$_{4}$ is only a marginal solvent for 
polystyrene.  In CCl$_{4}$, spectra become nonexponential.  The diffusion 
coefficient in nondilute CCl$_{4}$ 
solutions is reduced by more than three-fold relative 
to diffusion by the same concentrations of the two polymers in toluene, but 
the probe radius of gyration is only very slightly reduced by increasing the 
matrix concentration.        

Desbrieres and collaborators\cite{desbrieres1993a} applied QELSS to solutions 
of dextran and polyvinylpyrrolidone in water in solutions more concentrated 
than the overlap concentration.  Two modes whose properties are consistent with 
Benmouna\cite{benmouna1987a}-type models of diffusion in ternary polymer 
solutions were observed.  The mixtures have a phase separation at elevated 
concentration.  As polymer concentrations are increased towards the phase 
separation, a third slow mode whose various properties are consistent with the 
formation of aggregates was observed.  

Giebel and co-workers\cite{giebel1990a, giebel1992a} studied QELSS spectra of 
polydimethylsiloxane and polymethylmethacrylate in several solvents as a 
function of the relative concentration of the two polymers.   At the fixed 
total polymer concentration, the polymers were reasonably expected to be 
non-dilute.  Comparison was made with theoretical results of Benmouna, et 
al.\cite{benmouna1987a}, with particular attention to the "zero average 
contrast" condition.  A strong variation of some spectral parameters with 
composition was described well by Benmouna\cite{benmouna1987a}-type models and 
a small number of free parameters.  

Jamil, et al.\cite{jamil1994a} report QELSS measurements on the diffusion of 
random-coil polystyrene probes through solutions of the rigid-rod polymer 
poly-($\gamma$-stearyl $\alpha$,L-glutamate) in its isorefractive solvent 
toluene.  The polymers are incompatible, poor miscibility of rods with random 
coils being identified by Jamil, et al.\cite{jamil1994a} as a significant 
experimental challenge.  Experimentally, the light scattering spectrum in these 
systems is due entirely to the polystyrene probes.  

Jamil, et al.\cite{jamil1994a} found that the dominant slow mode of the QELSS 
spectra is strongly concentration-dependent.  At lower matrix concentration the 
diffusion coefficient corresponding to the slow mode increases with increasing 
probe concentration.  At elevated matrix concentration this diffusion 
coefficient instead decreases with increasing probe concentration.  
Extrapolating this diffusion coefficient to zero probe concentration gives the 
tracer-diffusion coefficient of the probe polymer through the rod matrix.  
Jamil et al.\ reported 
fitting their six values of $D_{s}$ to a stretched-exponential form.  They 
found $\alpha \approx 0.4$ and $\nu \approx 1.3$.  Jamil, et 
al.\cite{jamil1994a} interpret the finding $\nu > 1$ as arising from end-to-end 
aggregation of the matrix polymer; the matrix polymer increases its 
hydrodynamic radius as its concentration is increased.  

Konak, et al.\cite{konak1990a} report QELSS spectra of mixtures of polystyrene 
and polymethylmethacrylate in toluene.  Neither polymer was dilute.  Comparison 
was made for a limited number of concentrations with theoretical models arising 
from work of  Benmouna, et al.\cite{benmouna1987a}.  Treating spectra as 
bimodal, the ratio of relaxation times was predicted theoretically to better 
than 50\%, but predictions of the mode amplitude ratio were often inexact by 
factors of 2 or 3.  

Marmonier and Leger\cite{marmonier1985a} report extensive measurements 
using FRS on
tracer diffusion of labelled polystyrenes through polystyrene in a good 
solvent.  Unfortunately, the reported data were modified by dividing 
them by an unreported \emph{concentration-dependent} factor, namely the normalized 
concentration-dependent tracer diffusion coefficient of the free label in the 
same polymer solutions.  It is therefore impossible to compare these 
measurements with other papers analyzed here.

Numasawa, et al.\cite{numasawa1986b} report scattering spectra of a series of 
dilute polystyrenes in polymethylmethacrylate (in its isorefractive solvent 
benzene).  The major focus was the scattering-vector dependence of the 
linewidth $\Gamma$, especially at larger scattering vectors $q$. At larger $q$,
$\Gamma/q^2$ of the self-diffusive mode increases with increasing $q$. 
Measurements were only made at a single non-dilute matrix concentration, 
preventing further analysis of this data along the lines of this review.  

Russo, et al.\cite{russo1999a} report self-diffusion coefficients 
for poly($\gamma$-benzyl-$\alpha$,L-glutamate) in pyridine.  The rodlike 
polymer has an isotropic-cholesteric liquid crystal phase transition with 
increasing concentration.  Russo, et al.\ found that $D_{s}$ decreases with 
increasing polymer concentration until the phase transition is reached.  
At the phase trnsition,
$D_{s}$ increases adruptly; it then decreases again as the polymer 
concentration is further increased.  $D_{s}(c)$ is qualitatively consistent 
with a stretched exponential form in the isotropic phase, but the number and 
spacing of points in the isotropic regime limits the accuracy of a quantitative 
fit to this very interesting data.  

Scalettar, et al.\cite{scalettar1989a} used FRAP and FCS to study diffusion of 
phage $\lambda$ DNA solutions.  By comparing 
systems in which either few, or almost all, 
molecules were labelled, Scalettar, et al.\ were able to confirm the prediction 
of this author\cite{phillies1975a} that if very few macromolecules are 
labelled, fluorescence correlation spectroscopy determines their tracer 
diffusion coefficient, but if almost all macromolecules are labelled, 
fluorescence correlation spectroscopy determines their mutual diffusion 
coefficient.  

Sun and Wang\cite{sun1996a,sun1997a,sun1997b} report a series of studies of 
polystyrene/polymethylmethacrylate mixtures (in benzene, dioxane, and toluene, 
respectively) using QELSS as the major experimental technique.  Both polymers 
were in general nondilute.  Neither polymer is isorefractive with any 
of the solvents.  
The objective was to study the bimodal spectra that arise under these 
conditions and to show that the two relaxation times and the mode amplitude 
ratio can be used to infer diffusion and cross-diffusion coefficients of the 
two components.   Experimental series varied both the total polymer 
concentration and the concentration ratio of the two components.   The 
theoretical model predicts a biexponential spectrum; the experimental data was 
fit by a bimodal distribution of relaxation rates or by a sum of two 
Williams-Watts functions.  The inferred self-diffusion coefficients of both 
species fall with increasing polymer concentration, but the concentration 
ranges that Sun and Wang studied are too narrow for further interpretation.

\section{Analysis} 

The above sections have presented a detailed examination of nearly the entirety 
of the published literature on polymer self-diffusion and probe diffusion in 
polymer solutions.  The dependences of $D_{s}$ and $D_{p}$ on polymer 
concentration, probe molecular weight, and matrix molecular weight have been
determined.  Some features of this literature are incidental consequences of the chemical identity of the polymer being studied.  The objective of this Section is to extract systematic behaviors 
from the above particular results.  Here we ask: If we rise above particular features 
determined by the identity of the polymer under examination, what
are the ideal features common to self-- and probe--diffusion of all polymers in solution?  

In the following: First, the functional forms of the concentration and 
molecular weight dependences of the self- and probe diffusion coefficients are
considered.  Second, having found that $D_{s}$ and $D_{p}$ uniformly follow 
stretched exponentials in $c$, correlations of the stretched-exponential 
scaling parameters with other polymer properties are examined.  Third, for 
papers in which diffusion coefficients were reported for a series of homologous 
polymers, we examine a joint function of matrix concentration and 
matrix and probe molecular weights and what it 
reveals about polymer diffusion.  Fourth, we 
examine the few cases in which stretched-exponential behavior is not seen, or 
in which particular features of a given system clarify the systematic behavior 
of the phenomenological parameters used to describe $D_{s}$ and $D_{p}$.  
Finally, other results implying a generalized phenomenology for aspects of 
diffusion behavior are examined.

{\bf First}, in the above I have reviewed virtually the entirety of the 
published literature on self-diffusion and probe diffusion of random-coil 
polymers in solution.  As seen from the Figures in the preceding sections, the 
concentration dependences of $D_{s}$ and $D_{p}$ are essentially always 
described well by stretched exponentials (eq.\ \ref{eq:dsseeq}) in the matrix 
concentration $c$.  

Correspondingly, scaling (power-law) behavior is clearly rejected by almost the entire 
published literature on polymer self-- and probe--diffusion.  On a log-log plot, 
a stretched exponential appears as a smooth curve of monotonically varying
slope.  In contrast, on a log-log plot, scaling (power-law) 
behavior would appear as a straight line.  Almost without exception, log-log 
plots of real measurements of $D_{s}(c)$ give smooth curves, not straight lines. Power laws could be 
fit to reported data, but in almost every case the power law would only provide an accurate description of
a tangent to the data over some narrow range of concentrations.  

An observation that experimental data is described very well by a particular 
mathematical form does not prove that the form in question is physically 
significant.  In principle it may be the case that several different 
mathematical forms describe, to within the actual experimental error, the same 
data.  However, the observation that experimental data is uniformly \emph{not}
described by some mathematical form, to well beyond experimental error, is good 
evidence that models that predict the mathematical form are inadequate.  Power laws appear on log-log plots as straight lines. As seen from the above 70+ figures, straight lines are almost never observed in 
plots of $D_{s}$ or $D_{p}$ against $c$ or other physical variables.  Correspondingly, 
the concentration dependence of $D_{s}$ and $D_{p}$ almost never shows scaling (power-law) behavior.
Scaling models that predict or assume for polymer {\em solutions\/} that 
$D_{s}$ and $D_{p}$ follow power laws in $c$, $P$, and/or $M$ are very 
definitely rejected by almost the entirety of the published literature on 
$D_{s}$ and $D_{p}$.  

\begin{figure} 

\includegraphics{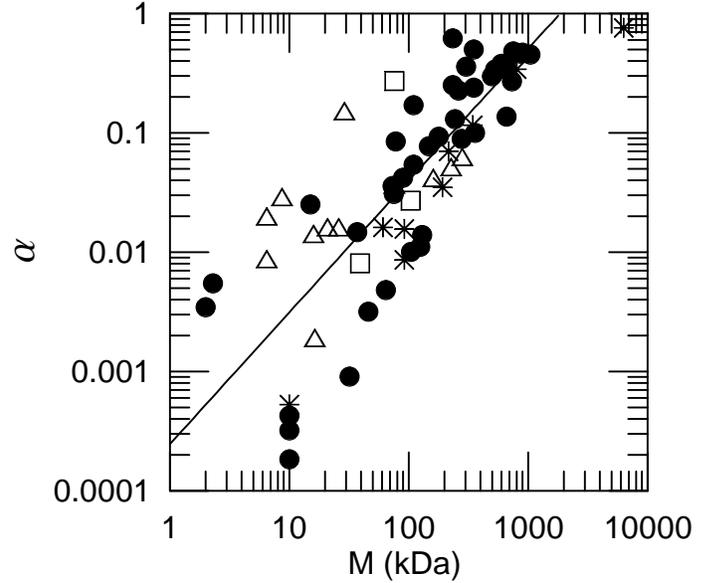} %revdsam1.eps
\caption{\label{figure73}  Scaling pre-factor $\alpha$ for linear polymers ($\bigcirc$) and 3-, 8-, and 18-
armed star polymers ($\bigtriangleup, \square, \ast$)
as listed Table I, plotted against polymer 
molecular weight $M$.  Solid line is a power law $\alpha \sim 
M^{1.1}$.}
\end{figure}

{\bf Second}, referring to eqs \ref{eq:dsseeq}, there are systematic 
correlations between scaling parameters 
$\alpha$ and $\nu$ and the solution variables $P$ and $M$.  These 
correlations reflect the concentration and molecular 
weight dependences of $D_{s}$ and $D_{p}$.  In particular:

{\em The scaling prefactor $\alpha$ depends strongly on $M$.}  
Figure \ref{figure73} shows the scaling prefactor $\alpha$ from measurements of 
the self-diffusion coefficient and fits to eq.\ \ref{eq:dsseeq}, as plotted 
against polymer molecular $M$.  The Figure, based on Table I, shows almost all data on 
linear and star polymers, with 
concentrations in g/L.
Results from the one system\cite{tinland1990a} with a large-concentration 
phase transition are omitted.  $M$ varies over nearly three orders of 
magnitude; $\alpha$ varies over almost four orders of magnitude. While there is 
substantial scatter, the figure is consistent a power-law 
correlation between $\alpha$ and $M$.  The 
solid line in the Figure 
is a best fit to measurements on linear polymers.  It shows $\alpha = 
2.45 \cdot 10^{-4} M^{1.10}$ with $M$ in kDa.  

It is also possible to examine the correlation of $\alpha$ with the size of the 
polymer, as reflected, e.g., by the diffusion coefficient $D_{o}$.
However, the variation in $D_{o}$ from system to system arises in part from 
differences in the measurement temperature and solvent viscosity.  To eliminate 
these effects, $D_{o}$ was used to compute a nominal chain hydrodynamic radius 
\begin{equation} 
      R = \frac{k_{B}T}{6 \pi \eta D_{o}}.
     \label{eq:Rdef}
\end{equation}
Here $k_{B}$ is Boltzmann's constant and $T$ is the absolute temperature.  The 
solvent viscosity $\eta$ was taken from standard tables and interpolated as 
necessary to actual temperatures.  In a few cases, reported experimental 
conditions do not permit an accurate conversion from $D_{o}$ to $R$; these 
cases are not considered further.  

\begin{figure} 

\includegraphics{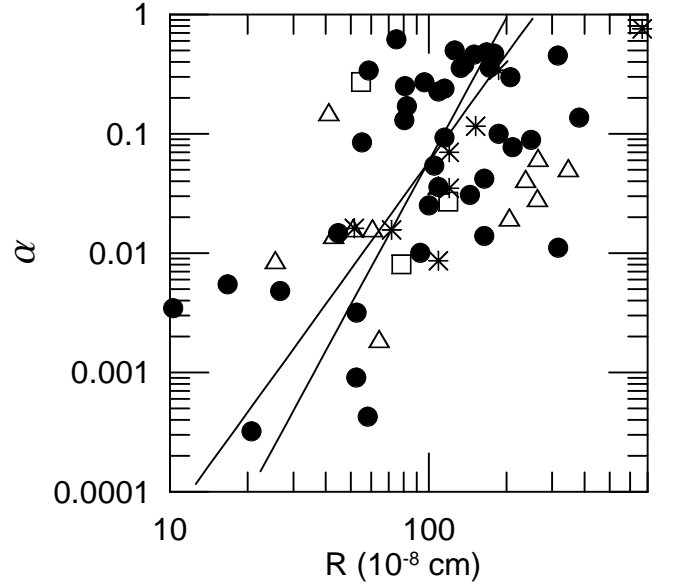} %revdsar1.eps
\caption{\label{figure74} Scaling pre-factor $\alpha$ for linear polymers, Table I,
plotted against the bare hydrodynamic radius $R$, eq.\ \ref{eq:Rdef}.
Solid lines represent power laws with exponents of 3 and 
4.}
\end{figure}

Figure \ref{figure74} plots $\alpha$ as a function of $R$.  In Fig.\ \ref{figure74}, 
the solid lines are best fits for the data on linear chains (filled circles) to 
 $\alpha \sim R^{n}$ for $n$ of 3 or 4.  If $\alpha \sim [\eta]$ as predicted by 
the Adler model\cite{adler1980a}, then $n=3$ is expected.  The hydrodynamic 
scaling model\cite{phillies1987a,phillies1988a} instead predicts $n=4$.  The 
fit with $n=3$ gives a modestly better fit to the data than does the $n=4$ fit, 
but neither fit is visibly inconsistent with experiment.  

Figures \ref{figure73} and \ref{figure74} also present $\alpha$ of 3-, 8- and 18-armed 
stars.  $\alpha$ of an 18-armed star polymer tends to be somewhat smaller than 
$\alpha$ for a representative linear chain.  However, $\alpha$ of star 
polymers almost always lies within the scatter in the values 
for $\alpha$ observed for the 
linear chains.  $\alpha$ for three-armed stars includes 
results\cite{vonmeerwall1983a} in which only a limited number of 
concentrations were studied for a given polymer.

\begin{figure} 

\includegraphics{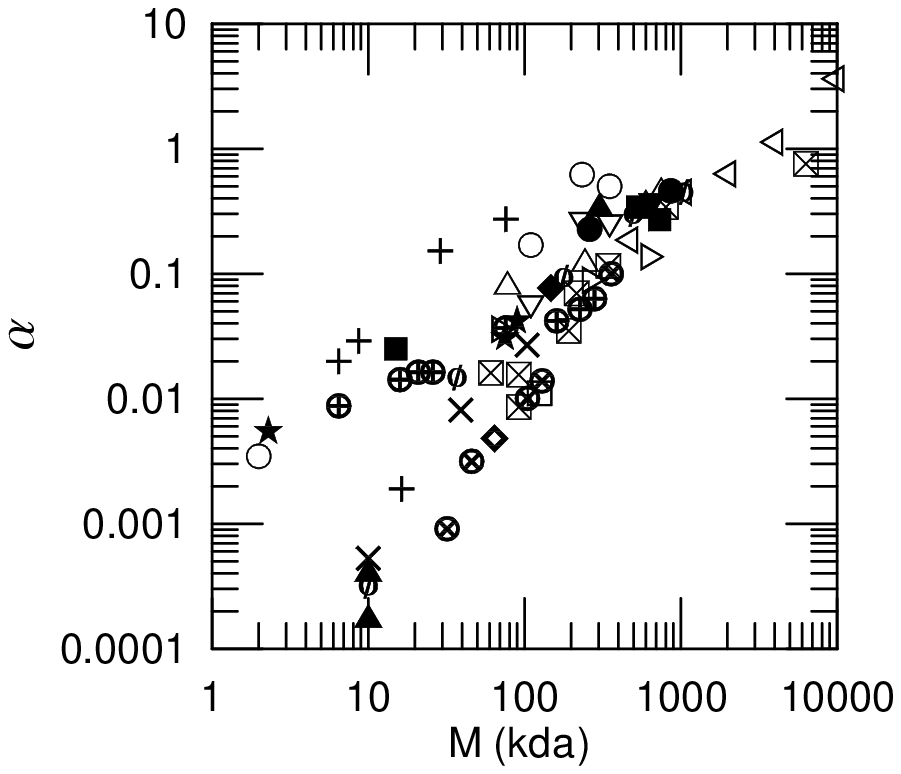} %revdsam4.eps
\caption{\label{figure75} Scaling pre-factor $\alpha$  as a function of $M$ for data from Refs.\  
($\bigcirc$) \cite{callaghan1980a} for polymers in CCl$_{4}$,
($\bullet$) \cite{deschamps1986a},
($\square$) \cite{fleischer1999a},
($\blacksquare$) \cite{giebel1993a,skirda1988a},
($\bigtriangleup$) \cite{leger1981a},
($\blacktriangle$) \cite{vonmeerwall1982a,vonmeerwall1984a} with linear chains,
($+$) \cite{vonmeerwall1982a,vonmeerwall1983a} with $f=3$,
($\times$) \cite{vonmeerwall1982a} with $f=8$,
($\boxtimes$) \cite{vonmeerwall1984a} with $f=18$,
($\lozenge$) \cite{brown1982a},
($\blacklozenge$) \cite{brown1983a},
($\bigtriangledown$) \cite{callaghan1980a} for polymers in C$_{6}$D$_{6}$,
($\blacktriangledown$) \cite{deschamps1986a},
($\otimes$) \cite{wesson1984a},
($\varnothing$) \cite{vonmeerwall1985a},
($\star$) \cite{vonmeerwall1983a} for linear polybutadiene,
($\oplus$) \cite{vonmeerwall1983a} for $f=3$ polybutadiene,
($\triangleright$) \cite{brown1983a} for PEO in water, and
($\triangleleft$) \cite{tinland1990a} for xanthan in water.
Other details as in Fig.\ \ref{figure73}}.
\end{figure}

\begin{figure} 

\includegraphics{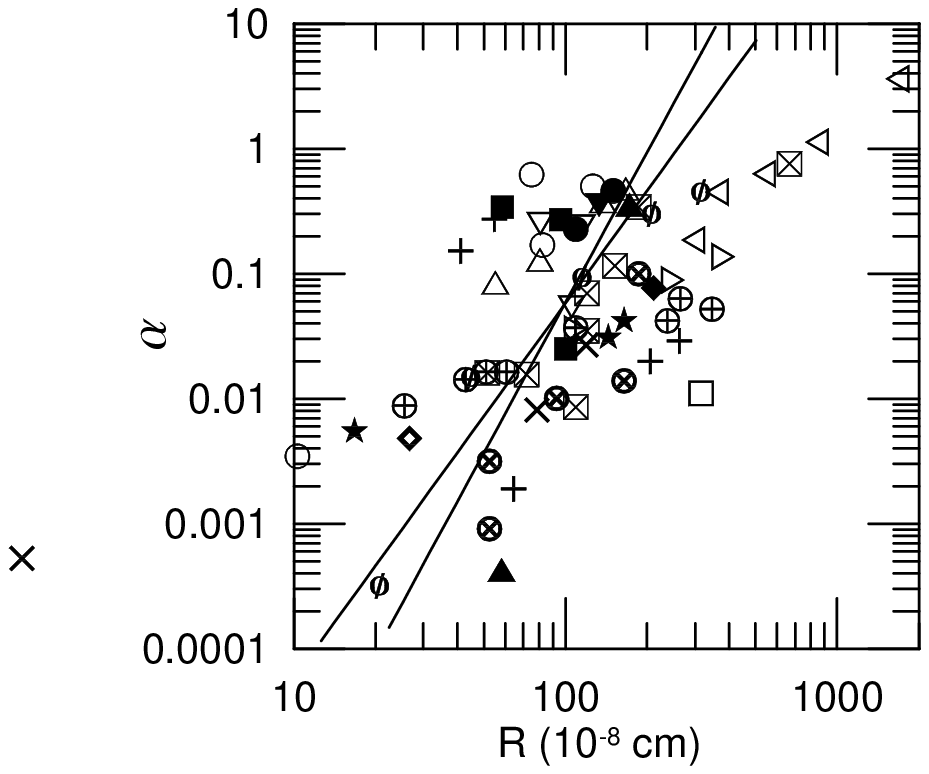} %revdsar4.eps
\caption{\label{figure76} 
Scaling pre-factor $\alpha$  as a function of $R$.
Other details as in Fig.\ \ref{figure75}.}
\end{figure}

Figures \ref{figure73} and \ref{figure74} emphasize polymer topology: one point style 
each for linear, 3-, 8-, and 18-armed chains.
For some purposes, identifying the points by reference (and, hence, by chemical 
system and experimental method) is more useful.  Figures \ref{figure75} and 
\ref{figure76} give $\alpha$ against $M$ and $R$, respectively, with points 
labelled by reference, this time including results of Tinland, et 
al.\cite{tinland1990a} on a system with a large-$c$ phase transition.  
The correlation between $\alpha$ and molecular weight is somewhat better that 
the correlation between $\alpha$ and the inferred -- in most cases, not 
directly measured -- chain radius.

\begin{figure} 

\includegraphics{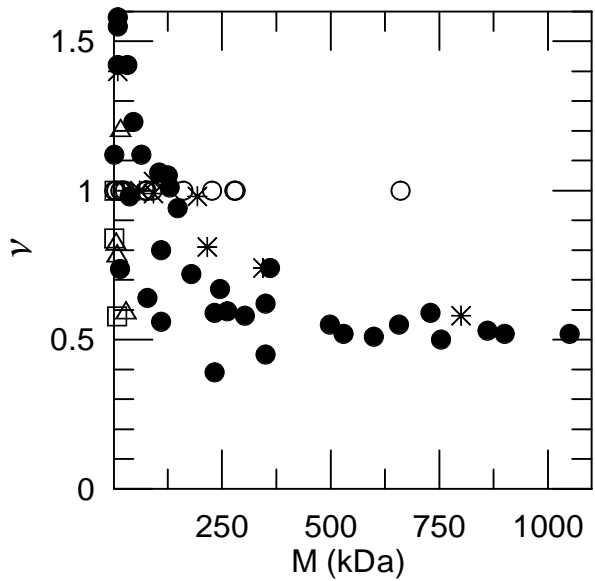} %revdsnm4.eps
\caption{\label{figure77}   Scaling exponent $\nu$ for polymers of
Table 1 plotted against the polymer molecular weight $M$.  A large-molecular-
weight asymptote $\nu \approx 0.5$, and a low-molecular-weight increase in 
$\nu$ toward $\nu = 1$ are apparent for linear polymers (filled circles) and 
for 3- (triangles), 8- (squares), and 18- (asterisks) armed star 
polymers.  Open circles refer to linear polymers in which an adequate fit was 
obtained with $\nu=1$ forced during the fitting process.  }
\end{figure}

{\em The scaling exponent $\nu$ depends on $M$ at low $M$.}
Figure \ref{figure77} shows the dependence of $\nu$ on polymer molecular weight.  
The filled circles refer to linear polymers as studied in refs. 
\cite{brown1982a,brown1983a,callaghan1980a,deschamps1986a,fleischer1999a,giebel1993a,skirda1988a,leger1981a,vonmeerwall1982a,vonmeerwall1984a,vonmeerwall1985a,wesson1984a}, but excludes systems\cite{tinland1990a} in which there is a phase transition at 
elevated polymer concentration.  As seen from the Figure, for polymers larger 
than 250 kDa $\nu$ approaches very closely to 0.5.  For smaller polymers, $\nu$ 
is substantially scattered, but increases with decreasing $M$.  

Sixteen systems in Table I, including linear and three-armed star 
polymers, had $\nu$ forced to 1.00 during the fitting process.   In these fits $\nu = 1$ 
was forced because the data would not support more free parameters, or because 
the fit with $\nu$ as a free parameter did not have a significantly better root 
mean square fractional error in the fit than did the fit with $\nu =1$ forced.  
All but four of these fits refer to polymers with $M < 200$ kDa.  The 
circumstance that successful fits with $\nu$ forced to unity are largely found 
with polymers of lower molecular weight is consistent with the interpretation 
that $\nu$ increases toward 1.0 at small $M$.  These fits are represented in 
Figure \ref{figure77} by the open circles.  

Figure \ref{figure77} also shows the molecular weight dependence of $\nu$ for 
star polymers in solution.  The data on stars 
are substantially confined to smaller 
polymers ($M < 250$ kDa).  However, a trend in $\nu$ with increasing $M$ 
is apparent.  At very small $M$, $\nu$ is substantially 
scattered around 1.0.  For $M \geq 200$ kDa, 
$\nu$ for star polymers appears from the few points to be 
trending toward 0.5.  

{\bf Third,} several papers 
\cite{brown1983a,brown1988a,martin1986a,hanley1985a,kent1992a,kim1986a,lodge1986a,lodge1987a,lodge1989a,martin1984a,nemoto1985a,nemoto1989a,nemoto1990a,numasawa1986a,pinder1990a,wheeler1987a,wheeler1989a} report $D_{s}$ and $D_{p}$ for a series of homologous polymers with different 
molecular weights.  Each of these papers describes a study made by a given 
method using consistent operating conditions and data analysis procedures, 
thereby avoiding data scatter arising from any practical experimental 
issues.  To use these studies to obtain information about the molecular weight 
dependences of $D_{s}$ and $D_{p}$, data from each paper were 
separately fit to 
eq \ref{eq:Dsseeq2}, which is a joint stretched exponential in $c$, $M$, and 
$P$. (For $D_{s}$, $P$ and $M$ are the same variable.)  
As seen Sections III and IV, eq \ref{eq:Dsseeq2} with parameters in Table 
\ref{table4} fits most of each data set well.  

Some studies were limited in the range of matrix concentrations that was 
explored, either by not going to very high concentration or by not making
measurements at low concentration.  
These\cite{brown1982a,hanley1985a,nemoto1985a,nemoto1989a,numasawa1986a} 
studies are 
excluded here from further consideration.  Some studies are subsets of other 
studies, for example cases in which measurements were first made at fixed $M$ 
for a range of $P$, and then extended to cover a range of $M$.  To avoid 
double-counting experiments, data sets that are subsets of other data are not 
considered further here.  

Section IV described fits of $D_{p}$ to eq \ref{eq:Dsseeq2}.  Much of the 
probe diffusion data is represented well by this joint stretched exponential 
in $c$, $P$, and $M$.  However, eq \ref{eq:Dsseeq2} is less accurate when 
$M$ and $P$ differ substantially.   In particular, for $M/P >3$ or for $P/M > 
3$, eq \ref{eq:dsseeq} tends to overstate the decrease in $D_{p}$ with 
increasing $c$.  Specific results of Section IV pertaining to the range of 
validity of eq \ref{eq:Dsseeq2} include: 

Brown, et al.\cite{brown1988a}'s data 
for $D_{p}/D_{p0} < 10^{-3}$ 
show an experimental 
$D_{p}$ that is smaller than the calculated value from a fit to eq 
\ref{eq:Dsseeq2}.  Also, for $P/M \approx 80$, 
the measured $D_{p}$ is larger than  
predicted.  Hanley, et al.'s data\cite{hanley1985a} and 
Kent, et al.'s\cite{kent1992a} data (
all follow well eq \ref{eq:Dsseeq2}'s $P$ 
and $M$ dependences.  Kim, et al.'s\cite{kim1986a} extensive results 
on the 
effects of varying $P$ and $M$ indicate that eq \ref{eq:Dsseeq2} generally
works well for $M/P \leq 3$, but overstates the dependence of $D_{p}$ on $M$ 
for $M/P >3$. From this data, it appears that as $c$ is increased 
there is a narrowing of the range of values of $M/P$ for which eq
\ref{eq:Dsseeq2} works well.

For $M/P$ never larger than 3 or so, Lodge and Wheeler\cite{lodge1986a} found 
that the joint stretched exponential worked well for 
linear polystyrene chains and also worked well for $f=3$ polystyrene stars, 
both  diffusing through linear-chain matrices.  For large $f=12$ polystyrene 
stars in a short linear polyvinylmethylether, Lodge and 
Markland\cite{lodge1987a}'s data follow a joint stretched 
exponential, even for $P/M \approx 12$.  If the comparison is made by arm 
rather than total molecular weights, Lodge and Markland's data is confined 
to a region $P_{a}/M_{a} < 2$.  Lodge, et al.\cite{lodge1989a} examined the diffusion of small $f=12$ stars through 
solutions of a large linear polyvinylmethylether. For $M/P \geq 3$ and elevated 
concentrations, the measured $D_{p}$ is larger than expected from the joint 
stretched exponential. Lodge, et al\cite{lodge1987a,lodge1989a} also studied 
three-armed stars diffusing through linear chains; with increasing matrix 
molecular weight, the joint stretched exponential predicts too small a $D_{p}$ 
for $M/P > 3$.  

Martin\cite{martin1984a,martin1986a} reported $D_{p}$ of linear polystyrenes 
diffusing through linear polyvinylmethylether.  Equation \ref{eq:Dsseeq2} fits well 
(Fig.\ \ref{figure55}) 
the joint data, but thoughout this data $M/P < 2.2$.  Nemoto, et 
al.\cite{nemoto1985a} studied diffusion of PMMA through polystyrene solutions; 
the joint fit (Fig.\ \ref{figure58}) 
of all data to eq \ref{eq:Dsseeq2} was more accurate for $2 
\geq M/P \geq 0.5$ then it was outside this range.  Numasawa, et al.'s 
data\cite{numasawa1986a} on polystyrene in polymethylmethacrylate:benzene,
Fig.\ \ref{figure61}, is 
described reasonably well by eq \ref{eq:Dsseeq2}.  However, for Numasawa, et 
al.'s data the joint 
stretched exponential clearly gives a better description of the concentration 
dependence of $D_{p}$ than it does of the $M$-dependence.  
Wheeler, et al.\cite{wheeler1987a,wheeler1989a} report on linear polystyrenes 
diffusing through polyvinylmethylethers.  Fits to their measurements 
(Figs.\ \ref{figure71} and \ref{figure72}) show that 
eq \ref{eq:Dsseeq2} underestimates $D_{p}$ for $M/P \geq 7$ and  
for $P/M \geq 3$.  

The joint stretched exponential thus describes well a great number of measurements 
of $D_{p}$ in systems with $0.3 \leq M/P \leq 3$.  Outside this range, eq \ref{eq:Dsseeq2} predicts a concentration dependence for $D_{p}$ that is stronger than that found experimentally.  Corresponding to each set of measurements are a set of scaling 
exponents and pre-factors.  To what extent do these parameters show universal
rather than system-specific behavior?  In eq \ref{eq:Dsseeq2}, data were 
parameterized as $D_{o} M^{-a} \exp(-\alpha c^{\nu} 
P^{\gamma} M^{\delta})$.  For each system, the exponent $\nu$ was 
treated as a single constant, the 
dependence of $\nu$ on $M$ at small $M$, discussed above, being
suppressed.  This suppression will reduce the quality of the fits when a very 
wide range of molecular weights is studied, as will be seen below in one 
system.

From Table \ref{table4}: (i) $a$ is almost always in the range 0.5 to 0.6.  (ii) $\nu$ is 
generally in the range 0.5-0.75.  In two cases, $\nu$ was found to be larger 
(0.86, 0.99) than this range; in one case it was smaller (0.43).  In a few 
cases, $\nu =1$ was forced during the fitting process, rather than being 
obtained from the fit.  (iii) For linear chains, $\gamma$ is almost always in the 
range 0.25-0.3.  There are outliers at 0.14, 0.19, and 0.43; each of these is 
associated with a system in which $\nu$ was unusually small (0.43) or unusually 
large (0.86, 0.99).  For star polymers, $\gamma$ is $\approx 0.15$.  (iv) For 
linear polymers and three-armed stars, $\delta$ is usually in the range 0.25--0.30.  

{\bf Fourth}, in all of two cases, 
concentration and molecular weight dependences were not, or were not 
necessarily, described by stretched exponentials.  First, Nemoto, et 
al.\cite{nemoto1991a} studied the $M$-dependence of $D_{s}$ at large fixed $c$.  
Their data show a transition from a stretched-exponential to a power-law 
molecular weight dependence for $M \geq 800$ kDa.  Second, Tao, et 
al.\cite{tao2000a} studied very large polymers in highly concentrated 
solutions.  Their data are almost equally well-described by power laws and by 
stretched exponentials, as seen in Figs.\ \ref{figure14} and \ref{figure15}.  For 
this data, $\alpha$ of eq.\ \ref{eq:dsseeq} is nearly independent of $M$, 
contrary to the behavior of $\alpha$ in other systems.  The lack of an $M$-dependence 
in $\alpha$, and the ability of a power-law to describe this data, 
are consistent with the proposal\cite{tao2000a} that polymer transport coefficients have different phenomenologies in the melt/near-melt 
domain and in the solution domain  here.

{\bf Fifth}, there is evidence for other correlations.

Several papers that report $D_{s}$ or $D_{p}$ also report the solution viscosity $\eta$. In 
particular,  Martin \cite{martin1984a,martin1986a} gives not only $D_{p}$ but 
also $\eta$ of solutions of his 110 kDa polyvinylmethylether matrix polymer.  
Martin reported for  50 kDa probe polystyrene that the product $D_{p} \eta$ 
increased dramatically (up to 6-fold) with increasing $c$.  For larger probe 
polymers (100 kDa, 420 kDa), the increase in $D_{p} \eta$ was less dramatic; 
with 900 kDa probes, "...this product is very nearly independent of 
concentration."  Martin \cite{martin1986a} interpreted the dramatic increase in 
$D_{p} \eta$ with increasing matrix concentration as arising from a crossover 
from Stokes-Einsteinian to reptational diffusion, the crossover occurring more 
readily for $P \leq M$ than for $P \gg M$.  

$D_{p}$, $\eta$ and probe radius of gyration $R_{g}$ for the same set of 
solutions were reported by Numasawa, et al.\cite{numasawa1986a}, who studied 
polystyrene diffusing through polymethylmethacrylate:benzene.  For a range of 
matrix concentrations and matrix and probe molecular weights, Numasawa, et al.\ 
identified a regime in which $D_{p} \eta R_{g}$ is approximately constant.  For 
small probe chains diffusing through larger matrix chains but not large probe 
chains diffusing through smaller matrix chains, the product $D_{p} \eta R_{g}$ 
increases markedly (up to 100-fold) with increasing $c$ and $M$.  Numasawa, et 
al.\cite{numasawa1986a} interpreted their data as showing Stokes-Einstein type 
diffusion and a crossover with increasing matrix molecular weight to a region 
in which reptational motion was dominant.  

Martin\cite{martin1986a} and Numasawa, et al.\cite{numasawa1986a} 
propose that the failure of the Stokes-Einstein equation, with $D_{p} \eta$ 
increasing markedly with increasing $c$, is associated with 
a transition to reptation dynamics for large polymer chains.  However, the diffusion of rigid spherical probes through high-molecular-weight polymers shows a phenomenology 
highly similar to the phenomenology for $D \eta$ is observed\cite{phillies1985a};
namely $D \eta$ increases dramatically with increasing $c$ at large $M$,
even though
there is no possibility that spheres have a transition permitting them to 
diffuse via reptation.

Like effects do not prove that like causes are at work.
The existence of non-reptating spherical probe-polymer systems in 
which $D \eta$ increases markedly with increasing $c$ does {\em not\/}
prove that polymer systems showing the same phenomenology are not reptating.
However, non-reptating sphere-probe systems can have a $D \eta$ 
that increases markedly at large matrix $c$.   Therefore, the observation that 
$D_{p} \eta$ for probe polymers in a polymer solution increases markedly at 
large matrix $c$ is not evidence that the polymer solution has had a transition 
to reptation dynamics.  

Xuexin, et al.\cite{vonmeerwall1984a} report $D_{p}$ for linear and $f =
18$ star polymers having very nearly the same $D_{o}$, i.e., very nearly
the same hydrodynamic radius.  For these two polymers, the scaling
parameters $\alpha$ and $\nu$ are also very nearly the same, consistent
with but not proving the interpretation that $\alpha$ and $\nu$ are
determined by chain size and not by chain topology.

Kent, et al.\cite{kent1992a} examined $D_{p}$ for a range of probe as well as 
matrix concentrations.  For much of their data, Kent, et al.'s probes were 
concentrated enough that $D_{p}$ manifestly depended on the probe concentration
$c_{p}$ as well as on the matrix concentration $c$.  At low $c$, $D_{p}$ 
increased with increasing $c_{p}$.  At elevated $c$, $D_{p}$ decreased with 
increasing $c_{p}$.  These results provide a specific target for theoretical 
investigation.   Just as good hydrodynamic models\cite{carter1985a} can predict 
quantitatively the concentration dependence of $D$ of hard spheres, so also a 
good hydrodynamic model for polymers should be able to predict the dependence 
of $d D_{p}/ d c_{p}$ on matrix concentration.  
 
\section{Conclusions and Discussion} 

In the above, virtually the entirety of the published literature on polymer 
self-diffusion and on the diffusion of chain probes in polymer solutions has 
been reviewed.  Studies that determined how $D_{s}$ and $D_{p}$ depend on 
polymer concentration and molecular weight were systematically re-analysed. Without exception the 
concentration dependences of $D_{s}$ and $D_{p}$ are described by stretched 
exponentials in polymer concentration.  The measured molecular weight 
dependences also compare favorably in most cases with the elaborated stretched
exponential, eq.\ \ref{eq:Dsseeq2}.  Only when $ P \gg M$ or $M \gg P$ is there 
a deviation from eq.\ \ref{eq:Dsseeq2}, that deviation referring only to the 
molecular weight dependences.

Contrarywise, almost without exception the experimental data is inconsistent 
with descriptions of concentration or molecular weight dependences of the single-particle diffusion coefficient in terms of the power laws predicted or assumed by scaling models.  The published data is sufficient to reject 
scaling model descriptions of polymer diffusion.

The concentration dependence $\exp(- \alpha c^{\nu})$ of exponential-type
models for $D_{s}$ is found to be valid for all ratios of the probe and matrix 
molecular weights.  This validity of the stretched-exponential 
concentration dependence over a wide range of $P/M$ is consistent with the 
mathematical structure of the renormalization-group 
derivation\cite{phillies1998a} of these forms.  Namely, the derivation begins with a low-concentration pseudovirial expansion 
\begin{equation} 
    D_{s} = D_{o} ( 1 + k_{2} c + k_{3} c^{2})
    \label{eq:Dslow}
\end{equation}
for the concentration dependence of $D_{s}$, $c$ here being the matrix 
concentration, and then uses the Altenberger-Dahler positive-function 
renormalization group\cite{altenberger1996a} method to extend the series from 
lower to larger concentration.  Over a modest range of $P/M$ around $P \approx 
M$, coefficients $k_{2}$ and $k_{3}$ with a simple dependence on $P$ and $M$ 
are adequate, leading to eq \ref{eq:Dsseeq2}.  For $P \gg M$ or $P \ll M$, 
the simple calculations of $k_{2}$ and $k_{3}$ of Ref.\ \onlinecite{phillies1998a} 
are inadequate, because they do not include polymer internal modes, so the 
values of the $k_{i}$ change.  However, for $P \gg M$ or $P \ll M$ the 
renormalization group method still receives as input a polynomial having the 
form of eq \ref{eq:Dslow}, so it will still generate as output a stretched 
exponential in $c$ for the concentration dependence of $D_{p}$.  

It is sometimes argued that the success of the stretched exponential form in 
describing $D_{s}(c)$ arises from a particular flexibility of the stretched 
exponential, so that the successes shown in the Figures are accidental.  Claims 
that $D_{o} \exp(-\alpha c^{\nu})$ is 'unusually flexible', relative to other 
functional forms, are inconsistent with basic mathematics:  The stretched 
exponential describes the concentration dependence with three free parameters.  
The function is not singular for real $c$ and positive $\nu$.  Therefore, the 
region of function space spanned by the set of all stretched exponentials can 
be no larger than the region of function space spanned by any other function 
with three free parameters.  Correspondingly, the functional form used here for 
$D_{s}(c)$ is no more flexible than other three-parameter forms, so its success in describing $D_{s}(c)$
cannot be ascribed to an unusual flexibility of the stretched exponential form.  

On the other hand, there is no significant region of any set of measurements in which concentration scaling is observed, other than as tangents to smooth curves of monotonically decreasing slope.  It can always be asserted that corrections to scaling, of whatever basis, mean that the predicted power-law slopes are only observed over narrow ranges of concentration, or are only true asymptotically in the limit of large molecular weight:

(A) Until the scaling models are refined sufficiently so as to predict the concentrations over which the hypothesized power-law slope should be approximately observed, it is impossible to tell whether the experimental data are consistent with a scaling prediction.  It is entirely inadequate to show that the experimental data has, in some region, the slope predicted by a particular power law model, because the observed $D_{s}(c)$ someplace agrees with a predicted $c^{-x}$ behavior for every single positive $x$ over a very wide range. If one is allowed to draw the asymptote where one chooses, the predicted $x$ thus ceases to be falsifiable. On the other hand, a refined model might very well predict the slope for a range of concentrations in which the data did not have the predicted slope.   
(B) There is no indication in the experimental data that exponential models cease to be adequate adequate as the polymer molecular weight is increased.

On the other hand, as shown by the extremely thorough experiments of Tao, et al.\cite{tao2000a}, as one moves from polymer solutions toward the melt, one encounters a region of large $c$ in which scaling models are at least approximately correct.  Scaling models therefore may very well--this review does not examine this question in detail, and there is very little data, even though it\cite{tao2000a} is very good data--be asymptotically valid in the near-melt regime.

\begin{acknowledgments}

The partial support of this work by the National Science
Foundation under Grant DMR99-85782 is gratefully acknowledged.

\end{acknowledgments}

\clearpage

\pagebreak

\begin{table*}
\caption{ \label{table1} Fits of the concentration dependence of the self-diffusion 
coefficient $D_{s}$ of polymers in solution to 
a stretched exponential $D_{o} 
\exp(-\alpha c^{\nu})$.  The Table gives the fitting parameters,
the percent  root-mean-square fractional 
error \%RMS, the polymer:solvent system, and the reference.  Square brackets  ``{[ $\cdots$ 
]}'' denote parameters that were fixed rather than floated during the 
least-mean-squares fitting process.  $D_{0}$ is in 
cm$^{2}$/s or in units of $D_{0}$. Concentrations are 
g/L, except $^{*}$ $c$ in volume fraction units. Polymers include 
dex--dextran, pB--polybutadiene, HpBD--hydrogenated polybutadiene, 
pDMS--polydimethylsiloxane, PEO--polyethylene oxide, pI--polyisoprene, 
pS--polystyrene, and xanthan.  Solvents include THF--tetrahydrofuran. }
\begin{ruledtabular} 
\begin{tabular}{|r|r|r|r|r|l|r|} 
$D_{o}$ & $\alpha$& $\nu$ & \%RMS & $M$(kDa) & System& Refs \\ \hline
$9.17 \cdot 10^{-7}$ & $4.82 \cdot 10^{-3}$ & 1.12 & 6.1 &64.2&       
    dex:H$_{2}$O  &\cite{brown1982a} \\
%$1.02 \cdot 10^{-6}$ & $9.90 \cdot 10^{-3}$ & %[1]  & 6.5 &64.2&       
%    dex:H$_{2}$O  & \cite{brown1982a}\\
$2.24 \cdot 10^{-7}$ & 0.036                & [1] & 5.2 &73  &       
    PEO:H$_{2}$O  &\cite{brown1983a}\\
$1.16 \cdot 10^{-7}$ & 0.077                & 0.94 & 7.5 &148 &       
    PEO:H$_{2}$O  &\cite{brown1983a}\\
$ 9.9 \cdot 10^{-8}$ & 0.089                & [1]  & 5.9 &278 &       
    PEO:H$_{2}$O  &\cite{brown1983a}\\
$6.42 \cdot 10^{-8}$ & 0.137                & [1]  & 3.4 &661 &       
    PEO:H$_{2}$O  &\cite{brown1983a}\\
$2.47 \cdot 10^{-6}$ & $3.47 \cdot 10^{-3}$ & 1.12 & 6.6 &  2 &       
    pS:CCl$_{4}$ &\cite{callaghan1980a,callaghan1981a,callaghan1984a} \\
$3.08 \cdot 10^{-7}$ & 0.17                 & 0.56 & 13  &110 &        
    pS:CCl$_{4}$ &\cite{callaghan1980a,callaghan1981a,callaghan1984a} \\
$3.38 \cdot 10^{-7}$ & 0.62                 & 0.39 & 7.4 &233 &  
    pS:CCl$_{4}$ &\cite{callaghan1980a,callaghan1981a,callaghan1984a} \\
$2.01 \cdot 10^{-7}$ & 0.50                 & 0.45 & 8.1 &350 & 
    pS:CCl$_{4}$ &\cite{callaghan1980a,callaghan1981a,callaghan1984a} \\
$3.62 \cdot 10^{-7}$ & 0.054                & 0.80 & 6.6 &110 & 
    pS:C$_{6}$D$_{6}$ &\cite{callaghan1980a,callaghan1981a,callaghan1984a} \\
$4.70 \cdot 10^{-7}$ &  0.25                & 0.59 & 5.8 &233 & 
    pS:C$_{6}$D$_{6}$ &\cite{callaghan1980a,callaghan1981a,callaghan1984a} \\
$3.30 \cdot 10^{-7}$ & 0.24                 & 0.62 & 6.9 &350 & 
    pS:C$_{6}$D$_{6}$ &\cite{callaghan1980a,callaghan1981a,callaghan1984a} \\
$4.50 \cdot 10^{-7}$ & 0.227                & 0.595& 11  &262 & 
    pS:C$_{5}$H$_{10}$ &\cite{deschamps1986a} \\
$[3.7 \cdot 10^{-7}]$ & 0.355                & 0.55 & 23  &657 & 
    pS:C$_{5}$H$_{10}$ &\cite{deschamps1986a} \\
$3.29 \cdot 10^{-7}$ & 0.462                & 0.53 & 12  &861 & 
    pS:C$_{5}$H$_{10}$ &\cite{deschamps1986a} \\
$1.20 \cdot 10^{-7}$ & 0.0111               & 1.05 & 3.4 &125 & 
    pS:tol &\cite{fleischer1999a} \\
$4.88 \cdot 10^{-6}$ & 0.025 & 0.737 & 3.4 & 15 & 
    pDMS:tol &\cite{giebel1993a,skirda1988a} \\
$8.30  \cdot 10^{-7}$ & 0.34                   & 0.52  & 9.5 &530 & 
    pDMS:tol &\cite{giebel1993a,skirda1988a} \\
$5.06 \cdot 10^{-7}$ & 0.27                   & 0.59 & 1.3 &730 & 
    pDMS:tol &\cite{giebel1993a,skirda1988a} \\
$6.21 \cdot 10^{-7}$ & 0.0848                 & 0.64 & 5.1 &78 & 
    pS:C$_{6}$H$_{6}$ &\cite{leger1981a} \\
$7.90 \cdot 10^{-7}$ & 0.21                 & 0.52 & 4.0 &123 & 
    pS:C$_{6}$H$_{6}$ &\cite{hervet1979a} \\
$4.26 \cdot 10^{-7}$ & 0.13                 & 0.67 & 24 &245 & 
    pS:C$_{6}$H$_{6}$ &\cite{leger1981a} \\
$2.33 \cdot 10^{-7}$ & 0.38                 & 0.51 & 22 &599 & 
    pS:C$_{6}$H$_{6}$ &\cite{leger1981a} \\
$2.03 \cdot 10^{-7}$ & 0.48                 & 0.50 & 13 &754 & 
    pS:C$_{6}$H$_{6}$ &\cite{leger1981a} \\
$8.15 \cdot 10^{-6}$ & 3.68$^{*}$    & 0.76 & 0.7& 2   & 
    pEO:CHCl$_{3}$        &\cite{skirda1987a} \\
$2.19 \cdot 10^{-6}$ & 9.07$^{*}$    & 0.61 & 3.7& 40  & 
    pEO:CHCl$_{3}$        &\cite{skirda1987a} \\
$2.69 \cdot 10^{-6}$ & 7.61$^{*}$    & 0.78 & 12 & 20  & 
    pEO:C$_{6}$H$_{6}$        &\cite{skirda1987a} \\
$6.03 \cdot 10^{-6}$ & 4.30$^{*}$    & 0.67 & 3.5& 2   & 
    pEO:dioxane           &\cite{skirda1987a} \\
$2.05 \cdot 10^{-6}$ & 7.75$^{*}$    & 0.62 & 4.9& 20  & 
    pEO:dioxane           &\cite{skirda1987a} \\
$2.71 \cdot 10^{-6}$ & 9.72$^{*}$    & 0.46 & 16 & 40  & 
    pEO:dioxane           &\cite{skirda1987a} \\
$2.30 \cdot 10^{-7}$ &21.9$^{*}$    & 0.42 & 2.3&3000&
    pEO:dioxane           &\cite{skirda1987a} \\
$5.00 \cdot 10^{-7}$ &17.1$^{*}$    & 0.81 & 5.1&240 &
    pS:C$_{6}$H$_{6}$     &\cite{skirda1987a} \\
$3.30 \cdot 10^{-7}$ &16.5$^{*}$    & 0.80 &11.2&240 &
    pS:CCl$_{4}$     &\cite{skirda1987a} \\
$1.65 \cdot 10^{-7}$ &23.5$^{*}$    & 0.67 &13.5&1300&
    pS:CCl$_{4}$     &\cite{skirda1987a} \\
$7.62 \cdot 10^{-8}$ &0.186   & 0.79 & 17 & 450 & 
    xanthan:H$_{2}$O         &\cite{tinland1990a} \\
$6.24 \cdot 10^{-8}$ &0.45    & 0.91 & 15 & 990 & 
    xanthan:H$_{2}$O         &\cite{tinland1990a} \\
$4.18 \cdot 10^{-8}$ &0.63    & 1.00 & 15 &1900 & 
    xanthan:H$_{2}$O       &\cite{tinland1990a} \\
$2.67 \cdot 10^{-8}$ &1.13    & 0.88 & 15 &3800 & 
    xanthan:H$_{2}$O       &\cite{tinland1990a} \\
$1.35 \cdot 10^{-8}$ &3.63    & [1]  & 27 &9400 & 
    xanthan:H$_{2}$O      &\cite{tinland1990a} \\
$6.25 \cdot 10^{-7}$ & $4.25 \cdot 10^{-4}$ & 1.42 & 7.7 & 10 & $f=2$ 
    pI:CCl$_{4}$ &\cite{vonmeerwall1982a} \\
$5.64 \cdot 10^{-7}$ & $1.91 \cdot 10^{-3}$ & 1.21 & 4.6 & 16.4 & $f=3$ 
    pI:CCl$_{4}$ &\cite{vonmeerwall1982a} \\
$4.63 \cdot 10^{-7}$ & $8.08 \cdot 10^{-3}$ & 1.00 & 2.9 & 39 & $f=8$ 
    pI:CCl$_{4}$ &\cite{vonmeerwall1982a} \\
$3.33 \cdot 10^{-7}$ & $8.61 \cdot 10^{-3}$ & 0.99 & 3.1 & 92 & $f=18$ 
    pI:CCl$_{4}$ &\cite{vonmeerwall1982a} \\ 
$3.06 \cdot 10^{-7}$ & $2.69 \cdot 10^{-2}$ & 0.84 & 2.3 &104 & $f=8 $ 
    pI:CCl$_{4}$ &\cite{vonmeerwall1982a} \\ 
$7.63 \cdot 10^{-7}$ & $1.83 \cdot 10^{-4}$ & 1.55 &13.4 & 10 & $f=2$ 
    pI:C$_{6}$F$_{5}$Cl &\cite{vonmeerwall1982a} \\ 
$3.35 \cdot 10^{-7}$ & $5.3 \cdot 10^{-4}$ & 1.40 & 7.8 & 10 & $f=8$ 
    pI:C$_{6}$F$_{5}$Cl &\cite{vonmeerwall1982a} \\ 
\end{tabular}
\end{ruledtabular}
\end{table*}

\begin{table*}
\begin{ruledtabular}
\begin{tabular}{|r|r|r|r|r|l|r|} 
$D_{o}$ & $\alpha$& $\nu$ & \%RMS & $M$(kDa) & System& Refs \\ 
$2.18 \cdot 10^{-6}$ & $5.49 \cdot 10^{-3}$ & [1]  & 3.4 &2.3 & $f=2$ 
     pB:CCl$_{4}$ &\cite{vonmeerwall1983a} \\ 
$1.42 \cdot 10^{-6}$ & $8.78\cdot 10^{-3}$ & 1.00 & 8.6 &6.5 & $f=3$ 
    pB:CCl$_{4}$ &\cite{vonmeerwall1983a} \\ 
$8.44 \cdot 10^{-7}$ & 0.0143              & [1]  & 7.8 & 16 & $f=3$ 
    pB:CCl$_{4}$ &\cite{vonmeerwall1983a} \\ 
$7.07 \cdot 10^{-7}$ & 0.0163              & [1]  & 10  & 21 & $f=3$ 
    pB:CCl$_{4}$ &\cite{vonmeerwall1983a} \\ 
$5.98 \cdot 10^{-7}$ & 0.0163              & [1]  & 7.5 & 26 & $f=3$ 
    pB:CCl$_{4}$ &\cite{vonmeerwall1983a} \\ 
$2.53 \cdot 10^{-7}$ & 0.0308              & [1]  & 8   & 75 & $f=2$ 
    pB:CCl$_{4}$ &\cite{vonmeerwall1983a} \\ 
$3.32 \cdot 10^{-7}$ & 0.0369              & [1]  & 15  & 76 & $f=3$ 
    pB:CCl$_{4}$ &\cite{vonmeerwall1983a} \\ 
$2.22 \cdot 10^{-7}$ & 0.0421              & [1]  & 1.4 & 90 & $f=2$ 
    pB:CCl$_{4}$ &\cite{vonmeerwall1983a} \\ 
$1.53 \cdot 10^{-7}$ & 0.042               & [1]  & 11  &161 & $f=3$ 
    pB:CCl$_{4}$ &\cite{vonmeerwall1983a} \\ 
$1.05 \cdot 10^{-7}$ & 0.052               & [1]  & 13 &227 & $f=3$ 
    pB:CCl$_{4}$ &\cite{vonmeerwall1983a} \\ 
$1.37 \cdot 10^{-7}$ & 0.0629              & [1]  & 6.8 &281 & $f=3$ 
    pB:CCl$_{4}$ &\cite{vonmeerwall1983a} \\ 
$1.77 \cdot 10^{-6}$ & 0.020 & 0.83 & 12  &6.5 & $f=3$ 
    pB:CCl$_{4}$ &\cite{vonmeerwall1983a} \\ 
$1.38 \cdot 10^{-6}$ & 0.029 & 0.79 & 9.0 &8.7 & $f=3$ 
    pB:CCl$_{4}$ &\cite{vonmeerwall1983a} \\ 
$8.81 \cdot 10^{-7}$ & $0.153            $ & 0.60 & 13  & 29 & $f=3$ 
    pB:CCl$_{4}$ &\cite{vonmeerwall1983a} \\ 
$6.63 \cdot 10^{-7}$ & $0.273            $ & 0.578& 8.5 & 87 & $f=3$ 
    pB:CCl$_{4}$ &\cite{vonmeerwall1983a} \\ 
$2.20 \cdot 10^{-6}$ & $3.20 \cdot 10^{-4}$ & 1.58 & 22  & 10 &       
    pS:THF       &\cite{vonmeerwall1985a} \\ 
$1.02 \cdot 10^{-6}$ & $1.47 \cdot 10^{-2}$ & 0.98 & 5.5 & 37 &       
    pS:THF       &\cite{vonmeerwall1985a} \\ 
$3.97 \cdot 10^{-7}$ & $9.29 \cdot 10^{-2}$ & 0.72 & 7.4 &179 &       
    pS:THF       &\cite{vonmeerwall1985a} \\ 
$2.20 \cdot 10^{-7}$ & 0.299                & 0.55 & 15  &498 &       
    pS:THF       &\cite{vonmeerwall1985a} \\ 
$1.45 \cdot 10^{-7}$ & 0.452                & 0.52 & 21  &1050&       
    pS:THF       &\cite{vonmeerwall1985a} \\ 
$9.12 \cdot 10^{-7}$ & $9.07 \cdot 10^{-4}$& 1.42   & 15  & 32 &        
    pS:THF&\cite{wesson1984a} \\ 
$9.09 \cdot 10^{-7}$ & $3.17 \cdot 10^{-3}$& 1.23   & 16  & 46 &        
    pS:THF&\cite{wesson1984a} \\ 
$5.16 \cdot 10^{-7}$ & $1.01 \cdot 10^{-2}$& 1.06   & 21  &105 &        
    pS:THF&\cite{wesson1984a} \\ 
$2.92 \cdot 10^{-7}$ & $1.39 \cdot 10^{-2}$& 1.01   & 17  &130 &        
    pS:THF&\cite{wesson1984a} \\ 
$2.58 \cdot 10^{-7}$ & 0.100              & 0.74   & 19  &360 &        
    pS:THF&\cite{wesson1984a} \\ 
$7.04 \cdot 10^{-7}$ & 0.0161            &  [1]   & 4.7    & 61 & $f=18$ 
    pI:CCl$_{4}$&\cite{vonmeerwall1984a} \\ 
$5.05 \cdot 10^{-7}$ & 0.0156            &  1.03  & 2.0    & 92 & $f=18$ 
    pI:CCl$_{4}$&\cite{vonmeerwall1984a} \\ 
$3.02 \cdot 10^{-7}$ & 0.035            &  0.98  & 2.7    &193 & $f=18$ 
    pI:CCl$_{4}$&\cite{vonmeerwall1984a} \\ 
$3.03 \cdot 10^{-7}$ & 0.070            & 0.81   & 2.8    &216 & $f=18$ 
    pI:CCl$_{4}$&\cite{vonmeerwall1984a} \\ 
$2.38 \cdot 10^{-7}$ & 0.116             & 0.74   & 2.9 &344 & $f=18$ 
    pI:CCl$_{4}$&\cite{vonmeerwall1984a} \\ 
$1.95 \cdot 10^{-7}$ & 0.341             & 0.58   & 8.1 &800 & $f=18$ 
    pI:CCl$_{4}$&\cite{vonmeerwall1984a} \\ 
$5.43 \cdot 10^{-8}$ & 0.76              &[0.5]   & 20  &6300& $f=18$ 
    pI:CCl$_{4}$&\cite{vonmeerwall1984a} \\ 
$2.11 \cdot 10^{-7}$ & 0.359             & 0.58   & 8.6 &302 & $f=2$ 
    pI:CCl$_{4}$&\cite{vonmeerwall1984a} \\ 
1.11 $D_{0}$        & 0.167             & 0.57   & 1.5 &70.8& $f=2$ 
    pI:CCl$_{4}$&\cite{vonmeerwall1984a} \\ 
0.991$D_{0}$         & 0.256             & 0.67   & 3.6 &251 & $f=2$ 
    pI:CCl$_{4}$&\cite{vonmeerwall1984a} \\ 
$[1.0 D_{0}]$           & 0.244             & 0.68   & 6.7 &302 & $f=2$ 
    pI:CCl$_{4}$&\cite{vonmeerwall1984a} \\ 
$2.07 \times 10^{-7}$      & 0.47             & 0.52   & 17 & 900 &  
    pS:tol &\cite{kim1986a} \\ 
%\hline
\end{tabular}
\end{ruledtabular}
\end{table*}

\begin{table*}
\caption{ \label{table2}  Concentration dependence of the probe diffusion 
coefficient $D_{p}$ for molecular weight $P$ probes in solutions of  molecular 
weight $M$ matrix polymers.  The fits are to stretched exponentials $D_{o} 
\exp(-\alpha c^{\nu})$ in the polymer concentration $c$.  The Table gives the 
fitting parameters, the percent 
root-mean-square fractional fit error \%RMS, the system, and the reference.  
Square brackets  ``{[ $\cdots$ ]}'' denote parameters that were fixed rather 
than floated during the non-linear least squares fits.  $D_{0}$ is in 
cm$^{2}$/s or in units of $D_{0}$; concentrations are in g/L.  Abbreviations 
include  dex--dextran, "$f = n$" for an $n$-armed star polymer, 
hyal--hyaluronic acid, mr--methyl red, oFT--orthofluorotoluene, 
pI--polyisoprene, 
PMMA--polymethylmethacrylate, PPO--polypropylene oxide,  
pS--polystyrene, pVME--polyvinylmethylether, 
PF--pulsed field gradient nuclear magnetic resonance, 
QE--quasielastic light scattering spectroscopy, thiop--thiophenol, and 
tol--toluene.  Parentheses indicate parameters from fits of limited accuracy.  }
\begin{ruledtabular} 
\begin{tabular}{|r|r|r|r|r|l|l|r|} 
$D_{o}$ & $\alpha$& $\nu$ & \%RMS & $P$(kDa)& $M$(kDa) & System & Refs. \\
$3.35 \cdot 10^{-7}$ & 0.107 & 0.70 & 8.9 & 245 & 598 &      
    pS:pS:C$_{6}$H$_{6}$& \cite{leger1981a} \\
$2.19 \cdot 10^{-7}$ & 0.453 & 0.50 & 25 & 598 & 1800 &      
    pS:pS:C$_{6}$H$_{6}$ &\cite{leger1981a} \\
1.13 & 0.38                    & 0.38  &7.6  &8000 & 101 &      
   pS:pMMa:tol           & \cite{brown1988a} \\
1.01& 0.61                    & 0.40  &3.6  &8000 & 163 &      
   pS:pMMa:tol            &\cite{brown1988a} \\
1.24 & 1.03                    & 0.34 &7.0  &8000 & 268 &      
   pS:pMMa:tol            &\cite{brown1988a} \\
1.47 & 1.39                    & 0.32  &5.4  &8000 & 445 &      
   pS:pMMa:tol            &\cite{brown1988a} \\
1.06 & 1.18                    & 0.38  &9.0  &8000 & 697 &      
   pS:pMMa:tol            &\cite{brown1988a} \\
1.00 & 1.48                    & 0.45  &14  &8000 &1426 &      
   pS:pMMa:tol            &\cite{brown1988a} \\
0.91 & 1.19                    & 0.34  &14 &2950 & 445 &      
   pS:pMMa:tol            &\cite{brown1988a} \\
0.603& 0.62                    & 0.46  &18   &15000& 445 &      
   pS:pMMa:tol            &\cite{brown1988a} \\
0.973& 0.0101     & \tablenotemark[1]  & 6.7   &73& 19 &      
   pEO:dex:H$_{2}$O            &\cite{brown1983a} \\
2.13& 0.0169     &  \tablenotemark[1]  & 1.9   &73& 110 &      
   pEO:dex:H$_{2}$O            &\cite{brown1983a} \\
1.05 & 0.0256     &  [1]  & 2.8   &73& 510 &      
   pEO:dex:H$_{2}$O            &\cite{brown1983a} \\
1.03& 0.0105    &  1]  & 4.1   & 278 & 19 &      
   pEO:dex:H$_{2}$O            &\cite{brown1983a} \\
0.976& 0.0164    & [1]  & 4.3   & 278 & 110 &      
   pEO:dex:H$_{2}$O            &\cite{brown1983a} \\
0.97 & 0.022     & [1]  & 4.4   & 278 & 510 &      
   pEO:dex:H$_{2}$O            &\cite{brown1983a} \\
0.317 & 0.034     & [1]  & 1.2   & 1200 & 110 &      
   pEO:dex:H$_{2}$O            &\cite{brown1983a} \\
$1.50\cdot 10^{-7}$ & 0.017 &  [1]  &13 &864  & 20.4&      
   dex:dex:H$_{2}$O   &\cite{daivis1984a} \\
$4.97\cdot 10^{-7}$ & 0.174               & 0.60  & 27  &110  & 110 &      
 pS:pVME:tol(QE)       &\cite{daivis1992a} \\
$4.88\cdot 10^{-7}$ & 0.097               & 0.71  & 21  &110  & 110 &      
 pS:pVME:tol(PF)         &\cite{daivis1992a} \\
$4.51\cdot 10^{-7}$ & 0.18                 & 0.59 & 1.9 & 71  & 680 &      
   dex:hyal:H$_{2}$O   &\cite{desmedt1994a} \\
$3.22\cdot 10^{-7}$ & 0.33                 & 0.59 & 5.4 &148  & 680 &      
   dex:hyal:H$_{2}$O   &\cite{desmedt1994a} \\
$1.75\cdot 10^{-7}$ & 0.44                 & 0.48 & 4.8 &487  & 680 &      
   dex:hyal:H$_{2}$O   &\cite{desmedt1994a} \\
$3.49 \cdot 10^{-7}$ & $3.58 \cdot 10^{-2}$ & 0.87 & 3.8 & 162 & 105 &      
    pS:pMMA:C$_{6}$H$_{6}$ &\cite{hadgraft1979a} \\
$(2.53 \cdot 10^{-7})$ & (0.142)                & (0.55) & (10)  & 410 & 105 &      
    pS:pMMA:C$_{6}$H$_{6}$ &\cite{hadgraft1979a} \\
$1.35 \cdot 10^{-7}$ & $6.35 \cdot 10^{-2}$ & 0.76 & 5.6 &1110 & 105 &      
    pS:pMMA:C$_{6}$H$_{6}$ &\cite{hadgraft1979a} \\
$6.70 \cdot 10^{-8}$ & $2.89 \cdot 10^{-2}$ & 0.95 & 6.0 &4600 & 105 &      
    pS:pMMA:C$_{6}$H$_{6}$ &\cite{hadgraft1979a} \\
$3.16 \cdot 10^{-7}$ & $2.26 \cdot 10^{-2}$ & 0.90 & 24  &  50 & 60  &      
    pS:pVME:oFT            &\cite{hanley1985a} \\
$2.15 \cdot 10^{-7}$ & $5.6  \cdot 10^{-2}$ & 0.80 & 17  & 179 & 60  &      
    pS:pVME:oFT            &\cite{hanley1985a} \\
$1.73 \cdot 10^{-7}$ & 0.164                & 0.65 & 9.4 &1050 & 60  &      
    pS:pVME:oFT            &\cite{hanley1985a} \\
$6.09 \cdot 10^{-8}$ & $2.63 \cdot 10^{-2}$ & 1.01 & 19  &1800 & 60  &      
    pS:pVME:oFT            &\cite{hanley1985a} \\
$1.33 \cdot 10^{-7}$ & $6.13 \cdot 10^{-3}$ &  [1]  & 3.8 & 930 & 840 &      
    pS:pMMA:tol        &\cite{kent1992a} \\
$2.99 \cdot 10^{-7}$ & $5.05 \cdot 10^{-3}$ &  [1]  & 8.9 & 233 & 840 &      
    pS:pMMA:tol        &\cite{kent1992a} \\
$2.70 \cdot 10^{-7}$ & $2.27 \cdot 10^{-3}$ &  [1]  & 2.2 & 233 & 66  &      
    pS:pMMA:tol        &\cite{kent1992a} \\
$1.31 \cdot 10^{-5}$ & $4.4 \cdot 10^{-3}$  &  [1] & 6 & 0   & \tablenotemark[1]  &      
    mr:pS:tol &\cite{kim1986a} \\
$1.26 \cdot 10^{-6}$ & $1.87\cdot 10^{-3}$  &1.28 &5.6  & 10 & \tablenotemark[1]  &      
    pS:pS:tol &\cite{kim1986a} \\
$7.89 \cdot 10^{-7}$ & $1.82\cdot 10^{-2}$  &0.92 &3.3  & 35 & \tablenotemark[1]  &      
    pS:pS:tol &\cite{kim1986a} \\
$4.40 \cdot 10^{-7}$ & $5.85\cdot 10^{-2}$  &0.80 &4.0  &100 & \tablenotemark[1]  &      
    pS:pS:tol &\cite{kim1986a} \\
$2.30 \cdot 10^{-7}$ & 0.27                 &0.60 &18   &390 & \tablenotemark[1]  &      
    pS:pS:tol &\cite{kim1986a} \\
$1.81 \cdot 10^{-7}$ & 0.70                 &0.48 &25   &900 & \tablenotemark[1]  &      
    pS:pS:tol &\cite{kim1986a} \\
$8.54 \cdot 10^{-8}$ & 0.64                 &0.55 &17   &1800&8400 &      
    pS:pS:tol &\cite{kim1986a} \\
$2.81 \cdot 10^{-7}$ & $5.4 \cdot 10^{-2}$  &0.82 & 12  & 179& 50  &      
    pS:pVME:oFT &\cite{lodge1983a} \\
$2.94 \cdot 10^{-7}$ & 0.26                 &0.58 & 11  &1050& 50  &      
    pS:pVME:oFT &\cite{lodge1983a} \\
$2.92 \cdot 10^{-7}$ & 0.368                & 0.59 & 10  & 422 &1300 &      
    pS:pVME:oFT            &\cite{lodge1986a} \\
$2.06 \cdot 10^{-7}$ & 0.439                & 0.60 & 9.8 &1050 &1300 &      
    pS:pVME:oFT            &\cite{lodge1986a} \\
\end{tabular}
\end{ruledtabular}
\end{table*}

\begin{table*}
\begin{ruledtabular}
\begin{tabular}{|r|r|r|r|r|l|l|r|} 
$2.97 \cdot 10^{-7}$ & 0.280                & 0.63 & 2.6 & 379 &1300 &      
 $f=3$ pS:pVME:oFT            &\cite{lodge1986a} \\
$1.84 \cdot 10^{-7}$ & 0.435                & 0.65 & 5.0 &1190 &1300 &      
 $f=3$ pS:pVME:oFT            &\cite{lodge1986a} \\
$8.46 \cdot 10^{-7}$ & 0.085                & 0.71 & 3.7 & 55  & 140 &      
 $f=12$ pS:pVME:oFT            &\cite{lodge1987a} \\
$3.01 \cdot 10^{-7}$ & 0.161                & 0.67 & 3.6 & 467 & 140 &      
 $f=12$ pS:pVME:oFT            &\cite{lodge1987a} \\
$1.89 \cdot 10^{-7}$ & 0.184                & 0.66 & 1.7 &1110 & 140 &      
 $f=12$ pS:pVME:oFT            &\cite{lodge1987a} \\
$1.60 \cdot 10^{-7}$ & 0.196                & 0.65 & 2.0 &1690 & 140 &      
 $f=12$ pS:pVME:oFT            &\cite{lodge1987a} \\
$2.09 \cdot 10^{-7}$ & 0.49                 &0.56 &5.8  &1690& 1300  &      
 $f=12$   pS:pVME:oFT &\cite{lodge1989a} \\
$2.52 \cdot 10^{-7}$ & 0.48                 &0.55 &4.7  &1110& 1300  &      
 $f=12$   pS:pVME:oFT &\cite{lodge1989a} \\
$4.07 \cdot 10^{-7}$ & 0.49                 &0.48 &4.9  &467 & 1300  &      
 $f=12$   pS:pVME:oFT &\cite{lodge1989a} \\
$(9.41 \cdot 10^{-7})$ & (0.35)                 &(0.34) &3.9  & 55 & 1300  &      
 $f=12$   pS:pVME:oFT &\cite{lodge1989a} \\
$1.44 \cdot 10^{-7}$ & 0.18                 &0.63 &4.7  &1190& 140   &      
 $f=3$   pS:pVME:oFT &\cite{lodge1989a} \\
$1.58 \cdot 10^{-7}$ & 0.25                 &0.65 &14.6 &1190& 630   &      
 $f=3$   pS:pVME:oFT &\cite{lodge1989a} \\
$2.18 \cdot 10^{-7}$ & 0.612                &0.55 &18   &1190&1300   &      
 $f=3$   pS:pVME:oFT &\cite{lodge1989a} \\
$2.69 \cdot 10^{-7}$ & 0.162                &0.63 &3.7  & 379& 140   &      
 $f=3$   pS:pVME:oFT &\cite{lodge1989a} \\
$2.57 \cdot 10^{-7}$ & 0.167                &0.68 &7.5  & 379& 630   &      
 $f=3$   pS:pVME:oFT &\cite{lodge1989a} \\
$2.87 \cdot 10^{-7}$ & 0.288                &0.61 &12   & 379&1300   &      
 $f=3$   pS:pVME:oFT &\cite{lodge1989a} \\
$5.27 \cdot 10^{-7}$ & 0.061                &0.74 &5.9  & 50 & 110   &      
         pS:pVME:oFT &\cite{martin1984a,martin1986a} \\
$3.65 \cdot 10^{-7}$ & 0.054                &0.80 &1.6  &100 & 110   &      
         pS:pVME:oFT &\cite{martin1984a,martin1986a} \\
$2.48 \cdot 10^{-7}$ & 0.145                &0.66 &3.6  &420 & 110   &      
         pS:pVME:oFT &\cite{martin1984a,martin1986a} \\
$1.49 \cdot 10^{-7}$ & 0.116                &0.71 &4.6  &900 & 110   &      
         pS:pVME:oFT &\cite{martin1984a,martin1986a} \\
$1.56 \cdot 10^{-7}$ & 0.32               &0.77 & 0.9 & 342&43.9 &      
 pMMA :pS:thiop& \cite{nemoto1985a} \\
$1.43 \cdot 10^{-7}$ & 0.54               &0.54 & 1.3  & 342& 186 &      
 pMMA :pS:thiop& \cite{nemoto1985a} \\
$1.18 \cdot 10^{-7}$ & 0.79                &0.68 & 6.1 & 342& 775 &      
 pMMA :pS:thiop& \cite{nemoto1985a} \\
$1.23 \cdot 10^{-7}$ & 1.03                 &0.67 & 8.4& 342&8420 &      
 pMMA :pS:thiop& \cite{nemoto1985a} \\
$1.51 \cdot 10^{-5}$ & $3.21 \cdot 10^{-3}$ & (1)  & 5.5& M  & \tablenotemark[1]  &      
 styrene:pS:CCl$_{4}$& \cite{pinder1990a} \\
$2.53 \cdot 10^{-5}$ & $3.34  \cdot 10^{-3}$ & (1)  & 3.0& M  & \tablenotemark[1]  &      
 styrene:pS:C$_{6}$H$_{12}$& \cite{pinder1990a} \\
$1.51 \cdot 10^{-5}$ & $3.15 \cdot 10^{-3}$ & (1)  & 5.8& M  &\tablenotemark[1]  &      
 styrene:pS:CCl$_{4}$& \cite{pinder1990a} \\
$4.74 \cdot 10^{-6}$ & $5.36 \cdot 10^{-3}$ & (1)  & 5.0&0.58&\tablenotemark[1]  &      
      pS:pS:CCl$_{4}$& \cite{pinder1990a} \\
$3.31 \cdot 10^{-6}$ & $6.53 \cdot 10^{-3}$ & (1)  & 4.8&1.2 &\tablenotemark[1]  &      
      pS:pS:CCl$_{4}$& \cite{pinder1990a} \\
$2.16 \cdot 10^{-6}$ & $8.08 \cdot 10^{-3}$ & (1)  & 4.1&2.47&\tablenotemark[1]  &      
      pS:pS:CCl$_{4}$& \cite{pinder1990a} \\
$5.13 \cdot 10^{-9}$ & 0.033     &0.69 & 4.3   & 33.6& 32 &      
    PPO:PPO:PPO &\cite{smith1986a} \\
$7.86 \cdot 10^{-11}$ & 5.23           &0.79 &10.8 &255 & 93  &      
    pS:pS:pS &\cite{tead1988a} \\
$9.17 \cdot 10^{-11}$ & 6.29           &0.51 & 23  &255 & 250 &      
    pS:pS:pS &\cite{tead1988a} \\
$1.24 \cdot 10^{-11}$ & 4.81           &0.47 & 8.7 &255 &20 000&      
    pS:pS:pS &\cite{tead1988a} \\
$1.93 \cdot 10^{-7}$ & $7.3 \cdot 10^{-2}$  &0.84 &10.3 &433 &310   &      
    dex:pVP:water &\cite{tinland1994a} \\
$7.1  \cdot 10^{-7}$ & 0.20                 &0.57 &3.1  & 65 &1300  &      
    pS:pVME:oFT       &\cite{wheeler1987a} \\
$5.42 \cdot 10^{-7}$ & 0.21                 &0.66 &5.4  &179 &1300  &      
    pS:pVME:oFT       &\cite{wheeler1987a} \\
$2.87 \cdot 10^{-7}$ & 0.35                 &0.60 &10.2 &422 &1300  &      
    pS:pVME:oFT       &\cite{wheeler1987a} \\
$2.19 \cdot 10^{-7}$ & 0.47                 &0.59 &10.1 &1050&1300  &      
    pS:pVME:oFT       &\cite{wheeler1987a} \\
$6.30 \cdot 10^{-7}$ & 0.101                &0.62 & 3.7 & 65 & 140  &      
    pS:pVME:oFT       &\cite{wheeler1989a} \\
$3.72 \cdot 10^{-7}$ & 0.150                &0.62 & 2.3 &179 & 140  &      
    pS:pVME:oFT       &\cite{wheeler1989a} \\
$2.34 \cdot 10^{-7}$ & 0.165                &0.63 & 4.6 &422 & 140  &      
    pS:pVME:oFT       &\cite{wheeler1989a} \\
$1.64 \cdot 10^{-7}$ & 0.194                &0.62 & 5.5 &1050& 140  &      
    pS:pVME:oFT       &\cite{wheeler1989a} \\
$6.32 \cdot 10^{-7}$ & 0.135                &0.58 & 4.5 & 65 & 630  &      
    pS:pVME:oFT       &\cite{wheeler1989a} \\
$3.99 \cdot 10^{-7}$ & 0.227                &0.57 & 4.1 &179 & 630  &      
    pS:pVME:oFT       &\cite{wheeler1989a} \\
$2.60 \cdot 10^{-7}$ & 0.29                &0.58 & 8.2 &422 & 630  &      
    pS:pVME:oFT       &\cite{wheeler1989a} \\
$1.80 \cdot 10^{-7}$ & 0.34                &0.58 &12.0 &1050& 630  &      
    pS:pVME:oFT       &\cite{wheeler1989a} \\
\end{tabular}
\end{ruledtabular}
\tablenotetext[1]{Various matrix molecular weights with $M \gg P$.}
%\tablenotetext[2]{Matrix concentration in volume fraction.}
\end{table*}

\begin{table*}
\caption{ \label{table4} Concentration and molecular weight dependences of $D_{s}$ 
and $D_{p}$ for molecular weight $P$ probes in solutions of molecular weight 
$M$ matrix polymers (for $D_{s}$, one has $P = M$) at concentration $c$.  The 
fits are to stretched exponentials $D_{o} P^{-a} \exp(-\alpha c^{\nu} 
P^{\gamma} M^{\delta})$, with the percent root-mean-square fractional fit error 
\%R, the material, and the reference.  Molecular weights are in kDa; 
concentrations except as noted are in g/L.  Square brackets  ``{[ $\cdots$ ]}'' 
denote parameters that were fixed rather than floated.  Abbreviations as in other
Tables.}
\begin{ruledtabular} 
\begin{tabular}{|r|r|r|r|r|r|r|r|l|l|l|r|} 
$D_{o}$ & $a$ & $\alpha$ & $\nu$ & $\gamma$& $\delta$ & \%R
& $P$ & $M$ & $c$ & System& Refs. \\ 
$6.27 \times 10^{-5}$    & [0.5]    & $5.96 \cdot 10^{-5} $&0.93 & 0.61 & [0]  
& 9 & \tablenotemark[2] & \tablenotemark[6]  & (2)  &  PEO:H$_{2}$O  &\cite{brown1983a} \\ 
$7.99 \times 10^{-5}$& 0.50 & $7.38 \cdot 10^{-4}$&0.58 & 0.46 & [0]  & 17 &
(2)  & \tablenotemark[6]  & \tablenotemark[2]  &  pS:CCl$_{4}$ &\cite{callaghan1981a} \\
$3.41 \times 10^{-4}$& 0.55& $2.10 \cdot 10^{-3} $&0.64 & 0.36 & [0]  & 7 &
\tablenotemark[2]  & \tablenotemark[6]  & \tablenotemark[2]  &  pS:C$_{6}$D$_{6}$ &\cite{callaghan1981a} \\
$2.94 \times 10^{-4}$  & 0.50    & $1.53 \cdot 10^{-2} $&0.52 & 0.25 & [0]  
& 24 & \tablenotemark[2] & \tablenotemark[6]  & \tablenotemark[2]  &  pS:cp  &\cite{deschamps1986a} \\
$7.46 \times 10^{-4}$  & 0.501    & $6.17 \cdot 10^{-3} $&0.48 & 0.33 & 
[0]  & &
\tablenotemark[2] & \tablenotemark[6]  & \tablenotemark[2]  &  pDMS:tol  &\cite{giebel1993a} \\
$1.16 \times 10^{-4}$    & [0.5]   & $7.55 \cdot 10^{-7} $&0.95 & 0.86 & [0]  
& 37 & \tablenotemark[2] & \tablenotemark[6]  & \tablenotemark[2]  &  pS:C$_{6}$H$_{6}$  &\cite{leger1981a} \\
$1.67 \times 10^{-4}$  & [0.5]    & $5.08 \cdot 10^{-4} $&0.75 & 0.48 
& [0]  & 17 &
\tablenotemark[2] & \tablenotemark[6]  & \tablenotemark[2]  &  pB:CCl$_{4}$  &\cite{vonmeerwall1983a} \\
$1.87 \times 10^{-4}$  & [0.5]    & $1.93 \cdot 10^{-3} $&0.91 & 0.24 & 
[0]  & 51 &
\tablenotemark[2] & \tablenotemark[6]  & \tablenotemark[2]  &  pS:THF  &\cite{wesson1984a} \\
$9.68 \times 10^{-4}$  & 0.65    & $7.37 \cdot 10^{-4} $&0.68 & 0.42 & 
[0]  & 5.7 &
193-800 & \tablenotemark[6]  & \tablenotemark[2]  &  pI:CCl$_{4}$  &\cite{vonmeerwall1984a} \\
0.891                & [0]    & $5.86 \cdot 10^{-4} $&0.43 & 0.43 & [0]  & 24 &
8000 & \tablenotemark[2]  & \tablenotemark[2]  &  pS:pMMA:tol  &\cite{brown1988a} \\
1.025                & [0]    & $1.53 \cdot 10^{-3} $&0.95 &-0.01 &0.236 &4.8 &
\tablenotemark[2]  & \tablenotemark[2]  & \tablenotemark[2]  &  PEO:dex:H$_{2}$O&\cite{brown1983b} \\
$7.46 \cdot 10^{-6}$ &0.36   & $2.84 \cdot 10^{-4} $&1.15 &0.26 &[0]   &34  &
\tablenotemark[2]  & 60   & \tablenotemark[2]  &  pS:pVME:oFT      &\cite{hanley1985a} \\
$8.91 \cdot 10^{-6}$ &0.34   & $6.80 \cdot 10^{-4} $&1.03 &0.25 &[0]   &24  &
\tablenotemark[5]  & 60   & \tablenotemark[2]  &  pS:pVME:oFT      &\cite{hanley1985a} \\
$3.34 \cdot 10^{-4}$ &0.57   & $1.82 \cdot 10^{-5} $&0.99 &0.139&0.287 &6.4 &
\tablenotemark[2]  & \tablenotemark[2]  & \tablenotemark[2]  &  pS:pMMA:tol      &\cite{kent1992a} \\
$1.86 \cdot 10^{-4}$ &0.52   & $4.45 \cdot 10^{-5} $&0.69 &0.30 &0.33  &17  &
\tablenotemark[2]  & \tablenotemark[2]  & \tablenotemark[5]  &  pS:pS:tol      &\cite{kim1986a} \\
$8.5  \cdot 10^{-4}$ & 0.61 &$2.99\cdot 10^{-2}$ & 0.61  &0.19 &[0]   & 12  &
\tablenotemark[2]  & 1300   & \tablenotemark[2]  &  $f=2$ pS:pVME:oFT      &\cite{lodge1986a} \\
$3.0  \cdot 10^{-4}$ & 0.54 &$1.07\cdot 10^{-3}$ & 0.66  &0.42 &[0]   & 6.6 &
\tablenotemark[2]  & 1300   & \tablenotemark[2]  &  $f=3$ pS:pVME:oFT      &\cite{lodge1986a} \\
$1.84 \cdot 10^{-4}$ & 0.50 &$1.77\cdot 10^{-2}$ & 0.68  &0.16 &[0]   & 12  &
\tablenotemark[2]  & 140    & \tablenotemark[2]  &  $f=12$ pS:pVME:oFT      &\cite{lodge1987a} \\
$2.20 \cdot 10^{-4}$ & 0.52 &$5.09\cdot 10^{-4}$ & 0.61  &0.16 & 0.32 & 32  &
\tablenotemark[2]  & \tablenotemark[2]    & \tablenotemark[2]  &  $f=3$ pS:pVME:oFT      &\cite{lodge1989a} \\
$2.93 \cdot 10^{-4}$ & 0.53 &$2.22\cdot 10^{-3}$ & 0.66  &0.15 & 0.19 & 15&
\tablenotemark[2]  & \tablenotemark[2]    & \tablenotemark[2]  &  $f=12$ pS:pVME:oFT      &\cite{lodge1989a} \\
$6.51 \cdot 10^{-4}$ & 0.67 &$9.49\cdot 10^{-3}$ & 0.86  &0.115& [0]  &     &
\tablenotemark[2]  & 132    & \tablenotemark[2]  &  pS:pVME:oFT      &\cite{martin1984a,martin1986a}\\
$6.67 \cdot 10^{-5}$  & [0.5] & $4.5  \cdot 10^{-3} $&0.68 & [0.3]&0.024 &20  &
342  & \tablenotemark[1]  & \tablenotemark[1]  & pMMA:pS:thiop&\cite{nemoto1985a} \\                     2
$7.9  \cdot 10^{-5}$  & [0.52] & 0.95                 &[0]  & [0.16]&0.028 &  
  & \tablenotemark[2] & \tablenotemark[2]  & \tablenotemark[2]  &   pS:pS:DBP       &\cite{nemoto1989a} \\
$7.64 \cdot 10^{-6}$ &  0.56  & $7.61 \cdot 10^{-4} $&  \tablenotemark[1] & 0.30 & 0.15 & 13 &
\tablenotemark[2]  & \tablenotemark[2]  & 130  &      pS:pS:DBP &\cite{nemoto1990a} \\
$7.89 \cdot 10^{-6}$ &  0.63  & $1.35 \cdot 10^{-3} $&  \tablenotemark[1] & 0.28 & 0.11 & 12 &
\tablenotemark[2]  & \tablenotemark[2]  & 180  &      pS:pS:DBP &\cite{nemoto1990a} \\
$8.41 \cdot 10^{-6}$ &  0.64  & $2.31 \cdot 10^{-4} $&0.74 & 0.26 & 0.29 & 31 &
\tablenotemark[2]  & \tablenotemark[2]  & \tablenotemark[2]  &      pS:pS:DBP &\cite{nemoto1990a} \\
$1.65 \cdot 10^{-4}$ &  0.52  & $1.56 \cdot 10^{-6} $&0.79 & 0.285& 0.501& 23 &
\tablenotemark[1]  & \tablenotemark[1]  & \tablenotemark[2]  & pS:pMMA:tol&\cite{numasawa1986a} \\
$3.98 \cdot 10^{-4}$ &  0.68  & $1.00 \cdot 10^{-3} $&[1.0]& 0.26 & [0]  & 7.9&
\tablenotemark[3]  & \tablenotemark[1]  & \tablenotemark[2]  & pS:pS:CCl$_{4}$&\cite{pinder1990a} \\
$1.31 \cdot 10^{-4}$ &  0.52  & $7.56 \cdot 10^{-4} $&[1.0]& 0.30 & [0]  & 5.1&
\tablenotemark[4]  & \tablenotemark[1]  & \tablenotemark[2]  & pS:pS:CCl$_{4}$&\cite{pinder1990a} \\
$2.01 \cdot 10^{-4}$ &  0.51  & $1.03 \cdot 10^{-2} $& 0.64 & 0.26 & [0]  & 14&
\tablenotemark[3]  & 1300  & \tablenotemark[2]  & pS:pVME:oFT &\cite{wheeler1987a} \\
$4.99\cdot 10^{-3}$ &  0.39  & $8.33 \cdot 10^{-4} $& 0.54 & 0.25 & 0.22 & 25&
\tablenotemark[3]  & \tablenotemark[2]  & \tablenotemark[2]  & pS:pVME:oFT &\cite{wheeler1989a} \\
\end{tabular}
\end{ruledtabular}
\tablenotetext[1]{ Various, with $M/P \geq 10$}
\tablenotetext[2]{ Various, see text.}
\tablenotetext[3]{ All four probes, see text.}
\tablenotetext[4]{ Excluding styrene monomer, see text.}
\tablenotetext[5]{ Not all data points, see text.}
\tablenotetext[6]{ $P = M$, self diffusion.}
\end{table*}

\begin{table*}
\caption{ \label{table3}  Molecular weight
dependence of the self and probe diffusion 
coefficients $D_{s}$ and
$D_{p}$ for molecular weight $P$ probes in solutions of
matrix polymers at a fixed concentration $c$.  
The fits are to stretched exponentials $D_{o} 
\exp(-\alpha M^{\gamma})$ in matrix molecular weight $M$.
The Table gives the best-fit parameters, the percent 
root-mean-square fractional fit error \%RMS, the system, and the reference.  
Square brackets  ``{[ $\cdots$ ]}'' denote parameters that were fixed rather 
than floated.  
Abbreviations as per previous Tables, and DBP--dibutylphthalate.}
\begin{ruledtabular} 
\begin{tabular}{|r|r|r|r|r|l|l|r|} 
$D_{o}$ & $\alpha$& $\gamma$ & \%RMS & $P$(kDa) & $c$ (g/L) &System & Refs. \\
$8.54 \cdot 10^{-8}$ & 0.64                 &0.55 &17   &1800& \tablenotemark[1] &      
    pS:pS:tol &\cite{kim1986a} \\
$2492 P^{-0.5} $ & $1.37 \cdot 10^{-2}$   &0.47 & 4.0 & $P$ & 130 &      
    pS:pS:dbp &\cite{nemoto1991a} \\
$1771 P^{-0.5} $ & $1.07 \cdot 10^{-2}$   & [0.5] & 6.9 & $P$ & 180 &      
    pS:pS:dbp &\cite{nemoto1991a} \\
\end{tabular}
\end{ruledtabular}
\tablenotetext[1]{Various, see text.}
\end{table*}

\begin{table*}
\caption{ \label{table5}
Concentration and molecular weight dependences of
the probe radius of gyration $R_{g}$
for molecular weight $P$ probes in solutions of 
molecular weight $M$ matrix polymers as functions of matrix concentration 
$c$.  The fits are to stretched exponentials 
$R_{g0}  \exp(-\alpha c^{\nu})$,
with the percent root-mean-square 
fractional fit error \%RMS, the materials, and the reference.  
Materials include EB--ethyl benzoate, 
PMMA--polymethylmethacrylate,
pS--polystyrene.}  
\begin{ruledtabular} 
\begin{tabular}{|r|r|r|r|r|r|l|r|} \hline 
$R_{g0} (\AA) $ &
$\alpha$ & $\nu$ & \%RMS & $P$(kDa) & $M$(kDa)  & System & Refs. \\ 
 \hline 
390 & $4.11 \cdot 10^{-3}$ & 0.99 & 2.2 & 930 & 1300 & pS: pMMA:  
EB&\cite{kent1992a} \\ 
397 & $8.06 \cdot 10^{-3}$ & 0.77 & 1.1 & 930 & 70
& pS: pMMA:  EB&\cite{kent1992a} \\ 
$[395]$ & $4.49 \cdot 10^{-3}$ & 0.64 &
1.9 & 930 & 7 & pS: pMMA: EB &\cite{kent1992a} \\
\hline
\end{tabular}
\end{ruledtabular}
\end{table*}

\end{document}